\documentclass[tightenlines,superscriptaddress,eqsecnum,floats,nofootinbib,twocolumn]
{revtex4-1}
\usepackage{amssymb,amsmath,amssymb,amsfonts,amsthm,stmaryrd,mathrsfs,physics,bbm}
\usepackage{graphicx}
\usepackage[T1]{fontenc}
\usepackage[normalem]{ulem}
\usepackage{enumerate} 
\usepackage{xcolor}

\usepackage{hyperref} 

\newtheorem{pro}{Proposition}

\newcommand{\sgn}{\mathrm{sgn}}
\newcommand{\sut}{\mathrm{SU(2)}}
\newcommand{\sltc}{\mathrm{SL}(2,\mathbb C)}

\newcommand{\Ks}{{\mathcal K}}
\newcommand{\arccosh}{\mathrm{arccosh}}
\newcommand{\se}{{\mathfrak{e}}}
\newcommand{\sres}{S_{\rm res}}
\newcommand{\sresc}{\mathring{S}_{\rm res}}

\begin{document}

\title{Spin foam amplitude of the black-to-white hole transition}

\author{Muxin Han}
\affiliation{Department of Physics, Florida Atlantic University, 777 Glades Road, Boca Raton, FL 33431-0991, USA}
\affiliation{Department Physik, Institut f\"ur Quantengravitation, Theoretische Physik III, Friedrich-Alexander Universit\"at Erlangen-N\"urnberg, Staudtstr. 7/B2, 91058 Erlangen, Germany}

\author{Dongxue Qu}
\affiliation{Perimeter Institute for Theoretical Physics, 31 Caroline St N, N2L 2Y5 Waterloo, ON, Canada}

\author{Cong Zhang}
\thanks{Corresponding author}
\email{zhang.cong@mail.bnu.edu.cn}
\affiliation{Department Physik, Institut f\"ur Quantengravitation, Theoretische Physik III, Friedrich-Alexander Universit\"at Erlangen-N\"urnberg, Staudtstr. 7/B2, 91058 Erlangen, Germany}

\begin{abstract}
It has been conjectured that  quantum gravity effects may cause the black-to-white hole transition due to quantum tunneling. The transition amplitude of this process is explored within the framework of the spin foam model  on  a 2-complex containing 56 vertices. We develop a systematic way to construct the bulk triangulation from the boundary triangulation to obtain the 2-complex.
By using Thiemann's complexifier coherent state as the boundary state to resemble the semiclassical geometry, we introduce a procedure to calculate the parameters labeling the coherent state from the continuous curved geometry. Considering that triad fields of different orientations, i.e., $e_i^a$ and $-e_i^a$, give the same intrinsic geometry of the boundary, we creatively adopt  the boundary state as a superposition of the coherent states associated with both orientations.  We employ the method of complex critical point to numerically compute the transition amplitude.  Despite the numerical results, it is interesting to note that the transition amplitude is dominated by the terms allowing the change in orientation. This  suggests that the black-to-white hole transition should be accompanied by the quantum tunneling process of a change in orientation. 
\end{abstract}


\maketitle


\section{Introduction}\label{sec:intro}
The black hole (BH), one of the most mysterious objects in our Universe, has drawn considerable attention in modern physics \cite{Maldacena:1996ky,Wald:1999vt,Page:2004xp,Dittrich:2005sy,Unruh:2017uaw,EventHorizonTelescope:2019dse,Gambini:2022hxr,Lan:2023cvz}. 
Along the way of our exploration on BHs, the Oppenheimer-Snyder model emerges as an important influence in our understanding of the formation of BHs \cite{PhysRev.56.455}.
{The Oppenheimer-Snyder model is a solution to the Einstein equation. It describes the collapse of a spherically symmetric, pressureless and homogeneous dust ball  into a BH. During the collapse, the geometry inside the dust ball is the Friedmann–Lema\^itre–Robertson–Walker cosmological metric, and the geometry outside is the Schwarzschild one. Consequently, the spacetime contains two singularities: the first one is the cosmological singularity emerging at the final stage of the matter collapse and the second one is the Schwarzschild singularity existing in the vacuum region outside the collapsing dust  \cite{Joshi:2011rlc}.
}
While this model plays a crucial role in transitioning BHs from abstract mathematical concepts to tangible physical realities, it faces challenges, particularly concerning the presence of the two singularities. Indeed, the existence of singularities is a common challenge within classical general relativity (GR), indicating the invalidity of the classical theory in the Planckian curvature region. This prompts the exploration of quantum gravity (QG) to refine our understanding of gravity in these high-curvature regions  \cite{polchinski1998string,Ambjorn:2001cv,Ashtekar:2004eh,Han:2005km,Thiemann:2007pyv,Rovelli:2014ssa,Zhang:2015bxa,Loll:2019rdj,Surya:2019ndm,Gambini:2013ooa,Ashtekar:2018lag,Song:2020arr,Zhang:2020qxw,Han:2020uhb,Zhang:2021wex,Zhang:2021xoa,Husain:2022gwp,Lewandowski:2022zce,Gan:2022mle,Giesel:2022rxi,Giesel:2023tsj}. 

\begin{figure}
\centering
\includegraphics[width=0.2\textwidth]{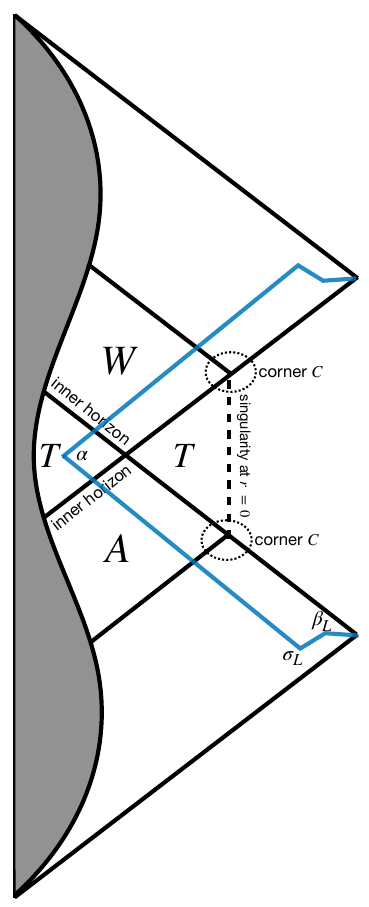}
\caption{The Penrose diagram of the quantum Oppenheimer-Snyder model with large $M$. $T$ is the transition region connecting the WH $W$ to the BH $A$. The blue dashed line is the cut and glue surface for constructing the spacetime with a single asymptotic region.  
}\label{fig:penrose1}
\end{figure}

Among various quantum Oppenheimer-Snyder models \cite{Tippett:2011hz,Piechocki:2020bfo,BenAchour:2020bdt,Lewandowski:2022zce,PhysRevD.107.126006,Bobula:2023kbo}, recent attention has been directed toward the one that incorporates loop quantum gravity (LQG) effects \cite{Lewandowski:2022zce}.  { This model is constructed based on the fact that the metric outside can be derived without applying the Einstein equation. As shown in \cite{poisson2004relativist}, once the interior metric describing the collapsing dust ball is given, the exterior metric can be obtained by assuming spherical symmetry and stationarity for the outside, and applying Israel junction conditions at the boundary surface between the inside and outside. Due to this fact, the quantum Oppenheimer-Snyder model  in \cite{Lewandowski:2022zce} considers the interior metric as the one in  loop quantum cosmology instead of the classical cosmology, and derives the corresponding spherically symmetric and stationary exterior metric using the junction condition. The metric reads}
\begin{equation}\label{eq:metric}
\begin{aligned}
\dd s^2=-f(r)\dd t^2+f(r)^{-1}\dd r^2+r^2\dd\Omega^2,
\end{aligned}
\end{equation}
where $f(r)$ is given by
\begin{equation}\label{eq:metricf}
f(r)=1-\frac{2G M}{r}+\frac{\alpha G^2M^2}{r^4},\ \alpha=2\sqrt{3}\beta^3 \kappa \hbar,
\end{equation}
with the Barbero-Immirzi parameter  $\beta$ and $\kappa=8\pi G$ (see e.g., \cite{Marto:2013soa,Parvizi:2021ekr,Husain:2022gwp,Bobula:2023kbo,Giesel:2023hys,Lin:2024flv} for similar works). 
{ As the dust ball follows the dynamics of loop quantum cosmology, which predicts a bounce when the density of the collapsing ball reaches the Planck scale, the cosmological singularity is prevented. This will influence the nature of the Schwarzschild singularity outside the dust ball. The Penrose diagram of the Oppenheimer-Snyder model for a large BH mass $M$ is presented in Fig. \ref{fig:penrose1}. As shown there, the classical singularity is replace by a transition region $T$ connecting a white hole (WH) with the BH. (see \cite{Lewandowski:2022zce} for discussion on Penrose diagrams for different value of $M$.) }

While the spacetime illustrated in Fig. \eqref{fig:penrose1} offers distinct advantages, based on which observational effects have been studied \cite{Yang:2022btw,Gingrich:2023fxu,Zhang:2023okw,Shao:2023qlt,Ye:2023qks,Chen:2023bao}, it is not without debates. For instance, the existence of the inner horizon implies that the spacetime could be unstable under perturbation \cite{Cao:2023aco,Cao:2024oud}. { In addition, the spacetime is not singularity free. As shown in  Fig. \ref{fig:penrose1}, there is a timelike  singularity at $r=0$. 
These unresolved issues highlight the ongoing complexity of our understanding and indicate that we have not reached a final conclusion in exploring the quantum Oppenheimer-Snyder model. 
An aspect that we may have  overlooked is what happens at the corner $C$ (see Fig. \ref{fig:penrose1}) where the QG effects may play a significant role.
 As proposed in \cite{Han:2023wxg}, the QG effects at this corner could potentially lead to quantum transition from the BH horizon to the WH horizon (see, e.g., \cite{Haggard:2014rza,DAmbrosio:2020mut,Rignon-Bret:2021jch,Soltani:2021zmv} for previous investigation on this topic).
 }

{As proposed by \cite{Han:2023wxg}, we cut and
glue the spacetime of the quantum Oppenheimer-Snyder model as follows. First, draw a blue line as shown in Fig.   \eqref{fig:penrose1}. This blue line is symmetric under time reversal, and has null portion joining $\alpha$ and
$\beta_L$  and a spacelike portion joining $\beta_L$ to spacelike infinity along a constant Schwarzschild time. Next,  delete the entire spacetime region enclosed
within the blue line, and glue its two space-like portions.  This is can be done since they are both constant Schwarzschild time surfaces. The resulting spacetime,  as shown in Fig. \ref{fig:regionB}, exhibits a global structure with a single asymptotic region.}

The metric outside the collapsing dust in revised spacetime is locally the same as given in \eqref{eq:metric}, except for a region, called the $B$ region, where quantum transition takes place. { This fact means that the dynamics outside the $B$ region is governed by the classical equations of motion with quantum corrections, as proposed by the LQG model for symmetric spacetimes. A question then arises regarding the dynamics within the $B$ region: is it governed by quantum gravity? Notably, it is not viable to describe the dynamics within the $B$ region as corrections to
the classical equations of motion, since quantum tunnelling, like the one occurring in the $B$ region, is not analytical in $\hbar$ (see \cite{Rovelli:2024sjl} for more discussion). A fundamental way to describe the dynamics of a purely quantum process is using the transition amplitude, which can be done in the framework of the covariant LQG, i.e., 
 spin foam model \cite{Engle:2007uq,Engle:2007wy,Rovelli:2014ssa}.  Our task is to study the spin foam amplitude for 
for the quantum transition from the geometry in the past boundary of the $B$ region to the geometry of its future boundary. 
} 

Several challenges prevent the direct application of the spin foam model to the $B$ region embedded in classical spacetime. The first issue is that the spin foam amplitude is defined based on discrete graphs, which are usually dual to a triangulation. This thus needs us to triangulate the $B$ region at first.  We will construct a triangulation adapting to the symmetries of the $B$ region, namely its spherical symmetry and time reversal symmetry.  This triangulation results in a dual 2-complex, denoted by $\Ks$. The intersection of $\Ks$ with the boundary $\partial B$ of the $B$ region yields a graph $\gamma$, on which one can define the boundary Hilbert space $\mathcal H_{\gamma}$ of spin network states \cite{Ashtekar:2004eh,Thiemann:2007pyv}.  The spin foam amplitude $A$ associated with the 2-complex is a specific functional on $\mathcal H_\gamma$, namely $A:\mathcal H_\gamma\to \mathbb C$ \cite{Rovelli:2014ssa}. To make the Hilbert space $\mathcal H_\gamma$ well-defined, it is required that the boundary $\partial B$ should be spacelike. However, the initial definition of $\partial B$ in \cite{Han:2023wxg} is null. To resolve this issue, we need to choose a slightly larger region $B'$ in the spacetime, ensuring that $B'$ contains $B$ and that the boundary $\partial B'$ is spacelike. Indeed, all our calculations, including the triangulation aforementioned, are performed in $B'$ rather than $B$. Throughout the subsequent discussion, we will continue to use $B$ to denote this slightly larger region $B'$ for consistency. 

Another noteworthy concern is the mismatch between the amplitude $A$, defined on the Hilbert space $\mathcal H_\gamma$, and the classical geometry specifying the boundary condition on $\partial B$. {This mismatch needs us to encode the curved continuous geometry on the discrete boundary graph $\gamma$, capturing both its intrinsic and extrinsic
curvature to properly describe the physics.} To this end, we choose a coherent state in $\mathcal H_\gamma$ to resemble the classical geometry. Then, the value of $A$ on this coherent state is interpreted as the probability amplitudes which describes the transition, occurring in the $B$ region, between the classical boundary geometries.

{In our current computation, the position of the $\partial B$ is set manually. The position of   $\partial B$ is interpreted as the moment when the QG effects cannot be neglected. Importantly, within a well-defined dynamics,  this moment should be determined by the theory rather than being fixed arbitrarily. Thus, we would expect the spin foam model to determine the position of the $\partial B$ by itself. This can be implemented as follows.  }
We calculate the amplitude $A$ for a family of $B$ regions with different boundaries, and pick up the one which maximizes the amplitude. Since  the value of $A$ is related to the probability of the transition, the $B$ region we pick up correspond to the moment when the transition is most likely to occur.  The above discussion is valid for any observables which should be determined by theory itself, and explains how the spin foam model makes prediction. 

The paper is constructed as follows: In Sec. \ref{sec:constructmodel}, the spin foam model in the $B$
region is constructed, where we introduce the systematical method to yield the bulk triangulation from the boundary triangulation, define the boundary coherent state, and introduce the spin foam amplitude.  Sec. \ref{sec:boundarydata} describes how to calculate the boundary data of the boundary state from the curved geometry outside the $B$ region. In Sec. \ref{sec:criticalPointComplex}, the algorithm for computing the complex critical point is introduced. In Sec. \ref{sec:num}, we present the numerical results. Finally, this work is summarized in Sec. \ref{sec:conclusion}. To keep the flow of the discussion, the theoretical derivations forming the foundation of this work are presented in the appendices.
 
\section{Spinfoam model in the ${B}$ region}\label{sec:constructmodel}

\begin{figure}
\centering
\includegraphics[width=0.3\textwidth]{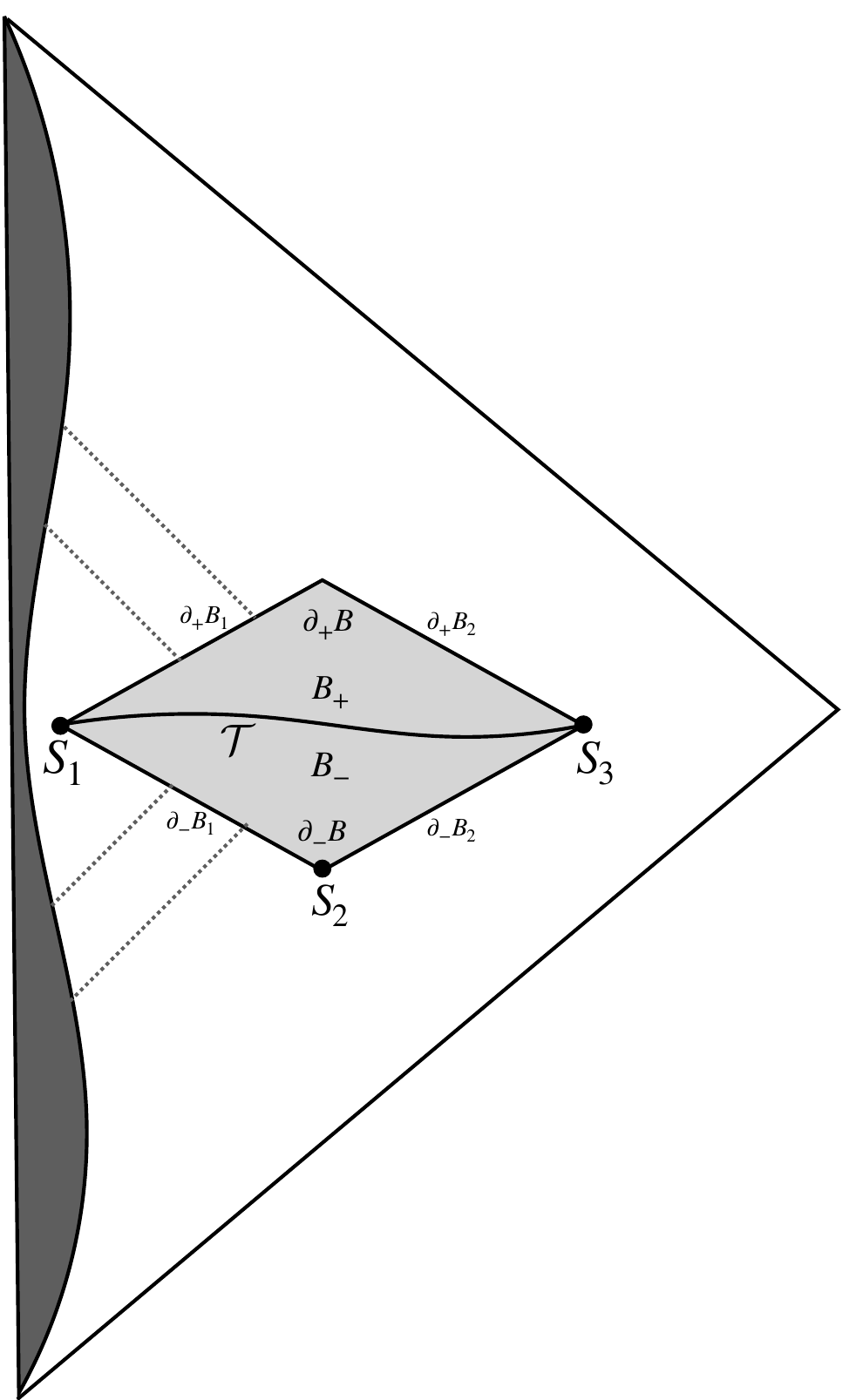}
\caption{The BH spacetime with the $B$ region. The gray shaded region represents the $B$ region with the boundaries $\partial B=\partial_{+}B\cup\partial_{-}B$. For our triangulation, we divide the boundary $\partial_-B$ into a union of two parts, $\partial_-B_1\cup\partial_-B_2$.}\label{fig:regionB}
\end{figure}

The spacetime structure proposed in \cite{Han:2023wxg} is presented  in Fig. \ref{fig:regionB}. The $B$ region is the gray shaded region in the spacetime. The metric outside the region is the same as the one in \eqref{eq:metric}. In the $B$ region, quantum  tunneling of black-to-white hole transition is {supposed to occur}, and the spin foam model is used to describe the dynamics. This section will focus on the $B$ region to construct the spin foam  model.

As mentioned in Sec. \ref{sec:intro}, the $B$ region has the time reversal symmetry. The time reversal symmetry is with respect to a hypersurface $\mathcal T$ within $B$. This hypersurface $\mathcal T$,  referred to as the transition surface, divides $B$ into a union of the past region $B_-$  and the future region $B_+$.  Let $\partial_-B$ and $\partial_+ B$ be the remaining boundary of $B_-$ and $B_+$ except $\mathcal T$, so that $\partial B=\partial_- B\cup\partial_+ B$. The time reversal symmetry gives a diffeomorphism between  $B_+$ and $B_-$ that preserves the transition surface $\mathcal T$ and maps between the boundary {data} on $\partial_+ B$ and $\partial_- B$ (see Fig. \ref{fig:regionB}).

The first step to construct the model is to triangulate the $B$ region, so that the spin foam amplitude can be defined on the 2-complex dual to the triangulation. { It has been noted that several previous works (e.g., \cite{DAmbrosio:2020mut,Soltani:2021zmv}) have proposed 2-complexes for this model. However, these 2-complexes are found to be degenerate, meaning they are not dual 2-complexes of a triangulation. }
In our work, we only consider the triangulation invariant under the time reversal diffeomorphism. To this end, we firstly triangulate the region $B_-$, and then apply the diffeomorphism of the time reversal symmetry to transform the triangulation from $B_-$ to $B_+$. 

{In the current work, we will apply a systematic way to construct the bulk triangulation for $B_-$ from triangulation for the boundary $\partial_-B_1$, where we divide $\partial_- B$ into the union $\partial_-B_1\cup \partial_-B_2$  as shown in Fig. \ref{fig:regionB}.   To illustrate how to construct the bulk triangulation from the boundary triangulation, in the next section, we will first introduce how the systematic way works, by using three simple examples closely related to the main problem of triangulating $B_-$.  
Then, in the subsequent section \ref{sec:triangulationBm}, we will directly apply the systematic way to construct the triangulation for $B_-$ from the boundary triangulation for $\partial_-B_1$. }

\subsection{Three examples}\label{sec:triangulationmethod}
For clarity, we use the word `point' to refer to the vertex of a triangulation in this paper. {Let us set aside the primary issue of triangulating $B_-$ and} focus on a spacetime region denoted as $R$, which is bounded by two spatial hypersurfaces $P$ and $P'$. The intersection of $P$ and $P'$ forms a 2-surface  $S$, as depicted in Fig. \ref{fig:triangulation1}.

\subsubsection{First example}

\begin{figure}
\centering
\includegraphics[width=0.4\textwidth]{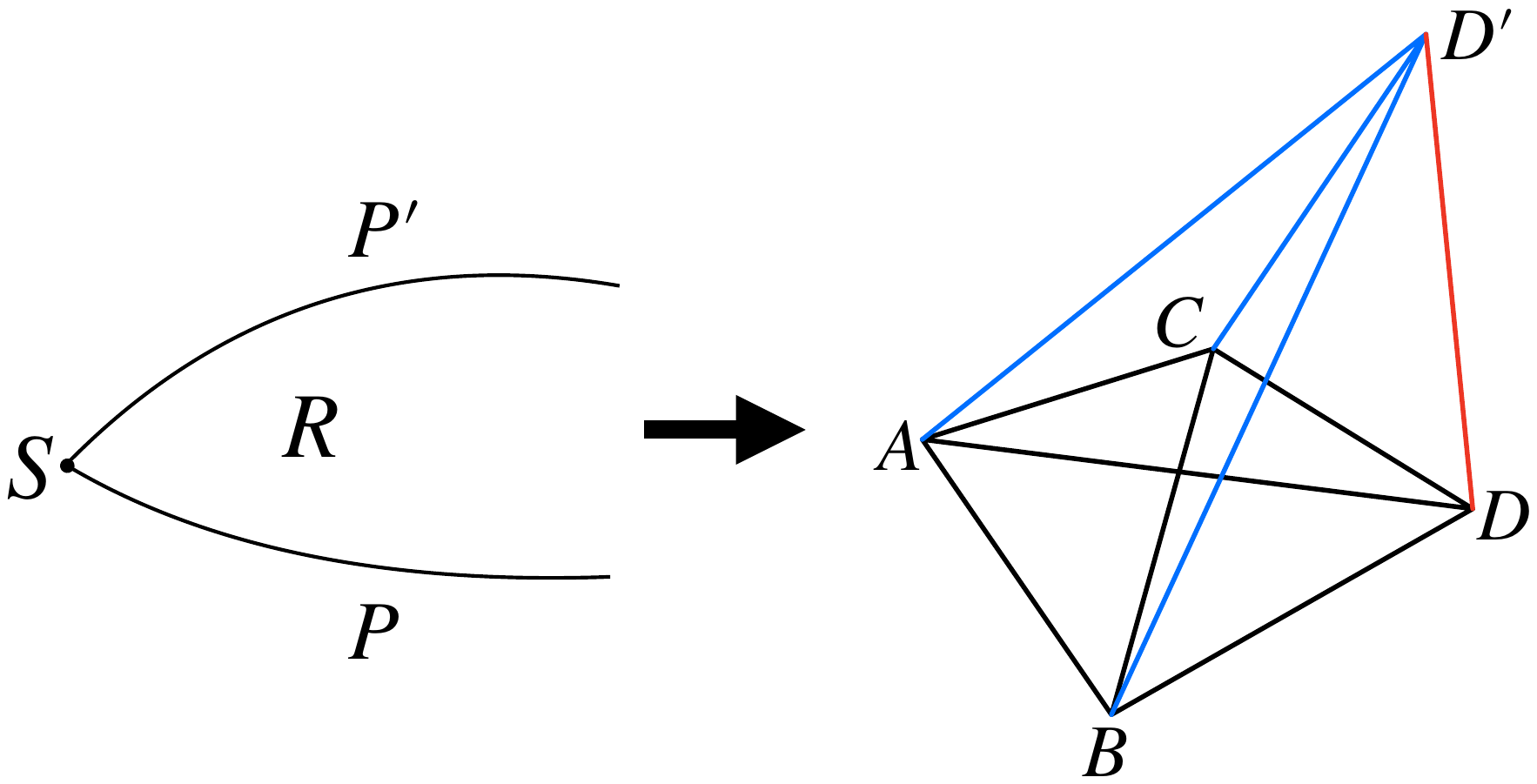}
\caption{Triangulation of the region $R$. The red line $DD'$ is the wordline of  the vertex $D$. The triangle $ABC$ lies in the 2-surface $S$. The initial tetrahedron $ABCD$ triangulates the hypersurface $P$, while the final tetrahedron $ABCD'$ results in a triangulation of the hypersurface $P'$. 
}\label{fig:triangulation1}
\end{figure}

In the first example, we consider a triangulation of the spatial surface $P$ using the tetrahedra $ABCD$, where the triangle $ABC$ lies in the 2-plane $S$,  as depicted in Fig. \ref{fig:triangulation1}. We call the  point $D$ as  the dynamical point, which moves from $P$ to $P'$ and eventually yield a new point $D'$ in $P'$. By connecting the point $D'$ with the points $A$, $B$, and $C$ respectively, we obtain the triangulation $ABCDD'$ of the region $R$. It is worth noting that the tetrahedron $ABCD'$ gives the triangulation of the hypersurface $P'$. We call this tetrahedron the final tetrahedron. Besides the final tetrahedron $ABCD'$,  the 4-simplex  contains four other boundary tetrahedra: the initial tetrahedron $ABCD$ and the tetrahedra $ACDD'$, $ABDD'$, and $BCDD'$. The last three tetrahedra  are viewed as the world volume of the triangles $ACD$, $ABD$ and $BCD$ respectively.

\subsubsection{Second example}
  
\begin{figure}
\centering
\includegraphics[width=0.4\textwidth]{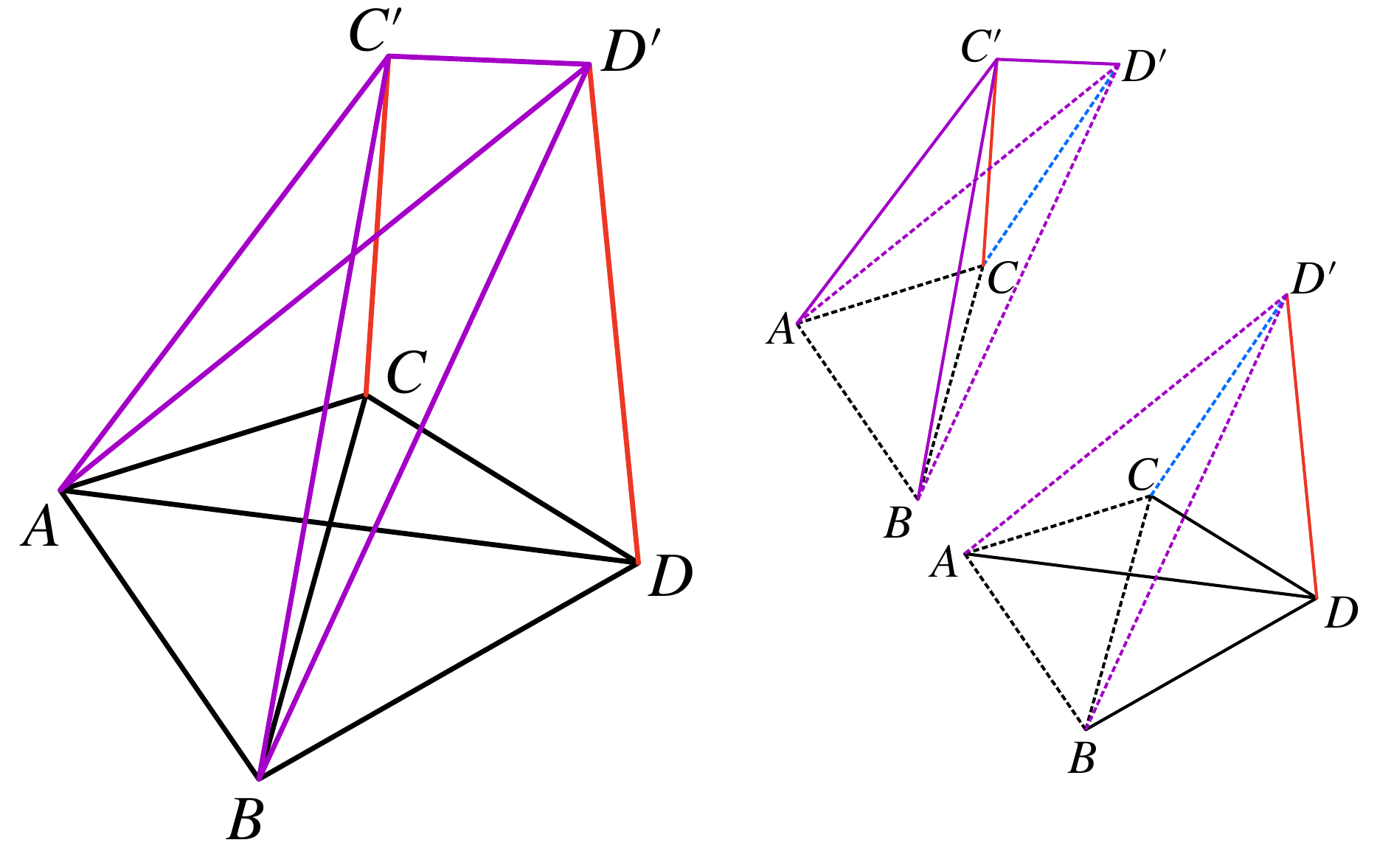}
\caption{The triangulation of the cell $ABCDD'C'$. The right side shows the 4-simplices triangulating the cell. The first 4-simplex $ABCDD'$ results from the evolution of the tetrahedron $ABCD$ with $D$ as the dynamical point. The second 4-simplex $ABCC'D'$ is generated by the evolution of the tetrahedron $ABCD'$  with $C$ treated as the dynamical point. 
}\label{fig:two_DV}
\end{figure}

In a more complicated example, we triangulate the spatial surface $P$ using the tetrahedron $ABCD$  where only the segment $AB$ lies in the 2-plane $S$. In this case, we have two dynamical points $C$ and $D$ as shown in Fig. \ref{fig:two_DV}.  Both $C$ and $D$ travel  from $P$ to $P'$, resulting in the points $C'$
 and $D'$. To establish the corresponding triangularization, we firstly examine the evolution of the tetrahedron $ABCD$ with $D$ treated as the dynamical point. This evolution yields a 4-simplex $ABCDD'$, as what we did in the previous case. Within this 4-simplex, the final tetrahedron $ABCD'$ represents the world volume of the triangle $ABC$. Subsequently, we explore the evolution of the tetrahedron $ABCD'$ with $C$ as the dynamical point, generating a new 4-simplex $ABCD'C'$. The cell $ABCDC'D'$ is then triangulated by the 4-simplices $ABCDD'$ and $ABCD'C'$.

\subsubsection{Third example}
\begin{figure}
\centering
\includegraphics[width=0.45\textwidth]{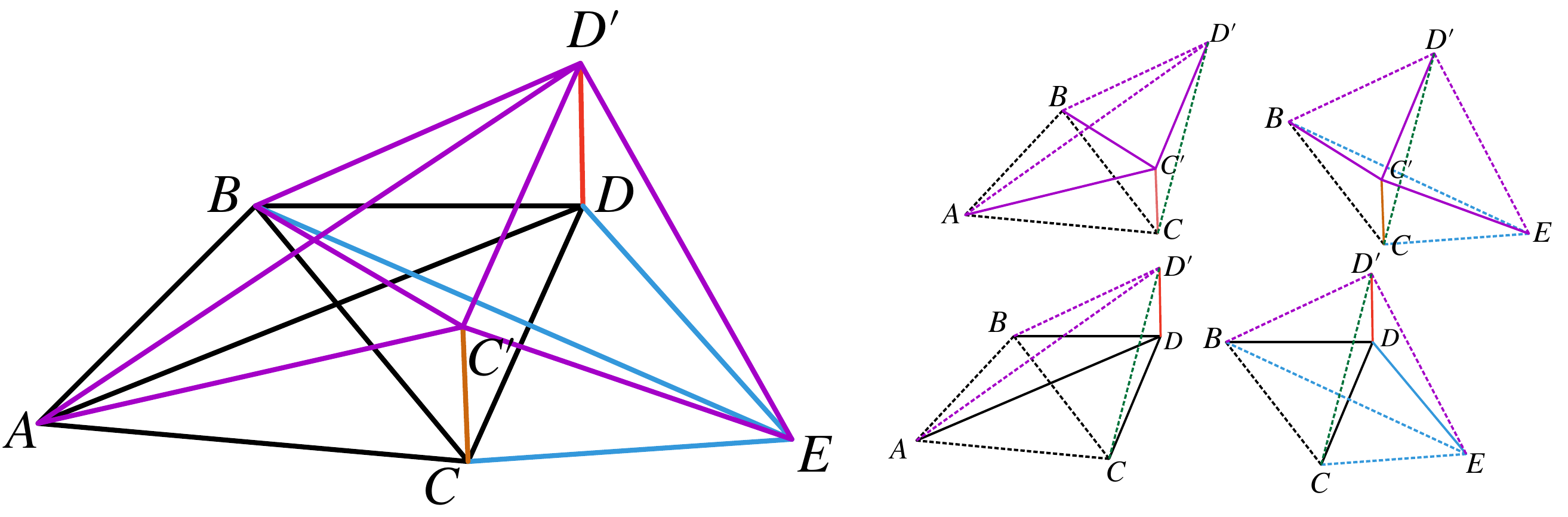}
\caption{The triangulation of the cell $ABCDED'C'$ (the left), where  $ABCDED'C'$ is triangulated into the four 4-simplices $ABCDD'$, $ABCD'C'$, $BCDED'$, and $BCED'C'$ (the right).
}\label{fig:last_case}
\end{figure}

 In the above two examples, the boundary triangulation for $P$ remains the same {where $P$ is triangulated with a single tetrahedron}, while the number of dynamical points varies.  In the subsequent example, we introduce a more complicated triangulation for $P$ to demonstrate how the procedure outlined in the previous examples can be applied. Specifically, we triangulate the spatial surface $P$ using two tetrahedra $ABCD$ and $BCDE$, where the segment $AB$ lies in the 2-sphere $S$. These tetrahedra share the triangle $BCD$ (Fig. \ref{fig:last_case}). Treating the points $C$ and $D$ as the dynamical ones, we obtain the resulting cell $ABCDEC'D'$, as illustrated in Fig. \ref{fig:last_case}. 
Since the cell  $ABCDEC'D'$ contains of two subcells $ABCDC'D'$ and $BCDEC'D'$, the triangulation of  $ABCDEC'D'$ is formed by combining the triangulations of these subcells. The triangulation for the subcells can be performed by following the procedure depicted in Figure \ref{fig:two_DV}. As a consequence, the cell $ABCDEC'D'$ emerges by gluing together the four 4-simplices $ABCDD'$, $ABCD'C'$, $BCDED'$, and $BCED'C'$, as illustrated in Figure \ref{fig:last_case}.

{ According  the three examples, our systematic way to triangulate a 4-dimensional region can be summarized as follows: first, we perform a 3-dimensional triangulation for the past boundary which is $P$ in the examples; second, select some points as the dynamical points and map these dynamical points to the future boundary which is $P'$ in the examples, with the worlds lines connecting each dynamical point with its image point; finally, connect the image points, the non-dynamical points and the dynamical points accordingly, as illustrated by the three examples, to construct the triangulation of a 4-dimensional region. }

\subsection{Triangulation of the region $B_-$}\label{sec:triangulationBm}

{ Now let us come back to the main problem of triangulating $B_-$. It is noted that the purpose of the triangulation is to construct the dual 2-complex to write down the spin foam amplitude. Thus, we only need to fix the  topology of  the region $B_-$ and ignore the geometric data such as the induced metric and extrinsic curvature on $\partial_-B$. Indeed,  one can always construct identical triangulation and, consequently, identical dual 2-complex in two manifolds having the same topology but different geometric data. Even though we have not specified $\partial_-B$ in the spacetime shown in Fig. \ref{fig:penrose1}, as far as $\partial_-B$ is selected as the one homotopic to the blue line, the topology of the corresponding $B_-$ does not changed.  
}

As shown in  Fig. \ref{fig:regionB}, the intersections between any two among the hypersurfaces $\partial_-B_1$, $\partial_-B_2$ and $\mathcal T$ are 2-spheres, denoted by $S_1$, $S_2$ and $S_3$ respectively (refer to Fig. \ref{fig:regionB}). Following the procedures outlined in Sec. \ref{sec:triangulationmethod}, the triangulation of $B_-$ is determined, if we do the following. { First, we perform a specific triangulation for $\partial_-B_1$ where the points of the triangulation lie either on $S_1$ or $S_2$, and the point on $S_2$ are chosen as the dynamical points; Second, we map the  dynamical  points on $S_2$ to points on $S_3$, with the world line of the dynamical points lying in $\partial_-B_2$.}

\subsubsection{triangulation for $\partial_-B_1$}\label{sec:triapB}

\begin{figure}
\centering
\includegraphics[width=0.45\textwidth]{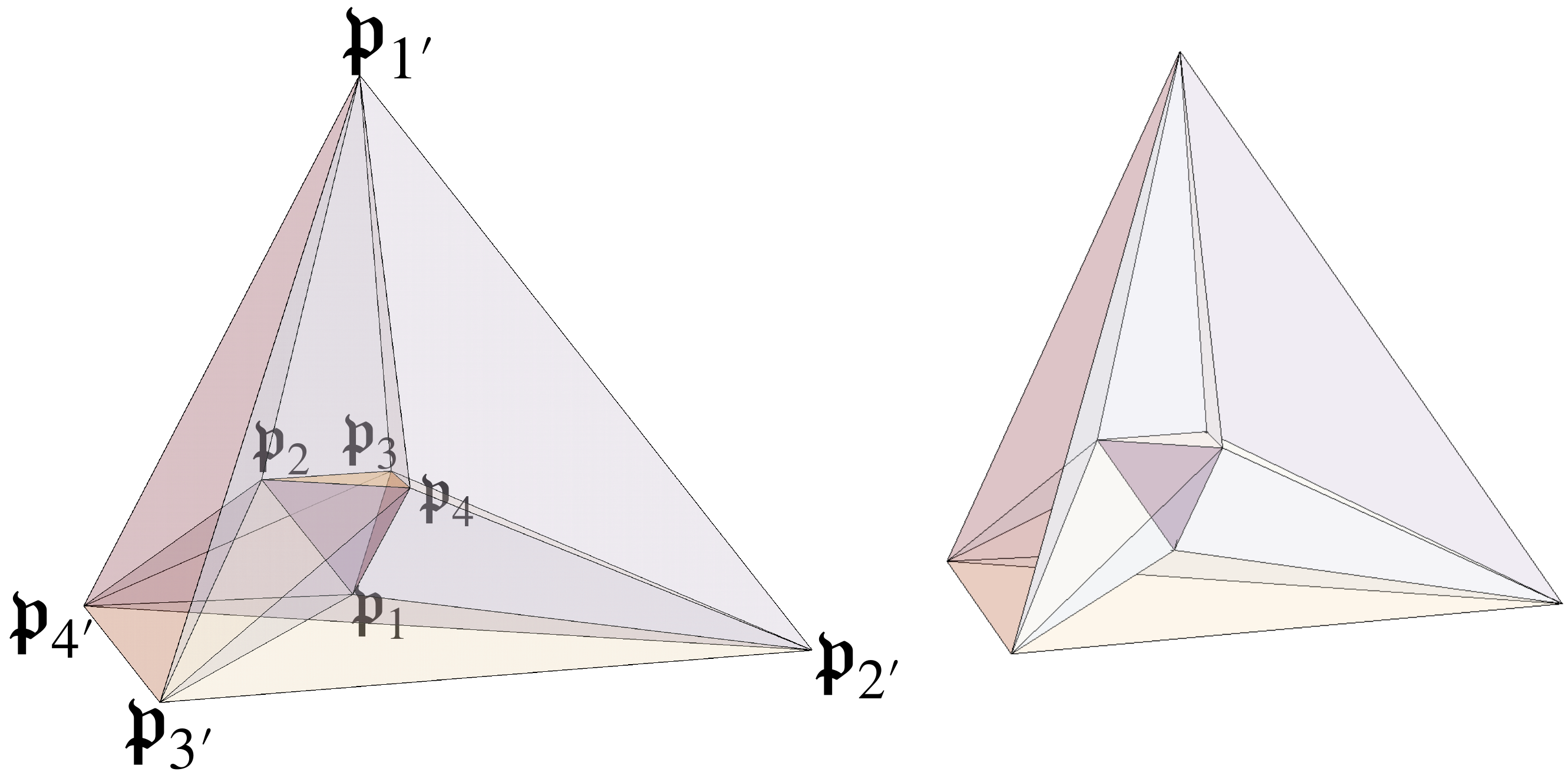}
\caption{A visualization for the triangulation of $\partial_-B_1$. To enhance clarity, both the transparent (left) and untransparent (right) versions are provided. }\label{fig:triangulationB}
\end{figure}

To construct the triangulation of $\partial_-B_1$, let us embed it into $\mathbb R^3\ni \vec x$ as a spherical shell, i.e., a 3D annulus, which is topologically homomorphism to $\partial_-B_1$. For convenience, the center of the spherical shell is put at the origin. The radii of the spheres $S_1$ and $S_2$ are $r_1$ and $r_2$ respectively, with $r_1<r_2$.
We adopt the triangulation of $\partial_-B_1$ proposed in \cite{DAmbrosio:2020mut,Soltani:2021zmv}, where one introduces a tetrahedron with points $\mathfrak p_a$ ($a=1,2,3,4$) to approximate the sphere $S_1$. 
Let $r_1 \vec x(a)\in\mathbb R^3$ for $a=1,2,3,4$ to be the coordinate of $\mathfrak p_a$. Then, select the points $\mathfrak p_{a'}\in S_2$ taking coordinates $-r_2\vec x(a)$. The connections between these points $\mathfrak p_{a'}$ ($a=1,2,3,4$) give rise to another tetrahedron that approximates the sphere $S_2$.  Finally, the triangulation of $\partial_-B_1$ is obtained by  drawing line segments connecting each point $\mathfrak p_a$ (for $a=1,2,3,4$) in $S_1$ to the corresponding points $\mathfrak p_{b'}$, $\mathfrak p_{c'}$, and $\mathfrak p_{d'}$ in $S_2$, where $b$, $c$, and $d$ represent labels that are different from $a$. Importantly, the minus sign in the coordinate of $\mathfrak{p}_{a'}$ ensures that this connectivity structure yields a well-defined triangulation.
The resulting triangulation can be visualized as shown in Fig. \ref{fig:triangulationB}. As illustrated in the figure, there are a total of $14$ tetrahedra, classified into three types:
\begin{itemize}
\item \textbf{A-type}: There are 4 tetrahedra having one point on $S_1$ and three points on $S_2$. The points of the tetrahedron are $\mathfrak p_{a}\mathfrak p_{b'}\mathfrak p_{c'}\mathfrak p_{d'}$.

\item \textbf{B-type}: There are 6 tetrahedra having two points in $S_1$ and two points in $S_2$. The points of the tetrahedron are $\mathfrak p_{a}\mathfrak p_{b}\mathfrak p_{c'}\mathfrak p_{d'}$.

\item \textbf{C-type}: There are 4 tetrahedra having three points on $S_1$ and one point on $S_2$. The points are $\mathfrak p_{a}\mathfrak p_{b}\mathfrak p_{c}\mathfrak p_{d'}$.
\end{itemize}
Here, $a,b,c,d$ represent distinct elements of the set $\{1,2,3,4\}$. {Once the triangulation is done in $\mathbb R^3$, one can pull back it to $\partial_-B_1$ to get the triangulation for $\partial_-B_1$. In what follows,  the pulling-back of the points $\mathfrak p_a$ and $\mathfrak p_{a'}$ will still be called $\mathfrak p_a$ and $\mathfrak p_{a'}$ respectively. }

\subsubsection{triangulation for $B_-$}

To obtain a triangulation for the 4-dimensional region $B_-$ from the triangulation of its boundary $\partial_-B_1$, we selected the points in $S_2$ to be dynamical.  Starting from each of these dynamical points $\mathfrak p_{a'}$ for $a=1,2,3,4$, a line $\mathfrak p_{a'}\mathfrak p_{a''}$ is drawn in $\partial_-B_2$, intersecting the sphere $S_3$ at a chosen point $\mathfrak p_{a''}$. $\mathfrak p_{a'}\mathfrak p_{a''}$ is viewed as the worldline of $\mathfrak p_{a'}$ in our final triangulation for $B_-$. Following the method outlined in Sec. \ref{sec:triangulationmethod}, one can construct the triangulation of $B_-$. Specially, each A-type tetrahedron contains three dynamical points, whose evolution results in a total of 12 4-simplices. Each B-type tetrahedron has two dynamical points, whose evolution leads to  a total of 12 4-simplices. Each C-type tetrahedron has one dynamical point, whose evolution  results in a total of 4 4-simplices. Consequently, there are a total of $28$ 4-simplices in the triangulation for the region $B_-$.

Consider embedding $S_3$ into $\mathbb R^3$ as a sphere centered at the origin with a radius $r_3>r_2$.  The spherical shell between the two spheres $S_2$ and $S_3$ is topologically the same as $\partial_-B_2$. Then, $\mathfrak p_{a''}$ for $a=1,2,3,4$ are chosen so that their coordinates are $r_3\vec x(a)\in\mathbb R^3$. The resulting triangulation for the transition surface $\mathcal T$, as the boundary of the triangulation of $B_-$, is the same as that for $\partial_-B_1$, i.e., the same as shown in Fig. \ref{fig:triangulationB}.

\begin{figure}[!htp]
\centering
\vspace{0.5cm}
\includegraphics[width=0.45\textwidth]{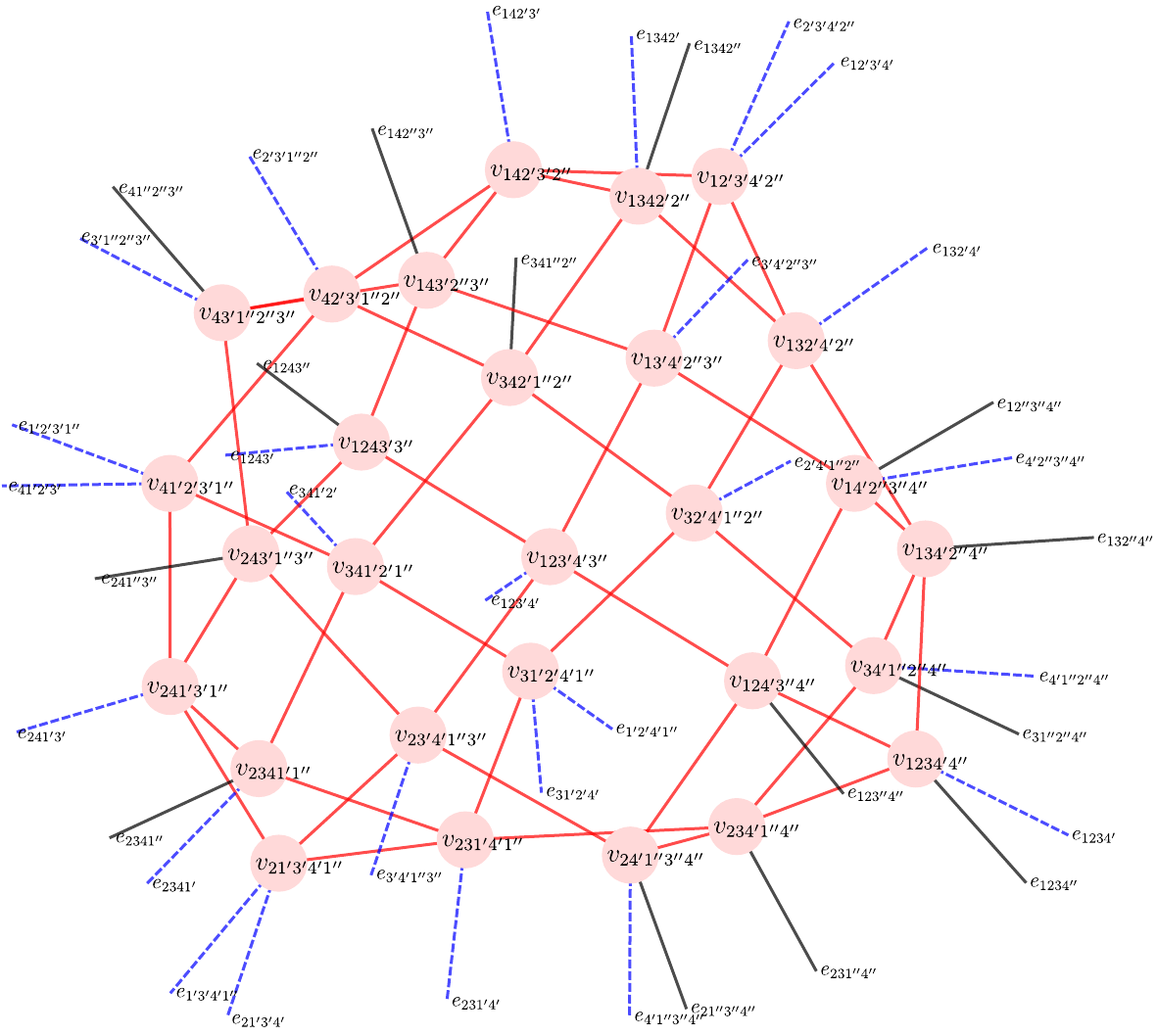}
\caption{2-complex dual to the triangulation of the region $B_-$. $v_{abcde}$ denotes the vertex dual to the 4-simplex $(\mathfrak p_a\mathfrak p_b\mathfrak p_c\mathfrak p_d\mathfrak p_e)$ and $e_{abcd}$ is the edge dual to the tetrahedron $(\mathfrak p_a\mathfrak p_b\mathfrak p_c\mathfrak p_d)$. In this figure, we use the black solid lines to represent the edges passing through the transition surface $\mathcal T$, and the blue dashed line for the edges in $\partial_-B$. 
}\label{fig:dualgraph}
\end{figure}

The 2-complex $\mathcal K$ dual to the triangulation of $B_-$ is shown in Fig. \ref{fig:dualgraph}. This dual complex contains 28 vertices, 90 edges, and 104 faces, which correspond to the 4-simplices, tetrahedra, and triangles in the original triangulation, respectively. The structure of the 2-complex is encoded in the following function, also denoted as $\mathcal K$, 
\begin{equation}
\mathcal K:f\mapsto \{(v_1,e_1),(v_2,e_2),\cdots,(v_n,e_n)\},
\end{equation}
where the domain of $\mathcal K$ consists of all faces $f$, and the sequence in the right-hand-side contains all half edges $(v,e)$ surrounding $f$.  Importantly, the order of the sequence coincides with the orientation of the face dual to $f$. 
Specific details regarding the assignment of the map $\mathcal K$ are provided in \cite{numericalResult}.


Taking advantage of the 2-complex for $B_-$ region, we construct the 2-complex for the entire $B$ region easily due to the time reversal symmetry.

\subsection{Boundary state}

The intersection between $\mathcal K$ and $\partial B$ gives the boundary graph $\gamma$. $\gamma$ contains the nodes $n_e$ and links $\ell_f$, which result from the intersections of edges $e$ and faces $f$, respectively, with $\partial B$. All nodes of $\gamma$ are 4-valence. The boundary states are vectors in the LQG Hilbert space $\mathcal H_\gamma$ on the graph $\gamma$. We will call $\mathcal H_\gamma$ the boundary Hilbert space in the current work. Specifically,  $\mathcal H_\gamma$ is given by 
\begin{equation}\label{eq:bdyHilberspace}
\mathcal H_\gamma=\bigotimes_{\ell \in\gamma}\left(\bigoplus_{j_{\ell}}\mathcal H_{j_\ell}\otimes\mathcal H_{j_\ell}^*\right)
\end{equation}
where $\mathcal H_{j}$ is the $\sut$ representation space with spin $j$ and $\mathcal H_{j}^*$ is its dual space. 
The spaces $\mathcal H_{j_\ell}$ and $\mathcal H_{j_\ell}^*$ are associated with the starting node $s_\ell$ and targeting node $t_\ell$ of $\ell$ respectively.

As discussed in Sec. \ref{sec:intro}, the current work uses the coherent states as the boundary states of spinfoam amplitude. { In LQG, there are several types of coherent state such as the coherent Livine-Speziale intertwiners \cite{LS} and the Thiemann's complexifier coherent state \cite{Thiemann:2000bw,Sahlmann:2001nv,Thiemann:2002vj}. They all have good properties and well studied in literate. However, the coherent Livine-Speziale intertwiners do not contain information about the embedding of the boundary. We thus adopt} the Thiemann's complexifier coherent state  (see, e.g. \cite{Bianchi:2009ky,Bianchi:2010mw} for the earlier applications to the spin foam model). Given a link $\ell\equiv \ell_f$, a Thiemann's complexifier coherent state labelled by $g_\ell\in\sltc$ can be expressed as a function of the holonomy $h_\ell\in \sut$ along $\ell$ as
\begin{equation}\label{eq:coherentstatebdy}
\psi_{g_\ell}^t(h_\ell)=\mathcal N \sum_{j_\ell}d_{j_\ell}e^{-\frac{t}{2}j_\ell(j_\ell+1)}\sum_{m}D^{j_\ell}_{mm}(g_\ell^{-1} h_\ell),
\end{equation}
with $\mathcal N$ denoting the normalization factor and {$t$ being the  classicality parameter which is dimensionless and  equal to $\beta\kappa\hbar$ divided by some area}.  {Here the Wigner-$D$ matrix $D^j_{nm}(g_\ell)$ for $g_\ell\in\sltc$ is defined as
\begin{equation*}
D^j_{mn}(g_{\ell})=\sum_{m'=-j}^j e^{-i(-ip_\ell+\phi_\ell)m'} D^j_{mm'}(n_{\ell s_\ell})D^j_{m'n}(n_{\ell t_\ell}^{-1})
\end{equation*}
where we} decompose $g_{\ell}\in\sltc$ as
\begin{equation}\label{eq:deccomposeg}
g_\ell=n_{\ell s_\ell}e^{(-ip_\ell+\phi_\ell)\tau_3} n_{\ell t_\ell}^{-1}
\end{equation}
with $n_{\ell s_\ell},\ n_{\ell t_\ell}\in\sut$ belonging to the starting and target points of $\ell$ and $p_\ell,\ \phi_\ell\in\mathbb R$. 
{ It can be checked that the state \eqref{eq:coherentstatebdy} is peaked on the elementary operators (refer to Appendix \ref{app:coherentstate} and \cite{Thiemann:2000ca,Dapor:2017rwv,Zhang:2021qul}):
}
\begin{equation}\label{eq:expectationvaluesbdypart}
\begin{aligned}
\langle\psi_{g_\ell}^t|\hat p_{s,\ell}^k\tau_k|\psi_{g_\ell}^t\rangle&\cong \frac{1}{t}p_\ell n_{\ell s_\ell}\tau_3 n_{\ell s_\ell}^{-1},\\
\langle\psi_{g_\ell}^t| \hat p_{t,\ell}^k\tau_k|\psi_{g_\ell}^t\rangle&\cong-\frac{1}{t} p_\ell n_{\ell t_\ell}\tau_3 n_{\ell t_\ell}^{-1},\\
\langle \psi_{g_\ell}^t| \hat h_\ell|\psi_{g_\ell}^t\rangle&\cong n_{\ell s_\ell} e^{\phi_\ell \tau_3}n_{\ell t_\ell}^{-1},
\end{aligned}
\end{equation}
where $\hat p_{s,\ell}^k$ and $\hat p_{t,\ell}^k$ are the flux operators and $\hat h_\ell$ denotes the holonomy operator (refer to Appendix \ref{app:holonomyflux}). {In addition, the spread of the state goes like $t$, because the uncertainties of the elementary operators are
\begin{equation*}
\langle \Delta \hat h_\ell\rangle= \frac{\langle\Delta p_{s,\ell}^k\rangle}{\langle \hat p_{s,\ell}^k\rangle}=\frac{\langle \Delta p_{t,\ell}^k\rangle}{\langle \hat p_{s,\ell}^k\rangle}=\mathcal O(\sqrt{t})
\end{equation*}
}

In Appendix \ref{app:eombdyaction}, the explicit expression of the spin foam amplitude with the boundary state \eqref{eq:coherentstatebdy} is derived.  Since the current work concerns with the asymptotics as $t\to 0$, we only need to focus on the critical equation. Then, as shown in Appendix \ref{app:eombdyaction}, the same critical equation can be obtained if we replace  the Thiemann's coherent state by
\begin{equation}\label{eq:stateLink}
\begin{aligned}
|g_\ell\rangle=&\mathcal N\sum_{j_f}\exp(-\frac{t}{2}\left(j_f-\frac{|p_\ell|}{t}\right)^2+ij_f\sgn(p_\ell)\phi_\ell)\times\\
&|j_f\xi_{\ell t_\ell}\rangle\otimes \langle j_f \xi_{\ell s_\ell}|,
\end{aligned}
\end{equation}
where  $\xi_x$ for $x=(\ell s_\ell)$ or $(\ell t_\ell)$ is the spinor given by
\begin{equation}
\xi_x=\left\{
\begin{array}{cc}
n_x(1,0)^T,&p_\ell>0,\\
n_x(0,1)^T,&p_\ell<0,
\end{array}
\right.
\end{equation}
{with $n_x$ for $x=(\ell s_\ell)$ or $(\ell t_\ell)$ given  in \eqref{eq:deccomposeg},}
and $|j_\ell,\xi_{\ell t_\ell}\rangle\in \mathcal H_{j_\ell}$ and $\langle j_\ell,\xi_{\ell s_\ell}|\in \mathcal H_{j_f}^*$ are the $\sut$  Perelomov coherent states \cite{klauder1985coherent,Perelomov:1986tf,Alesci:2016dqx}. {Geometrically, the $\sut$ coherent state $|j\xi\rangle\in\mathcal H_j$ corresponds to the vector $j\xi^\dagger\vec\sigma\xi\in\mathbb R^3$. }The state \eqref{eq:stateLink} is the similar coherent state used in the usual context of the spin foam model \cite{Bianchi:2009ky}, and its generalization to higher dimensional LQG is also studied \cite{Long:2021xjm,Long:2021lmd}. In terms of the state \eqref{eq:stateLink}, \eqref{eq:expectationvaluesbdypart} becomes 
\begin{equation}\label{eq:expectationvaluesbdypart2}
\begin{aligned}
\langle g_\ell|\hat p_{s,\ell}^k|g_\ell\rangle\cong& \frac{1}{t}|p_\ell| \xi_{\ell s_\ell}^\dagger\sigma^k\xi_{\ell s_\ell},\\
\langle g_\ell| \hat p_{t,\ell}^k|g_\ell\rangle\cong&-\frac{1}{t} |p_\ell| \xi_{\ell t_\ell}^\dagger\sigma^k\xi_{\ell t_\ell},\\
\langle g_\ell | \hat h_\ell|g_\ell\rangle \cong&n(\xi_{\ell s_\ell})
 e^{\sgn(p_\ell) \phi_\ell \tau_3} n(\xi_{\ell t_\ell})^{-1},
\end{aligned}
\end{equation}
where $n(\xi)$ for $\xi=(\xi_1,\xi_2)^T\in\mathbb C^2$ is given by
\begin{equation}
n(\xi)=\frac{1}{\sqrt{\xi^\dagger \xi}}\begin{pmatrix}
\xi_1&-\overline{\xi_2}\\
\xi_2&\overline{\xi_1}
\end{pmatrix}
\in\sut.
\end{equation}
In terms of these parameters, $g_\ell$ is expressed as
\begin{equation}\label{eq:gell2}
g_\ell=n(\xi_{\ell s_\ell}) e^{(-i|p_\ell|+\sgn(p_\ell) \phi_\ell )\tau_3} n(\xi_{\ell t_\ell})^{-1}.
\end{equation}

We get the coherent state on the entire graph $\gamma$ as their tensor product, 
\begin{equation}\label{eq:bdystatesTotal}
\left|\{g_\ell\}_{\ell\subset\gamma}\right\rangle=\bigotimes_{\ell \in\gamma} |g_\ell\rangle.
\end{equation}
Even though the boundary state given in \eqref{eq:bdystatesTotal}  lacks gauge invariance, the spin foam amplitude will inherently projects it into the gauge invariant subspace. As a consequence, the boundary state utilized in the spin foam amplitude is in essence the projected gauge  invariant the state. 

According to our discussion in Appendix \ref{app:A}, the parameters $p_\ell$ and $\xi_\ell$ in the coherent state carry the information about the densitied triad $E^a_i$ on $\partial B$, while $\phi_\ell$ is relevant to the Ashtekar connection $A_a^i$ on $\partial B$ (see also e.g. \cite{Rovelli:2010km}). Therefore, the coherent state carries the information about the 4-dimensional geometry of $\partial B$, relating the boundary geometry to the boundary coherent state labels. 

The intrinsic geometry of the boundary is characterized by the 3-metric $q_{ab}$, which is related to the triad components through $q_{ab}=\delta_{ij}e^i_ae^j_b$. This relationship suggests an $O(3)$ gauge transformation, wherein the triad undergoes a transformation $e^i_a\mapsto \mathcal{O}^i{}_je^j_a$ with $\mathcal{O}\in O(3)$. However, in LQG, only $\mathrm{SO(3)}$ gauge transformations are considered. Consequently, the triads $e^i_a$ and $-e^i_a$, while yielding the same intrinsic geometry, are not considered gauge-equivalent in the context of LQG. Based on this discussion, it seems reasonable to consider a superposition of states associated with both triads $e^i_a$ and $-e^i_a$ to capture a more comprehensive description of the boundary geometry.  

In the current work, the boundary $\partial B$ contains  $\partial_-B$ and $\partial_+B$. Let $\tilde e^i_a$ represent a triad in $\partial_+B$, and $\bar e^i_a$ represent a triad in $\partial_-B$. Then, across the entire boundary $\partial B$, the triad fields can take the forms $(\tilde e_a^i, \bar e^i_a)$, $(\tilde e_a^i, -\bar e^i_a)$, $(-\tilde e_a^i, \bar e^i_a)$, and $(-\tilde e_a^i, -\bar e^i_a)$. Here, we employ pairs like $(e_+,e_-)$ to denote the piecewise triad field, where $e(x)=e_+(x)$ for $x\in\partial_+B$ and $e(x)=e_-(x)$ for $x\in\partial_-B$. 
Let us use $g_\ell^{(\tilde s\bar s)}$, for each link $\ell$, to denote the label of the coherent state constructed from the triad $(\tilde s \tilde e_a^i, \bar s\bar e^i_a)$ with $\tilde s, \bar s = \pm$.  To keep the discussion flow,  we  will momentarily skip detailing the explicit relation between $g_\ell^{(\tilde s\bar s)}$ and $(\tilde s \tilde e_a^i, \bar s\bar e^i_a)$.
The above discussion motivates us to consider the boundary state as the superposition 
\begin{equation}\label{eq:superpositionPsi}
|\Psi_\gamma\rangle=\sum_{\tilde s,\bar s=\pm}\left|\{g_\ell^{(\tilde s\bar s)}\}_{\ell\subset\gamma}\right\rangle.
\end{equation}
Then the spin foam transition amplitude is a sum
\begin{equation}
\langle A|\Psi_\gamma\rangle=\sum_{\tilde s,\bar s=\pm}\left\langle A\middle |\{g_\ell^{(\tilde s\bar s)}\}_{\ell\subset\gamma}\right\rangle.
\end{equation}

\subsection{Spin foam amplitude and action} \label{se:sfA}

The spin foam amplitude is a specific linear functional acting  in the boundary Hilbert space \eqref{eq:bdyHilberspace} \cite{Rovelli:2014ssa}. The spinfoam amplitude with the boundary state \eqref{eq:bdystatesTotal} has an integral expression. This expression assigns an $\sut$ representation $j_f$ to each face $f$, an $\sltc$ element $g_{ve}$ to each half edge $(v,e)$ of the edge $e$ taking the vertex $v$ as its end point, and a spinor $z_{vf}\in\mathbb{C}^2$  to each corner $(v,f)$ of the face $f$ at the vertex $v$. The amplitude reads 
\begin{equation}\label{eq:amplitude}
\begin{aligned}
A=&\sum_{\{j_f\}}\int \prod_{f}(d_{j_f})^{|V_f|+1} \prod_{ve}\dd g_{ve} \prod_{vf}\dd^2\Omega_{vf} \times\\
&\exp[S(\{j_f\},\{g_{ve}\},\{z_{vf}\})]. 
\end{aligned}
\end{equation}
where $|V_f|$ is the number of vertices surrounding the face $f$, $\dd g_{ve}$ is the Haar measure, $\dd^2\Omega_{vf}$ is a measure on $\mathbb{C}\mathbb{P}^1$, and $S$ depending on the tuples $(\{j_f\},\{g_{ve}\},\{z_{vf}\})$ is the effective action given by the spin foam model. The precise definition of the measures $\dd g_{ve}$ and $\dd^2 \Omega_{vf}$ can be found in \cite{Han:2013gna}. The integration for the spinor $z_{vf}\in\mathbb C^2$ is over $\mathbb{CP}^1$, relying on the scaling invariance of the action $S$. 
The action $S$ can be written as the following summation
\begin{equation}\label{eq:actiontotal}
S(\{j_f\},\{g_{ve}\},\{z_{vf}\})=\sum_fS_f(j_f,g_{ve},z_{vf}),
\end{equation}
where $S_f$ is the action associated with the face $f$. 
Specifically, for an internal face $f$, $S_f$ reads
\begin{equation}\label{eq:action}
\begin{aligned}
S_f=&\sum_{e\in\partial f}j_f\Bigg[2\ln\frac{\langle Z_{s_e e f},Z_{t_eef}\rangle}{\left\|Z_{s_eef}\right\| \left\| Z_{t_e ef}\right\| }+i\beta \ln\frac{ \left\|Z_{s_eef}\right\|^2}{\left\| Z_{t_e ef}\right\|^2} \Bigg],
\end{aligned}
\end{equation}
where $Z_{vef}:=g_{ve}^\dagger z_{vf}$, $s_e$ and $t_e$ represent the source and target points of edge $e$ respectively, $\|Z\|$ denotes $\sqrt{\langle Z,Z\rangle}$ for all $Z\in \mathbb C^2$, {and $\beta$ is the Barbero-Immirizi parameter as introduced in \eqref{eq:metricf}.}  For a boundary face $f$, as shown in \eqref{eq:actionBoundary2} (see  Appendixes \ref{app:sfBdyAction} and \ref{app:eombdyaction} for more details), we have
\begin{equation}\label{eq:boundarySf}
\begin{aligned}
S_f=&-\frac{t}{2}\left(j_f-\frac{|p_{\ell_f}|}{t}\right)^2+i\,\sgn(p_{\ell_f}) j_f \phi_{\ell_f}+\\
&\sum_{e\in\partial f}j_f\Bigg[2\ln\frac{\langle Z_{s_e e f},Z_{t_eef}\rangle}{\left\|Z_{s_eef}\right\| \left\| Z_{t_e ef}\right\|}+i\beta \ln \frac{ \left\|Z_{s_eef}\right\|^2}{\left\| Z_{t_e ef}\right\|^2}\Bigg]
\end{aligned}
\end{equation}
where the extra term compared to \eqref{eq:action} comes from the coherent state. Moreover, in \eqref{eq:boundarySf}, $Z_{s_eef}=\xi_{\ell_f s_{\ell_f}}$ when  the boundary of the face $f$ starts from $e$, and  $Z_{t_eef}=\xi_{\ell_f t_{\ell_f}}$ when  the boundary of the face $f$ ends at $e$, with $\xi_{\ell_f s_\ell}$ and $\xi_{\ell_f t_\ell}$ introduced in \eqref{eq:stateLink}.

Let $F$, $E$, and $V$ denote the number of faces, \emph{half} edges, and vertices in the dual 2-complex $\mathcal K$.  The action $S$ is a function on the  configuration space  $$\Gamma=\left(\mathbb Z_{\geq 0}/2\right)^F\times \sltc^{E}\times (\mathbb C^2)^{10 V},$$
with $j_{f}\in \mathbb Z_{\geq 0}/2$, $g_{ve}\in \sltc$  and $z_{vf}\in\mathbb C^2$.

Considering arbitrary $\sltc$-valued field $v\mapsto h_v$ at the vertices, arbitrary complex field $(v,f)\mapsto \lambda_{vf}$ with $\lambda_{vf}\neq 0$, and arbitrary $\sut$-valued field $e\mapsto u_e$ on the edges, we observe that $S$ exhibits invariance under the transformation:\begin{equation}\label{eq:gaugetransformation}
\Big(j_f,g_{ve},z_{vf}\Big) \mapsto \Big(j_f,h_v^{-1} g_{ve} u_e,\lambda_{vf}h_v^\dagger z_{vf}\Big).
\end{equation}
The integration in \eqref{eq:amplitude} is carried out over the quotient space $\Gamma/\sim$, where the equivalence relation $\sim$ is defined by the gauge transformation \eqref{eq:gaugetransformation}.

\section{Boundary data}\label{sec:boundarydata}

Let us focus on specifying the boundary data for further calculations.

{Before going to the specific calculation, let us outline the logic of this section. The final goal of the current work is to numerically solve the equation of motion $\delta S=0$ with the action given in \eqref{se:sfA}. To this end, a starting point for numerical root finding is needed. This starting point can be constructed based on a flat geometry which has the intrinsic and extrinsic boundary geometries approximating those of $\partial B$. Since the boundary $\partial B$, as a  hypersurface in the spacetime depicted in Fig. \ref{fig:penrose1}, is not fixed so far, we can thus choose it such that $\partial B$ takes the same intrinsic geometry as the hypersurface surface $t=a r$, with an arbitrary real number  $a$ satisfying $|a|<1$, in the Minkowski spacetime $\mathbb R^4$. This choice will help us achieve the flat geometry needed for constructing the starting point for root finding.  Here we use the spherical coordinate $(t,r,\theta,\phi)$ in the Minkowski spacetime such that the Minkowski metric takes $\dd s^2=-\dd t^2+\dd r^2+r^2\dd\theta^2+r^2\sin^2\theta\dd\phi^2$.
}

With $\tilde v:=t+r$, the Minkowski metric becomes 
\begin{equation}
\dd s^2_{\rm M}=-\dd \tilde v^2+2\dd \tilde v\dd r+r^2\dd\Omega^2.
\end{equation}
On the surface $t=a r$, the reduced 3-metric is 
\begin{equation}\label{eq:3metric}
\dd s_{\rm M}^2=(1-a^2)\dd r^2+r^2\dd\Omega^2.
\end{equation}
Consider the Eddtington-Finkelsein coordinate where the BH metric reads
\begin{equation}
\dd s_{\rm BH}^2=-f(r)\dd v^2+2\dd v\dd r +r^2\dd\Omega^2.
\end{equation}
Assume that the surface in the BH spacetime taking the same 3-metric as \eqref{eq:3metric} is given by
\begin{equation}
v+F(r)=0
\end{equation}
{for some function $F$.} With a straightforward calculation, we get  
\begin{equation}\label{eq:eqF}
\frac{\dd F}{\dd r}=-\frac{1\pm \sqrt{1-(1-a ^2) f(r)}}{f(r)},
\end{equation}
where $1-(1-a^2)f(r)>0$ for all $r\geq r_b$ and $a^2<1$. Here, $r_b$ as the radius of the dust ball at the moment of bounce will be chosen as the radius of $S_1$. { It is reminded that Region $B$ is outside the dustball.}

If the positive sign is selected, the right hand side of \eqref{eq:eqF} is singular at $r=r_\pm$ where $f(r_\pm)=0$. Thus, we should choose
\begin{itemize}
\item[(1)] $F(r)=F_1(r)$ for $\partial_-B_1$ so that
\begin{equation}
\frac{\dd F_1}{\dd r}=-\frac{1- \sqrt{1-(1-a_1 ^2) f(r)}}{f(r)},
\end{equation}
with some $a_1$; and
\item[(2)] $F(r)=F_2(r)$ for $\partial_-B_2$ with 
\begin{equation}
\frac{\dd F_2}{\dd r}=-\frac{1-\sqrt{1-(1-a_2^2) f(r)}}{f(r)},
\end{equation}
for some $a_2$. 
\end{itemize}
With the past boundaries $\partial_-B_1$ and $\partial_-B_2$, we  get the corresponding future boundaries $\partial_+B_1$ and $\partial_+B_2$ by the time reversal symmetry. 

It is easy to check the outwards normal $\mathfrak n_i$ of $\partial_-B_i$ for $i=1,2$ is
\begin{equation}
\begin{aligned}
\mathfrak n_i=&-\frac{1- \sqrt{1-(1-a_i^2)f(r)}}{f(r)}\,  \partial_v+\\
& \sqrt{1-(1-a_i^2)f(r)}\,\partial_r. 
\end{aligned}
\end{equation}
Their lengths for $\mathfrak n_i$ is 
\begin{equation}
\|\mathfrak n_i\|^2=-(1-a_i^2)<0.
\end{equation}
Thus, the boundary $\partial_-B$ is spacelike.

Consider the intersection of $\partial_-B_1$ and $\partial_+B_1$. With their dihedral angle denoted by $\theta_1$, we have
\begin{equation}\label{eq:theta1}
\cosh(\theta_1)=-1+\frac{2}{(1-a_1^2)f(r_1)}.
\end{equation} 
Similarly, we get
\begin{equation}\label{eq:theta2}
\cosh(\theta_2)=-1+\frac{2}{(1-a_2^2)f(r_3)},
\end{equation}
where $\theta_2$ is the dihedral angle between $\partial_-B_2$ and $\partial_+B_2$. Eqs. \eqref{eq:theta1} and \eqref{eq:theta2} establish a connection between the parameters $a_i$ and geometrical properties of the region $B$. In our practical numerical computation, we will select $\theta_1$ and $\theta_2$ as the parameters to set, rather than $a_1$ and $a_2$.

\subsection{Values of $p_\ell$ and $\xi_{\ell n}$}\label{sec:valuepxi}

In \eqref{eq:expectationvaluesbdypart2}, the quantities $p_\ell$ and $\xi_{\ell n}$ are related to the expectation value of $\hat p^k_{s,\ell}$, which denotes the area vector of the surface for the flux. In our work, this surface corresponds to a face within the triangulation of $\partial B$, which serves as a discrete approximation of the geometry of $\partial B$. Therefore, determining $p_\ell$ and $\xi_{\ell n}$ involves computing the area vectors of surfaces within the discrete geometry, constructed as outlined below.

Our choice of the boundary $\partial B$ ensures that $\partial_-B_i$ for $i=1,2$ have the same intrinsic geometries as that of the hypersurfaces $t=a_i r$ in the Minkowski spacetime. This fact implies the existence of an embedding $\mathfrak i$ of $\partial_-B$ into the Minkowski spacetime, such that the 3-metric of $\mathfrak i(\partial_-B)$ reduced from the Minkowski metric is the same as the 3-metric of $\partial_-B$ reduced from the BH metric. {Since $\partial_-B$ and $\mathfrak i(\partial_-B)$ have the same intrinsic geometry, we can calculate the $p_\ell$ and $\xi_{\ell n}$, encoding the information of the intrinsic geometry of $\partial_-B$, in the Minkowski spacetime. 
}

{To calculate $p_\ell$ and $\xi_{\ell n}$, we first embed the triangulation for $B_-$ constructed in Sec. \ref{sec:triangulationBm} into $\mathfrak i(B_-)$, a region in the Minkowski spacetime.  As discussed at the beginning of Sec.  \ref{sec:triangulationBm}, we initially focused on the topological structure of the triangulation. Now, the geometric details,  such as the shape of the tetrahedra in the embedded triangulation, must be focused on, to calculate $p_\ell$ and $\xi_{\ell b}$. It is observed that the triangulation constructed in Sec.  \ref{sec:triangulationBm} is completely determined by the points $\mathfrak p_{a}$, $\mathfrak p_{a'}$ and $\mathfrak p_{a''}$ for all $a=1,2,3,4$. Thus, we need to specify the position of points $\mathfrak i(\mathfrak {p}_{a})$, $\mathfrak i(\mathfrak p_{a'})$ and $\mathfrak i(\mathfrak p_{a''})$ in the embedded triangulation corresponding to $\mathfrak p_{a}$, $\mathfrak p_{a'}$ and $\mathfrak p_{a''}$. Specifically, we require that the tetrahedra $\mathfrak i(\mathfrak p_{1})\mathfrak i(\mathfrak p_{2})\mathfrak i(\mathfrak p_{3})\mathfrak i(\mathfrak p_{4})$, $\mathfrak i(\mathfrak p_{1'})\mathfrak i(\mathfrak p_{2'})\mathfrak i(\mathfrak p_{3'})\mathfrak i(\mathfrak p_{4'})$,and $\mathfrak i(\mathfrak p_{1''})\mathfrak i(\mathfrak p_{2''})\mathfrak i(\mathfrak p_{3''})\mathfrak i(\mathfrak p_{4''})$, are equilateral. In addition, Let $r_1$, $r_2$ and $r_3$ be the radii of the spheres $\mathfrak i(S_1)$, $\mathfrak i(S_2)$ and $\mathfrak i(S_3)$, and $t_1$, $t_2$ and $t_3$ be the time coordinate of the spheres. 
Under the Cartesian coordinate system, we require that $\mathfrak i(\mathfrak p_{a})=(t_1,r_1\vec R_a)$, $\mathfrak i(\mathfrak p_{a'})=(t_2,-r_2\vec R_a)$, and $\mathfrak i(\mathfrak p_{a})=(t_3,r_3\vec R_a)$,  for some vector $\vec R_a\in\mathbb R^3$. The above requirements maximally respect the spherically symmetry. 
}  The embedded  boundary triangulation provided a discrete geometry approximating the intrinsic geometry of $\mathfrak i(\partial_-B)$. Using the discrete geometry, we can calculate $p_\ell$ and $\xi_{\ell f}$ by
\begin{equation}\label{eq:pellandA}
p_\ell \xi_{\ell n}^\dagger \vec\sigma\xi_{\ell n}=t\frac{\vec A_{\ell}}{\beta\kappa \hbar}
\end{equation}
where $\vec A_\ell$ is the area vector of the triangle dual to the link $\ell$. Since $p_\ell$ and $\xi_{\ell n}$ for all links $\ell$ connecting to each node $n$ are determined by a tetrahedron, the closure condition 
\begin{equation}
\sum_{\ell \text{ at } n}\epsilon_{\ell n} p_\ell \xi_{\ell n}^\dagger \vec\sigma\xi_{\ell n}=0
\end{equation}
 must be satisfied, where the sign factors $\epsilon_{\ell n}$ depends on the direction chosen for $\vec A_\ell$.

Let $\mathfrak r:\partial_-B\to\partial_+B$ denote the time reversal symmetry. Owing to this symmetry, $\partial_+B$ exhibits the same intrinsic geometry as that of $\partial_-B$. 
Thus, if $\bar e_a^i$ is the cotriad field in $\partial_-B$, the cotriad field in $\partial_+B$ could be $\mathfrak r^*(\bar e_a^i)$, i.e., the push-forward of $\bar e_a^i$. Here the push-forward of the forms $\bar e_a^i$ is regarded as their pull-back by $\mathfrak r^{-1}$.
Despite the time reversal symmetry, the 3-orientation of $\partial_+B$, induced by the outward normal field, is opposite to $\mathfrak r^*(o_-)$, namely, the push forward of the reduced orientation $o_-$ on $\partial_-B$.
Furthermore, as discussed in Appendix \ref{app:A}, in the spin foam model, the densitized triad should be related to the triad by $E^a_i=e e^a_i$ with $e\equiv \det(e_b^j)$. It is dual to a 2-form  $E^a_i\varepsilon_{abc}$, whose surface integration, defining the flux, is 3-orientation dependent (refer to \eqref{eq:flux}). Consequently, the flux at each node  $\mathfrak{r}(\ell)\mathfrak{r}(n)\subset \partial_+B$ becomes the opposite of the flux at the corresponding $(\ell, n)$.  In other words,
\begin{equation}\label{eq:minusp}
\langle g_\ell|\hat p_{n,\ell}^k|g_\ell\rangle=-\langle g_{\mathfrak r(\ell)}|\hat p_{\mathfrak r(n),\mathfrak r(\ell)}^k|g_{\mathfrak r(\ell)}\rangle
\end{equation}
with $\hat p_{n,\ell}^k=\hat p_{s,\ell}^k$ for $n=s_\ell$ and  $\hat p_{n,\ell}^k=\hat p_{t,\ell}^k$ for $n=t_\ell$.
Combining \eqref{eq:minusp} with \eqref{eq:expectationvaluesbdypart2}, we get 
 \begin{equation}\label{eq:pxiBp}
 |p_{\mathfrak{r}(\ell)}|  \xi_{\mathfrak{r}(\ell)\mathfrak{r}(n)}^\dagger \vec\sigma  \xi_{\mathfrak{r}(\ell)\mathfrak{r}(n)}=- |p_\ell| \xi_{\ell n}^\dagger \vec\sigma\xi_{\ell n},
\end{equation}
where $p_\ell$ and $\xi_{\ell n}$ are the data associated with $\ell,n\subset \partial_- B$, and  $p_{\mathfrak{r}(\ell)}$ and $ \xi_{\mathfrak{r}(\ell)\mathfrak{r}(n)}$ are the corresponding data for $\mathfrak{r}(\ell)$ and $\mathfrak{r}(n)$. The equation \eqref{eq:pxiBp} implies 
\begin{equation}\label{eq:pxiBp2p}
|p_{\mathfrak{r}(\ell)}| =|p_\ell|,\quad  \xi_{\mathfrak{r}(\ell)\mathfrak{r}(n)}=e^{i x}J \xi_{\ell n},
\end{equation}
where $x$ is an arbitrary phase, and $J\xi$ for $\xi=(\xi_1,\xi_2)^T\in \mathbb C^2$ is defined by
\begin{equation}
J\xi=(-\overline{\xi_2},\overline{\xi_1})^T.
\end{equation}
 Given that the phase $x$ in \eqref{eq:pxiBp2p} does not affect the outcomes in our paper, we can set it to zero, namely, we have
\begin{equation}\label{eq:pxiBp2}
|p_{\mathfrak{r}(\ell)}| =|p_\ell|,\quad  \xi_{\mathfrak{r}(\ell)\mathfrak{r}(n)}=J \xi_{\ell n}.
\end{equation}

In our computation, we choose the surfaces $\mathfrak i(\partial_-B)=\mathfrak i(\partial_-B_1)\cup \mathfrak i(\partial_-B_2)$ as 
\begin{equation*}
\begin{aligned}
\mathfrak i(\partial_-B_1):\ g_1(t,r)&:=t-a_1(r-r_1)=0,\\
\mathfrak i(\partial_-B_2):\ g_2(t,r)&:=t-a_2(r-r_2)-a_1(r_2-r_1)=0.
\end{aligned}
\end{equation*}
It  is reminded that $\mathfrak i$, as defined in the paragraph before \eqref{eq:pellandA}, denotes the embedding of $\partial_-B$ into the Minkowski spacetime. { Let $\Sigma_1$, $\Sigma_2$ denote the embedded 3-dimensional hypersurfaces of $\partial_-B_1$ and $\partial_-B_2$, i.e., $\Sigma_1\equiv \mathfrak i(\partial_-B_1)$ and $\Sigma_2\equiv \mathfrak i(\partial_-B_2)$.}
Then, we have 
\begin{equation}
\begin{aligned}
\mathfrak i(S_1)&=\{(t_1,r_1,\theta,\phi)\in\Sigma_1,\theta\in (0,\pi),\phi(0,2\pi)\}\\
\mathfrak i(S_2)&=\{(t_2,r_2,\theta,\phi)\in\Sigma_1,\theta\in (0,\pi),\phi(0,2\pi)\}\\
\mathfrak i(S_3)&=\{(t_3,r_3,\theta,\phi)\in\Sigma_2,\theta\in (0,\pi),\phi(0,2\pi)\}
\end{aligned}
\end{equation}
with
\begin{equation}\label{eq:a12aaa}
\begin{aligned}
t_1&=0,\\
t_2&=a_1(r_2-r_1),\\
 t_3&=a_2(r_3-r_2)+a_1(r_2-r_1). 
\end{aligned}
\end{equation}
Introducing 
\begin{equation}
\begin{aligned}
\vec R_1=&\left( \sqrt{\frac{8}{9}}, 0 , -\frac{1}{3}\right), \vec R_2=\left(-\sqrt{\frac{2}{9}}, \sqrt{\frac{2}{3}}, -\frac{1}{3}\right), \\
 \vec R_3=&\left( -\sqrt{\frac{2}{9}}, -\sqrt{\frac{2}{3}} , -\frac{1}{3} \right), \vec R_4=\left(0, 0, 1\right),
\end{aligned}
\end{equation}
we set the Cartesian  coordinates of $\mathfrak i(\mathfrak p_a)$, $\mathfrak i(\mathfrak p_{a'})$ and $\mathfrak i(\mathfrak p_{a''})$ in Minkowski spacetime to be
\begin{equation}
\begin{aligned}
\mathfrak i(\mathfrak p_a)&=(t_1,r_1\vec R_a),\ \mathfrak i(\mathfrak p_{a'})=(t_1,-r_2\vec R_a),\\
 \mathfrak i(\mathfrak p_{a''})&=(t_1,-r_3\vec R_a),\ \forall a=1,2,3,4. 
\end{aligned}
\end{equation}

\subsection{Relate $\phi_\ell$ to extrinsic curvature}\label{sec:xiell}

According  to \eqref{eq:expectationvaluesbdypart}, we need to calculate the holonomy within the boundary geometry to get $\phi_\ell$.  Appendix \ref{app:A} provides a detailed computation, with key aspects worth noting. At first, as  shown in \cite{Han:2011re}, the value of the spin foam action at the  real critical points differs from the Regge action by a factor $\sgn(V_4(v))$. This extra factor suggests that the volume form in classical action corresponding to the spin foam model is $\det(\mathfrak e)\dd^4 x$ where $\mathfrak e$ denotes the coframe field.  Starting from the Einstein-Hilbert action with this volume form, we establish the corresponding Ashtekar variables and finally determine the holonomy. Additionally, the link $\ell$ in the holonomy is chosen to be a geodesic for simplicity. Finally,  we assume that the extrinsic curvature $K_{ab}$ along the link $\ell$ takes the form 
\begin{equation}\label{eq:assumeextrinsicCurvatureBody}
K^{ab}(\tau)=\frac{\xi(\tau)}{\|\dot\ell_\tau\|^2}\dot\ell^a_\tau\dot\ell^b_\tau
\end{equation}
with $\xi$ being a field on $\ell$ and $\tau$ being the parameter of $\ell$. This assumption makes the computation tractable.  In LQG, quantities are often smeared by integration. Therefore, the simplified expression  \eqref{eq:assumeextrinsicCurvatureBody} can be interpreted as an approximation of the true behavior of the extrinsic curvature, which reflects the integrated properties of the actual extrinsic curvature over the link. In the context described above, we obtain the expression for $\phi_\ell$ as follows: 
\begin{equation}\label{eq:phiell}
\phi_\ell=\theta_\ell-\beta \sgn(p_\ell)\sgn(e)\Xi_\ell,
\end{equation}
where $\theta_\ell$ is the twist angle, $\Xi_\ell$ is related to the extrinsic curvature and $\sgn(e)\equiv \sgn(\det(e_a^i))$ is computed with any right-hand coordinate system. 

The twist angle $\theta_\ell$ will be fixed by choosing some certain frame for the two tetrahedra sharing the triangle dual to $\ell$. 
The way to fix this angle will be introduced in more details during the discussion of our numerical computation. For now, we will focus only on  $\Xi_\ell$ which, due to the assumption  \eqref{eq:assumeextrinsicCurvatureBody}, reads
 \begin{equation}\label{eq:xiKinmainpart}
\Xi_\ell=\int_\ell K_{ab}(s)\dot\ell^a_s \dot\ell^b_s \dd s
\end{equation}
where $s$ is the parameter of length of $\ell$, $\dot\ell_s^a$ is the tangent vector $\ell$ with respect to the parameter $s$. 

Turning our attention to the continuous geometry on $\partial_- B$, a naive  approach to relate $\Xi_\ell$ to this geometry involves a direct application of \eqref{eq:xiK} where 
one would integrate the extrinsic curvature $K_{ab}$ of the continuous geometry along a single edge.  However, this approach cannot fully encode the information of $K_{ab}$ within $\Xi_\ell$. Indeed, in the continuous geometry, the field $K_{ab}$ undergoes variations at distinct points, in contrast to the Regge geometry where the extrinsic curvature behaves as a distribution with a vanishing value everywhere except on the interfacing triangle $t_\ell$, where it remains constant. Thus, integrating $K_{ab}$ along a single edge, as in \eqref{eq:xiK}, is limited to capturing the information of $K_{ab}$ solely on that specific edge. Recognizing the limitations of this naive approach, we will abandon it and adopt a more refined strategy. The new approach takes advantage of the triangulation for $\partial_-B$. Consider two tetrahedra $T_1=\mathfrak q_1\mathfrak q_2\mathfrak q_3\mathfrak q_4$ and  $T_2=\mathfrak q_2\mathfrak q_3\mathfrak q_4\mathfrak p_5$, sharing the interfacing triangle $t_\ell=\mathfrak q_2\mathfrak q_3\mathfrak q_4$. Here 
$q_i$ for $i=1,2,3,4,5$ represents a point in the triangulation of $\partial_-B$. 
Introducing the ``centroids'' $\mathfrak c_1\in T_1$ and $\mathfrak c_2\in T_2$, we define a convex region $\mathfrak c_1 \mathfrak q_2\mathfrak q_3\mathfrak q_4 \mathfrak c_2$. This region is foliated into a family of two surfaces $\tau\mapsto A(\tau)\subset \mathbb R^2$ with $A(0)=t_\ell$. The average extrinsic curvature $\overline K(\tau)$ is defined on each surface as 
\begin{equation}\label{eq:averageK}
\overline{K}(\tau):=\|A_\tau\|^{-1}\int_{A_\tau}( K_{ab}n^an^b) n^c \varepsilon_{cde} \dd x^d\dd x^e,
\end{equation}
where $\|A_\tau\|$ is the area of $A_\tau$ and $n^a$ is the unit normal of $A_\tau$.  $\overline K(\tau)$ in \eqref{eq:averageK} captures the information  $K_{ab}n^an^b$ on the surface $A_\tau$. Choosing a link $\ell$ connecting $\mathfrak q_1$ and $\mathfrak q_2$  and transversely intersecting each $A_\tau$, we relate $\Xi_\ell$ to the extrinsic curvature by 
\begin{equation}\label{eq:xifromavergeK}
\Xi_\ell:=\int_\ell \overline K(\tau(s))\dd s.
\end{equation}
This equation \emph{defines} a connection between the extrinsic curvature and the data $\Xi_\ell$ inherent in the coherent state.   this definition reverts to \eqref{eq:xiKinmainpart} in the specific case of Regge geometry (refer to the discussion in Appendix \ref{app:Adc}). Importantly, as \eqref{eq:xifromavergeK} defines $\Xi_\ell$ through the average extrinsic curvature, it effectively encapsulates information about the extrinsic curvature across the entire region $\mathfrak c_1 \mathfrak q_2\mathfrak q_3\mathfrak q_4 \mathfrak q_2$.
 
Our actual computation adopts the coordinate give by $(x,y,z)=(r\sin\theta\cos\phi,r\sin\theta\sin\phi, r\cos\theta)$ for simplicity. Within this coordinate, we choose $\mathfrak c_1:=\frac{1}{4}(\mathfrak q_1+\mathfrak q_2+\mathfrak q_3+\mathfrak q_4 )$ and $\mathfrak c_2:=\frac{1}{4}(\mathfrak q_2+\mathfrak q_3+\mathfrak q_4 +\mathfrak q_5)$. The region $\mathfrak c_1\mathfrak q_2\mathfrak q_3\mathfrak q_4 \mathfrak c_2$ is defined as the polyhedron bounded by plane faces. These plane faces are expressed as surfaces in the coordinate system $(x, y, z)$ through linear equations. In the foliation of $\mathfrak c_1 \mathfrak q_2\mathfrak q_3\mathfrak q_4 \mathfrak cc_2$, the slices $A_\tau$ are chosen as plane surfaces parallel to $t_\ell$. The link $\ell$, connecting $\mathfrak c_1$ and $\mathfrak c_2$, is selected as the straight line $t\mapsto (1-t) \mathfrak c_1 +t \mathfrak c_2$ in this coordinate system. 

Due to the time reversal symmetry $\mathfrak r:\partial_-B\to\partial_+B$, the extrinsic curvature $K_{ab}^+$ in $\partial_+B$ is just $\mathfrak r^* K_{ab}^-$, i.e., the push forward of the extrinsic curvature $K_{ab}^-$ in $\partial_-B$. Therefore, for a given $\ell\subset\partial_-B$ carrying $\Xi_\ell$, the value of $\Xi_{\mathfrak r(\ell)}$ for $\mathfrak r(\ell)\subset\partial_+B$ is the same as $\Xi_\ell$, i.e.,
\begin{equation}\label{eq:XiinBp}
\Xi_{\mathfrak r(\ell)}=\Xi_\ell.
\end{equation}

\subsection{Superpose the coherent states}\label{sec:superposing}

Let us firstly summarize the results given in Secs. \ref{sec:valuepxi} and \ref{sec:xiell}. Given a link $\ell$ in $\partial_-B$, we can calculate $p_\ell$, $\xi_{\ell s_\ell}$, $\xi_{\ell t_\ell}$, $\theta_\ell$ and $\Xi_\ell$, to get the coherent state label $\bar g_\ell$ as
\begin{equation}\label{eq:gellbm}
\bar g_\ell=n(\xi_{\ell s_\ell})e^{(-ip_\ell +\theta_\ell-\beta\sgn(\bar e)\Xi_\ell)\tau_3} n(\xi_{\ell t_\ell})^{-1}
\end{equation}
where we choose $p_\ell>0$ without loss of generality and $\bar e_a^i$ denotes the triad field in $\partial_-B$. Then, by choosing $$p_{\mathfrak r(\ell)}=-p_\ell$$ that satisfies \eqref{eq:pxiBp2}, we get the coherent state label $\tilde g_{\mathfrak r(\ell)}$ 
\begin{equation}\label{eq:grell}
\tilde g_{\mathfrak r(\ell)}=n(J\xi_{\ell s_\ell})e^{(-ip_\ell -\theta_\ell+\beta\sgn(\bar e)\Xi_\ell)\tau_3}n(J\xi_{\ell t_\ell})^{-1}
\end{equation}
where we combine \eqref{eq:pxiBp2}, \eqref{eq:phiell}, \eqref{eq:XiinBp} and \eqref{eq:gell2}, and consider the discussion in the paragraph before \eqref{eq:minusp} showing that $\sgn(\mathfrak r^*\bar e)=-\sgn(\bar e)$.

To maintain a consistent global orientation for the two complexes dual to the triangulation of $B$, the boundary face in $\partial_+B$ corresponding to a face $f\subset \partial_-B$ is represented by $\mathfrak r(f)^{-1}$. In simpler terms, this means that the corresponding face includes the same points as $\mathfrak r(f)$ but with the opposite orientation. As a result, for a link $\ell\subset\partial_-B$, its counterpart in $\partial_+B$ is given by $\mathfrak r(\ell)^{-1}$.  By \eqref{eq:grell}, the coherent state associated with $\mathfrak r(\ell)^{-1}$ is labelled by
\begin{equation}\label{eq:grell2}
\begin{aligned}
&\tilde g_{\mathfrak r(\ell)^{-1}}=\tilde g_{\mathfrak r(\ell)}^{-1}\\
=&n(\xi_{\ell t_\ell}) e^{(-ip_\ell -\theta_\ell+\beta\sgn(\bar e)\Xi_\ell)\tau_3} n(\xi_{\ell s_\ell})^{-1}.
\end{aligned}
\end{equation}

As discussed in \eqref{eq:superpositionPsi}, we are considering superposition of various coherent states resembling the same geometry. Following the convention introduced in \eqref{eq:superpositionPsi}, the values for $g_\ell^{(++)}$ are obtained from \eqref{eq:gellbm} and \eqref{eq:grell2}.  Now, let us shift our attention to $g_\ell^{(s_1s_2)}$ for other signs of $s_1$ and $s_2$. Consider the transformation $\bar e_a^i\mapsto -\bar e_a^i$ in $\partial_-B$. This transformation does not alter the densitized triad field $E^a_i$ in $\partial_-B$, and consequently, it does not change $\xi_{\ell n}$ and $p_\ell$. Furthermore, the spin connection $\Gamma_a^i$ and the extrinsic curvature $K_{ab}$ remain unaffected. As a result, $\theta_\ell$ and $\Xi_\ell$ remains constant under this transformation.
However, due to the relationship $\sgn(\det(\bar e_a^i))=-\sgn(\det(-\bar e^i_a))$, $\phi_\ell$ will be transformed to 
\begin{equation}
\phi_\ell\mapsto \phi_\ell^{(-)}=\theta_\ell+\beta\sgn(\bar e) \Xi_\ell.
\end{equation}
This leads to the following transformation  for the coherent state label $\bar g_\ell$, 
\begin{equation}\label{eq:gellbmm}
\bar g_{\ell}\mapsto \bar g_{\ell}^{(-)}:=n(\xi_{\ell s_\ell})e^{(-ip_\ell +\theta_\ell+\beta\sgn(\bar e)\Xi_\ell)\tau_3} n(\xi_{\ell t_\ell})^{-1}.
\end{equation}
Similarly, under the transformation $\mathfrak r^*(\bar e_a^i)\mapsto -\mathfrak r^*(\bar e_a^i)$, $\tilde g_{\mathfrak r(\ell)^{-1}}$ for all $\mathfrak r(\ell)$ transformed as
\begin{equation}\label{eq:grellm}
\begin{aligned}
\tilde g_{\mathfrak r(\ell)^{-1}}\mapsto \tilde g_{\mathfrak r(\ell)^{-1}}^{(-)}:=n(\xi_{\ell t_\ell}) e^{(-ip_\ell -\theta_\ell-\beta\sgn(\bar e)\Xi_\ell)\tau_3} n(\xi_{\ell s_\ell})^{-1}.
\end{aligned}
\end{equation}
As a summary, the list of coherent state labels $\{g_\ell^{(\tilde s\bar s)}\}$ for links lying entirely in either $\partial_+B$ or $\partial_-B$ is given by 
\begin{equation}\label{eq:listlabels}
\{\tilde g_{\mathfrak r(\ell)^{-1}}^{(\tilde s)}\}_{\mathfrak r(\ell)\subset \partial_+B}\cup \{\bar g_{\ell}^{(\bar s)}\}_{\ell\subset \partial_-B},
\end{equation}
where we defined $\tilde g_{\mathfrak r(\ell)^{-1}}^{(+)}=\tilde g_{\mathfrak r(\ell)^{-1}}$ and $\bar g_{\ell}^{(+)}=\bar g_{\ell}$. 

\subsection{Boundary states associated with links at corners} 

It is crucial to highlight that the links passing through the spheres $S_1$ and $S_3$ are excluded from \eqref{eq:listlabels}. Now, let us discuss the coherent state labels for these links individually. To maintain the time reversal symmetry, we will designate such a link in a way that it can be divided into two parts, $\ell=\bar \ell\cup \tilde \ell$, where $\tilde \ell=\mathfrak r(\bar \ell)^{-1}$. In considering the values of $g^{(\tilde s\bar s)}_\ell$ for different signs of $\tilde s,\bar s$, we correspond to intrinsic geometries given by $(\tilde s\tilde e_a^i,\bar s\bar e_a^i)=(\tilde s\mathfrak{r}^*(\bar e_a^i),\bar s\bar e_a^i)$ (refer to \eqref{eq:superpositionPsi} for the convention). For each of the segments $\bar \ell$ and $\tilde \ell$, there exist associated coherent states $\psi^{t}_{\tilde g_{\tilde\ell}^{(\tilde s)}}$ and $\psi^{t}_{\bar g_{\bar\ell}^{(\bar s)}}$. Since $\bar \ell$ and $\tilde \ell$ entirely lie in $\partial_-B$ and $\partial_+B$ respectively, the coherent state labels $\tilde g_{\tilde\ell}^{(\tilde s)}$ and $\bar g_{\bar\ell}^{(\bar s)}$ can be obtained applying the algorithm introduced in Secs. \ref{sec:valuepxi}, \ref{sec:xiell}, and \ref{sec:superposing}. The results of the labbels can be parametrized as
\begin{equation}
\begin{aligned}
\tilde g_{\tilde\ell}^{(\tilde s)}=&n(\xi_{\tilde\ell s_{\tilde\ell}})e^{(-ip_{\tilde \ell}+\phi_{\tilde\ell}^{\tilde s})}n(\xi_{\tilde\ell t_{\tilde\ell}}),\\
\bar g_{\bar \ell}^{(\bar s)}=&n(\xi_{\bar\ell s_{\bar\ell}})e^{(-ip_{\bar \ell}+\phi_{\bar\ell}^{\bar s})}n(\xi_{\bar\ell t_{\bar\ell}}),
\end{aligned}
\end{equation}
where it is considered that the signs $\tilde s$ and $\bar s$ only affect the value of the angle $\phi$, as indicated by \eqref{eq:gellbm}, \eqref{eq:grell}, \eqref{eq:gellbmm}, and \eqref{eq:grellm}. Due to he time reversal symmetry, we have 
\begin{equation}
\begin{aligned}
p_{\tilde\ell}&=p_{\bar\ell},\ \xi_{\bar\ell s_{\bar\ell}}=\xi_{\tilde\ell t_{\tilde\ell}},\ \xi_{\bar\ell t_{\bar\ell}}=\xi_{\tilde\ell s_{\tilde\ell}}.
\end{aligned}
\end{equation}

Thanks to the coherent states associated to the segments, it is natural to consider the boundary state associated with $\ell$ as the combination of these individual coherent states. Specifically, the boundary state associated with $\ell$ is given by
\begin{equation}\label{eq:statecorner}
\begin{aligned}
\int_{\sut}\dd\mu_H(h)\psi^{t}_{\bar g_{\bar\ell}^{(\bar s)}}(h_{\bar\ell} h^{-1}) \psi^{t}_{\tilde g_{\tilde\ell}^{(\tilde s)}}(h h_{\tilde\ell})=\psi^{2t}_{\bar g_{\bar\ell}^{(\bar s)}\tilde g_{\tilde\ell}^{(\tilde s)}}(h_\ell).
\end{aligned}
\end{equation}
where $h_\ell=h_{\bar\ell}h_{\tilde\ell}$ is the holonomy for the entire link. 
Noticing the change of parameter $t\mapsto 2 t$, we find that the state \eqref{eq:statecorner} is approximated by $|g_\ell^{(\tilde s\bar s)}\rangle$ (refer to \eqref{eq:stateLink}) with
\begin{equation}
\begin{aligned}
g_\ell^{(\tilde s\bar s)}=&n(\xi_{\ell s_\ell})e^{(-ip_\ell+\phi_\ell^{(\tilde s\bar s)})}n(\xi_{\ell t_\ell})^{-1},\\
\xi_{\ell s_\ell}=&\xi_{\bar\ell s_{\bar\ell}}=\xi_{\ell t_\ell},\\
p_\ell=&\frac{1}{2}(p_{\tilde\ell}+p_{\bar\ell})=p_{\tilde\ell},\\
\phi_\ell^{(\tilde s\bar s)}=&\phi_{\bar\ell}^{(\bar s)}+\phi_{\tilde \ell}^{(\tilde s)}.
\end{aligned}
\end{equation}

Indeed, rather than the state \eqref{eq:statecorner}, we prefer to take  the boundary state associated with $\ell$ as $\psi^t_{g_\ell^{(\tilde s\bar s)}}$.   Even though both of the two state can be approximated by $|g_\ell^{(\tilde s\bar s)}\rangle$, the choice of $\psi^t_{g_\ell^{(\tilde s\bar s)}}$ aligns with the states chosen for other edges lying entirely in $\partial_\pm B$. Finally, by definition of $\phi_{\bar\ell}^{(\bar s)}$ and $\phi_{\tilde \ell}^{(\tilde s)}$, we get
\begin{equation}\label{eq:phiellc}
\begin{aligned}
\phi_\ell^{(\tilde s\bar s)}=\left\{
\begin{array}{cl}
0,&\tilde s=\bar s,\\
2\tilde s\beta\sgn(\bar e)\Xi_{\bar\ell},&\bar s=-\tilde s.
\end{array}
\right.
\end{aligned}
\end{equation}

\section{Computing the critical point}\label{sec:criticalPointComplex}
In the previous sections, we define the spin foam amplitude and construct the boundary data to study the dynamics of the $B$ region. In this section, we will explain how to apply the method of \emph{complex critical point} to numerically compute the spin amplitude \cite{Han:2020fil,Han:2021kll,Han:2023cen, Han:2024ydv}. This method involves obtaining the complex critical point for each set $\{g_\ell^{(\tilde s\bar s)}\}_{\ell\subset\partial B}$ with $\tilde s,\,\bar s=\pm 1$. Since the procedure for each $\{g_\ell^{(\tilde s\bar s)}\}_\ell$ is the same, we will use $g_\ell$ to represent either of $g_\ell^{(\tilde s\bar s)}$ in the following.

{By definition, the parameter $t$ is equal to $\beta\kappa\hbar$ divided  by some unit of area. In our computation, the unit, denoted by $\Delta$, will be chosen as some typical area in the triangulation. Then,} in the computation of the spin foam action \eqref{eq:boundarySf}, the parameter $t$ becomes
\begin{equation}
t=\frac{\beta\kappa\hbar}{\Delta}.
\end{equation}  
 Then, the effective action for a boundary face can be expressed in terms of $\|\vec A\|$ as
\begin{equation}\label{eq:boundarySfA}
\begin{aligned}
S_f=&-\frac{\beta\kappa\hbar}{2\Delta}\left(j_f-\frac{\|\vec A_{\ell_f}\|}{\beta\kappa\hbar}\right)^2\\
&+i\,\sgn(p_{\ell_f}) j_f \phi_{\ell_f}+j_f \mathcal S_o^{(f)}
\end{aligned}
\end{equation}
where \eqref{eq:pellandA} is applied, and all of the remaining terms are absorbed into $\mathcal S_o^{(f)}$ for convenience. 
Replacing $j_f\mapsto t^{-1} j_f$, we get
\begin{equation}
\begin{aligned}
S_f=&\frac{1}{t}\Bigg[-\frac{1}{2}\left(j_f-\frac{\|\vec A_{\ell_f}\|}{\Delta}\right)^2\\
&+i\,\sgn(p_{\ell_f}) j_f \phi_{\ell_f}+j_f \mathcal S_o^{(f)}\Bigg].
\end{aligned}
\end{equation}
This allows the action for a boundary face to take the form $t^{-1} S_f$. It is noteworthy that the substitution $j_f\to t^{-1} j_f$ can similarly modify the action for an internal face, rendering it in the form $t^{-1} S_f$. Consequently, the amplitude $A$ becomes 
\begin{equation}\label{eq:resumA}
A=\sum_{j}\int\dd g\dd ze^{\frac{1}{t} S[j,g,z]}\cong \int\dd j\dd g\dd z e^{\frac{1}{t}S[j,g,z]},
\end{equation}
where the Poisson resummation formula is employed to approximate the summation by an integral (see \cite{Han:2021kll} for more details). Thank to \eqref{eq:resumA}, we are allowed to discuss the asymptotic behavior of $A$ as $t^{-1}\to\infty$. This involves the complex critical point method.

\subsection{Complex critical point method}
As shown in \eqref{eq:resumA}, we are concerned with the integral of the form
$$
A_t=\int_{\mathcal C} e^{\frac{1}{t}S(x)}\dd\mu(x)
$$
over some $N-$dimensional configuration space $\mathcal C$ with some parameter $t$. The behavior of $A_t$ as $t\to 0$ is guided by the critical points of $S$, where a critical point is a point $x_o\in\mathcal C$  satisfying $$\partial_xS(x_o)=0.$$ 
It is worth noting that, in general, the equations $\partial_xS=0$ may lack solutions when $x$ is restricted to real values.  In such cases, the search extends to complex critical points. A complex critical point lies in the complexification $\mathcal C^{\mathbb C}$ of $\mathcal C$ and satisfies $\partial_z S=0$, where $S$ needs to be analytically continued to $\mathcal C^{\mathbb C}\ni z$. 

\begin{figure}[h]
    \centering
    \includegraphics[scale=0.35]{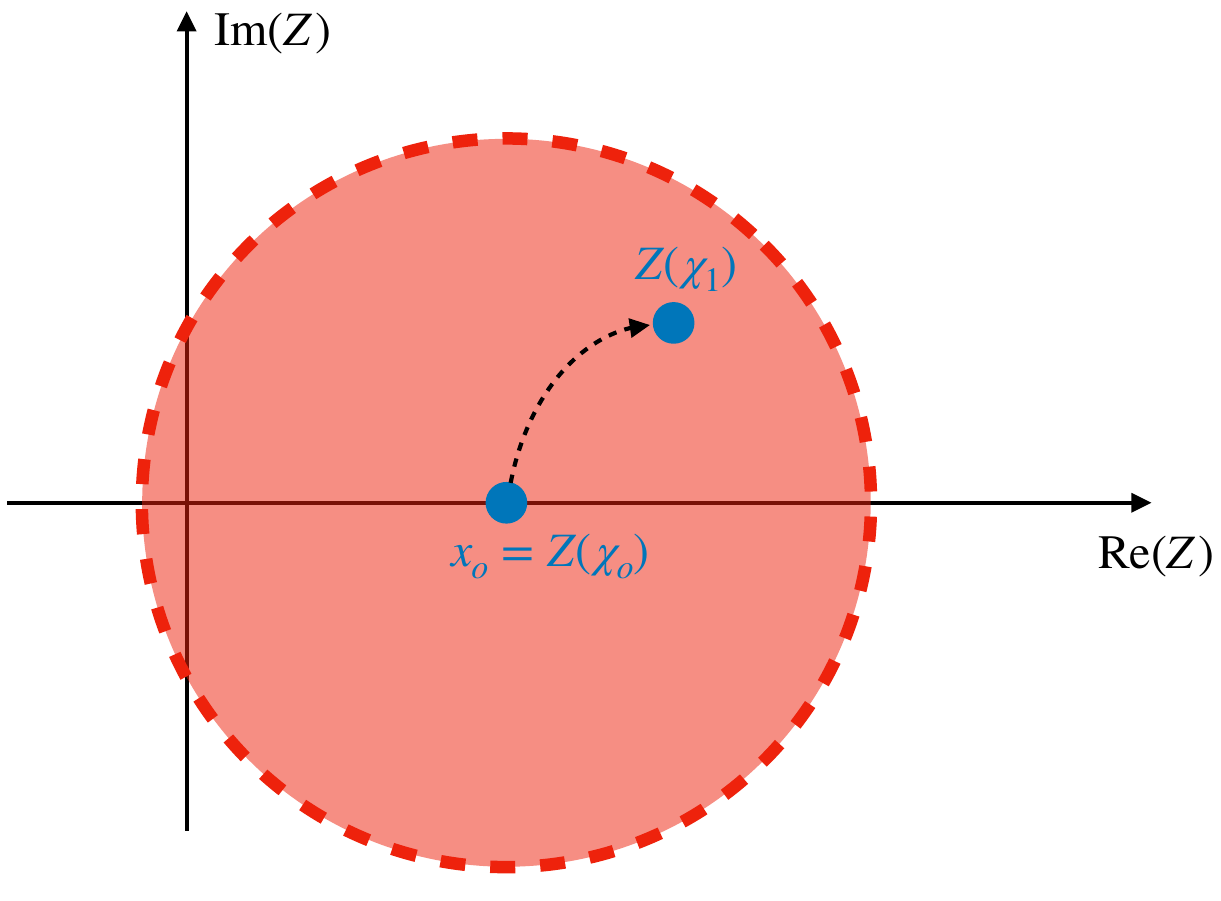}
    \caption[Caption for LOF0]{The real and complex critical points $x_o$ and $Z(\chi)$. $S(z,\chi)$ is analytic extended from the real axis to the complex neighborhood illustrated by the red disk.}
    \label{Figure0}
\end{figure}

In our computation, we will introduce some external parameters $\chi\in [\chi_o,\chi_1]$ as in \cite{Han:2021kll}, to rewrite the integral as 
\begin{equation}\label{eq:flowparamerer}
A_t(\chi)=\int_{\mathcal C} e^{\frac{1}{t} S(x,\chi)}\dd\mu(x).
\end{equation}
Here $S(x,\chi)$ is assumed to be an analytical function with respect to $\chi$. Additionally, $\chi$ should be selected in a way that $S(x,\chi_1)$ precisely describes our model of interest. Simultaneously, for $\chi=\chi_o$, $S(x,\chi_o)$  may represent a distinct model, but it possesses a real critical point\footnote{As we will see, in the concrete example, $\chi$ will be related to boundary data. }. Let $Z(\chi)$ be the solution to $\partial_x S(x,\chi)=0$ such that $Z(\chi_o)$ is the real critical point of $S(x,\chi_o)$. Here, the real critical point is defined as the solution to both $\partial_x S(x,\chi_o) = 0$ and $\mathrm{Re}(S(x,\chi_o)) = 0$. 
As $\chi$ deviates away from $\chi_o$, $Z(\chi)$ moves away from $\mathcal C$ to $\mathcal C^{\mathbb C}$ analytically. Consequently, $Z(\chi_1)\in \mathcal C^{\mathbb C}$ will be a complex critical point contribute the integral in our initial model (see Fig. \ref{Figure0} for the illustration). 
With the complex critical point, the small-$t$ asymptotic expansion for the integral in $A_t$ can be established as
\begin{equation}\label{eq:expandAlambda}
\begin{aligned}
A_t= \left(2\pi t\right)^{\frac{N}{2}} \frac{e^{\frac1t  S(Z(\chi),\chi)}}{\sqrt{\det[-H(Z(\chi),\chi)]}} \left[1+O(t)\right],
\end{aligned}
\end{equation}
where the right hand of \eqref{eq:expandAlambda} involves the action $S(Z(\chi),\chi)$ and Hessian $H(Z(\chi),\chi))$ at the complex critical point. Furthermore, the real part of $S(Z(\chi),\chi)$ satisfies the condition
\begin{equation}\label{eq:negativeReS}
\operatorname{Re}\left(S(Z(\chi),\chi)\right) \leq-C|\operatorname{Im}(Z(\chi))|^{2}.
\end{equation}
where $C$ is a positive constant \cite{Hormander}. 
This condition implies that in \eqref{eq:flowparamerer}, the purely imaginary phase can only occur at the real critical point, where $\operatorname{Im}(Z)=0$ and $\chi=\chi_o$. When $\chi$ deviates from $\chi_o$, causing $\operatorname{Im}(Z)$ finite and $\operatorname{Re}(\mathcal{S})$ to become negative, the result in \eqref{eq:flowparamerer} is exponentially suppressed as $t$ approaches $0$.

Now let us return to the integral \eqref{eq:resumA} to apply this method.
As discussed in Sec. \ref{se:sfA}, the spin foam action is a function on the space $\Gamma\ni (j_f,g_{ve},z_{vf})$. However, the integral in  \eqref{eq:resumA} is merely performed over the quotient space $\Gamma/\sim$, where the gauge orbit given by \eqref{eq:gaugetransformation} is divided. This fact implies that when considering the critical equations,, the derivatives along the gauge orbit should not be taken into account. Namely, the critical point is a solution to 
\begin{eqnarray}
\partial_{j_{f_i}}S&=&0,\quad \partial_{g_{ve}}'S=0,\quad \partial_{z_{vf}}'S=0\label{eq:SFeq1}\\
\partial_{j_{f_b}}S&=&0.\label{eq:SFeq2}
\end{eqnarray}  
where $\partial'$ denotes the derivative only along the direction transversal the gauge orbit.


{As illustrated in Fig. \ref{Figure0}, the process of obtaining a solution is divided into two steps. The first step is to find a real critical point, $x_o$. In the second step, $x_o$ is perturbed along the flow parameterized by $\chi$ until reaching $\chi_1$, resulting in a solution to \eqref{eq:SFeq1} and \eqref{eq:SFeq2}. 
In the spin foam model, a real critical point can be constructed from flat geometry. 
To construct such a critical point, it is necessary to modify the boundary data of our model of interesting to ensure that the modified boundary data permit a flat bulk geometry. The critical point $x_o$ can then be constructed from this flat geometry. To better understand the role played by $x_o$, the  following observation is helpful. Consider selecting the boundary coherent state corresponding to the boundary data of flat geometry to obtain the spin foam effective action, denoted by $\mathring{S}$.  In this case, $x_o$ becomes the solution to the critical equations \eqref{eq:SFeq1} and \eqref{eq:SFeq2}, with $S$ replaced by $\mathring{S}$. It is important to note that the critical equation \eqref{eq:SFeq1} associated with $\mathring{S}$ remains the same as the original one associated with $S$, while \eqref{eq:SFeq2} differs between $\mathring{S}$ and $S$. Therefore, $x_o$ is a solution to \eqref{eq:SFeq1} but not to \eqref{eq:SFeq2}. To obtain a solution to both \eqref{eq:SFeq1} and \eqref{eq:SFeq2}, we need to perturb $x_o$. This resulting solution is complex. 
This observation also helps guide the choice of the parameter $\chi$. By definition, $\chi$ is chosen to connect the concerning model with a model possessing a real critical point. Given the above understanding, we need to establish a relationship between the parameter $\chi$ and the boundary data, so that $S(x, \chi_o)\equiv \mathring S$.
}

To maintain the flow of the discussion, the subsequent subsections will disregard the parameter $\chi$ and focus on obtaining $x_o$ from geometry, as well as searching for complex critical point from $x_o$. When the parameter $\chi$ is considered, the procedure remains similar. We still need to obtain $x_o$ from geometry initially, but search for the critical point $Z(\chi)$ associated with $S(x,\chi)$ from $x_0$ for all $\chi\in [\chi_o,\chi_1]$\footnote{In the practical computation, we search for the critical point $Z(\chi_o+\delta\chi)$ associated with $S(x,\chi_o+\delta\chi)$ with $\delta\chi\ll 1$ from $x_0$, then search for the critical point $Z(\chi_o+2\delta\chi)$ associated with $S(x,\chi_o+2\delta\chi)$ from $Z(\chi_o+\delta\chi)$, and so forth, until we obtain the critical point associated with $S(x,\chi_1)$, i.e., the solution to both \eqref{eq:SFeq1} and \eqref{eq:SFeq2}.}. The exact definition of $\chi$ will be provided in Sec. \ref{sec:actionAtC}. 
From now on, $x_o$ will be referred to as the real critical point, while it does not fully satisfy the  critical equations  \eqref{eq:SFeq1} and \eqref{eq:SFeq2}. 

\subsection{Real  critical point}
Let us begin by explaining the choice of units in the numerical computation. in our numerical computations, when referring to an area, it is expressed as the value taking the typical area $\Delta$ as the unit. Under this unit, the parameter $t$ is equivalent to the Planck area $\beta\kappa\hbar$.



{As discussed below \eqref{eq:SFeq1} and \eqref{eq:SFeq2}, the real critical point is a solution only to \eqref{eq:SFeq1}. The absence of \eqref{eq:SFeq2} allows us to disregard the boundary conditions arising from the coherent state data $p_\ell$, $\xi_\ell$, and $\phi_\ell$. In other words, with any boundary data that permits a flat geometry, one can construct a corresponding real critical point $x_o$, which will serve as a solution to \eqref{eq:SFeq1} but not to \eqref{eq:SFeq2}. However, since the goal of introducing $x_o$ is to solve both \eqref{eq:SFeq1} and \eqref{eq:SFeq2}, we will select the boundary data corresponding to $x_o$ as closely as possible to that of our model, ensuring that $x_o$ is an approximate solution to \eqref{eq:SFeq2}. Specifically, the boundary data for $x_o$ will have the same intrinsic geometry as our model, while differing in extrinsic geometry. This means that the area vectors for each link $\ell$ will remain as $p_\ell$ and $\xi_\ell$, but the dihedral angles associated with each link will differ from $\phi_\ell$. By adjusting the dihedral angles, we can achieve a flat geometry.
}

Given a set of boundary data $\{(p_\ell,\xi_\ell)\}_{\forall \ell\in \partial B}$, the typical procedure to find the real critical point involves constructing a flat geometry that is compatible with the boundary data and extracting the spin foam data from this geometry. However, since the boundary is closed, a concern arises regarding whether the geometry satisfying the boundary condition given by $\{(p_\ell,\xi_\ell)\}_{\forall \ell\in \partial B}$ is flat. In the case at hand, the answer is negative. {As we will see, it is possible to construct a flat geometry in the region $B_-$ that satisfies the boundary data on $\partial_-B$. However, extending this flat geometry into the region $B_+$ results in a mismatch with the boundary conditions on $\partial_+B$.
Now, a tension arises: we are faced with two options. We can either construct a flat geometry throughout the entire region $B$, which will not match the intrinsic boundary geometry, or we can construct a geometry that matches the intrinsic boundary geometry but is not flat throughout the entire region. In this work, we choose the latter option. Specifically, we will construct a flat geometry in the $B_-$ region that matches the intrinsic boundary geometry of $\partial_-B$, and then reflect this geometry into the $B_+$ region, due to the time reversal symmetry, to obtain a geometry for the entire $B$ region. However, this resulting geometry is only partially flat, as there will be a deficit angle present on the transition surface $\mathcal{T}$.
}

In Sec. \ref{sec:valuepxi}, to construct  the discrete geometry approximating the intrinsic geometry of $\partial_-B$,  we introduced the points $\mathfrak i(\mathfrak p_a)$, $\mathfrak i(\mathfrak p_{a'})$ and  $\mathfrak i(\mathfrak p_{a''})$ in the Minkowski spacetime for all $a=1,2,3,4$. Additionally, in Sec. \ref{sec:constructmodel}, we introduce the triangulation of the region $B_-$ that encapsulates the connecting structure of the points $\mathfrak p_a$, $\mathfrak p_{a'}$ and $\mathfrak p_{a''}$. We thus replicate the connections between the points $\mathfrak p_a$, $\mathfrak p_{a'}$ and $\mathfrak p_{a''}$, to connect the points $\mathfrak i(\mathfrak p_a)$, $\mathfrak i(\mathfrak p_{a'})$ and  $\mathfrak i(\mathfrak p_{a''})$. This process results in a flat Regge geometry in the region $B_-$. Utilizing the geometric interpretation of the real spin-foam critical points \cite{Barrett:2009gg,Barrett:2009mw,Han:2011re}, we derive a set of data $(j_f,g_{ve},z_{vf})$ from this geometry for the $B_-$ region.
According to the definition of the boundary data $p_\ell$ and $\xi_\ell$, these data are clearly compatible with the boundary data on $\partial_-B$. Nevertheless, these data are specifically defined on the 2-complex in $B_-$ and solely solve the equations in \eqref{eq:SFeq1} associated with $j_{f_i}$, $g_{ve}$, and $z_{vf}$ within $B_-$. They even do not solve the equations $\partial_{j_{f_i}}S=0$ for the faces $f_i$ across the transition surface $\mathcal T$.

 Given a vertex $v$, an edge $e$ and a face $f$ contained in $B_-$, the time reversal symmetry $\mathfrak r:B_-\to B_+$ defines the corresponding objects $\mathfrak r(v),\mathfrak r(e),\mathfrak r(f)$ in $B_+$. With the spin foam data on the 2-complex in $B_-$, we assign the spin foam data to the 2-complex in $B_+$ as
 \begin{equation}\label{eq:extendingdata}
 (j_{\mathfrak r(f)},g_{\mathfrak r(v)\mathfrak r(e)},z_{\mathfrak r(v)\mathfrak r(f)})=\left(j_f,g_{ve}, z_{vf}\right).
 \end{equation}
 Then, we get a set of spin foam data for all $v,e,f$ in the entire $B$ region, referred to as the extended spin foam data. As shown in Appendix \ref{app:proofExtended}, it is proved that the extended  spin foam data are compatible with the boundary data on the entire boundary $\partial B$ and  solve the equations in \eqref{eq:SFeq1} for all relevant variables in the entire $B$ region. 
 
 The above discussion outlines the process of obtaining the  real  critical point. Nest, let us focus on finding the complex critical point.
 
 \subsection{Complex critical point}\label{sec:complexCCri}
 The absence of \eqref{eq:SFeq2} indicates that $x_o$ does not fully satisfy the critical equations. In other words, the geometry derived from the real   critical point does not align with the boundary data provided by the coherent state. More specifically, the resulting geometry exhibits a different dihedral angle compared to $\Xi_\ell$ computed in Sec. \ref{sec:xiell}. Introducing the dihedral angles $\Xi_\ell$ as an additional boundary condition prevents the construction of a flat geometry, even locally in $B_-$. Therefore, following the flatness property of spin foam model \cite{Dona:2020tvv}, we cannot obtain a real solution to \eqref{eq:SFeq1} and \eqref{eq:SFeq2}, and have to find complex solutions. 

 The first step in seeking a complex solution involves parameterizing $(j_f, g_{ve}, z_{vf})$. 
 The intricacies of this parametrization arise from that \eqref{eq:SFeq1} contains only the derivative along the direction transversal to the gauge orbits. This fact needs us to introduce a gauge fixing in the parametrization, to  ensure that the degrees of freedom along the gauge orbits are not considered. We denote the real   critical point as $(\mathring j_f,\mathring g_{ve}, \mathring z_{vf})$. Then the following parametrization of $(j_f, g_{ve}, z_{vf})$, with a gauge fixing condition taken into account, is applied. 

At first, for each vertex $v$, we choose an edge $e_o(v)$ and fix $g_{v e_o(v)}$ as
\begin{equation}
g_{v e_o(v)}=\mathring g_{v e_o(v)}. 
\end{equation}
This fixes the $\sltc$ gauge transformation in \eqref{eq:gaugetransformation}. Note that in the current work we make the choice of $e_o(v)$ such that $e_o(v)\neq e_o(v')$ for all pairs of $v$ and $v'$ sharing the same edge. Second, for each internal edge $e$, we choose a vertex $v_o(e)$ and parametrize $g_{v_o(e)e}$ as
\begin{equation}\label{eq:sutgauge}
g_{v_o(e)e}=\mathring g_{v_o(e)e}
\begin{pmatrix}
\frac{x_{v_o(e)e}^1}{\sqrt{2}}&\frac{x_{v_o(e)e}^2+y_{v_o(e)e}^2}{\sqrt{2}}\\
0&\mu_{v_o(e)e}
\end{pmatrix}.
\end{equation} 
Here, $\mu_{v_o(e)e}$, along with analogous parameters introduced subsequently, is determined by $\det(g_{v_o(e)e})=1$. This fixes the $\sut$ gauge transformation in \eqref{eq:gaugetransformation}. Note that \eqref{eq:sutgauge} needs to be applied to all internal edges, including those $e_o(v)$ that were chosen for gauge fixing the $\sltc$ transformation. For each $e_o(v)\equiv \tilde e$, its corresponding $v_o(\tilde e)$  is selected as the other endpoint compared to $v$. This choice is feasible due to the conditions we imposed on selecting $v_o(e)$. Moreover, all of the other $g_{ve}$ are parameterized as
\begin{equation}
g_{ve}=\mathring g_{ve}
\begin{pmatrix}
1+\frac{x_{ve}^1+i y_{ve}^1}{\sqrt{2}}&\frac{x_{ve}^2+y_{ve}^2}{\sqrt{2}}\\
\frac{x_{ve}^3+y_{ve}^3}{\sqrt{2}}&\mu_{ve}
\end{pmatrix}
.
\end{equation} 
Furthermore, the spinors $z_{vf}$ are parametrized as
\begin{equation}
z_{vf}=\left\{
\begin{array}{ll}
(1,\frac{\mathring z_{vf}^2}{\mathring z_{vf}^1}+ x_{vf}),& \text{ if } \mathring z_{vf}^1\neq 0,\\
( x_{vf},1),&  \text{ if } \mathring z_{vf}^1= 0.
\end{array}
\right.
\end{equation}
which fixes the rescaling transformation of $z_{vf}$ in  \eqref{eq:gaugetransformation}. Finally $j_f$ is parametrized as 
\begin{equation}
j_f=\mathring j_f(1+x_f).
\end{equation}

Substituting the above parametrization into the action \eqref{eq:actiontotal}, we write the action as a function of the real variables $x^i_{ve}$, $y^i_{ve}$, $x_{vf}$ and $x_f$. We then extend the domain of these variables from the real numbers to complex numbers. 
The critical equations with respect to these complex variables are given by $\partial_zS=0$ and $\partial_{\overline{z}}S=0$ for $z$ taking the complexification of $x^i_{ve}, y^i_{ve},x_{vf}$ or $x_f$, where $\overline z$ denote the complex conjugate of $z$.

A technique issue appearing here is the incorporation of the value $\Xi_\ell$ provided in Sec. \ref{sec:xiell} into our actual computation. This issue is noteworthy due to the lack of explanation regarding the definition of the twist angle $\theta_\ell$.  According to the expression of the action given in \eqref{se:sfA}, $S$ takes the form 
\begin{equation}
S=S_o+\sum_{f\subset\partial B} i\sgn(p_{\ell_f})j_f\phi_{\ell_f} , 
\end{equation}
where $\phi_\ell$ contains the twist angle $\theta_{\ell}$ and the dihedral angle $\Xi_{\ell}$.  We introduce $\mathring\phi_{\ell}$ as the coherent label given by
\begin{equation}\label{eq:involveXi}
\mathring\phi_{\ell_f}:=\left.i \partial_{j_f}S_o\right|_{x_o},\quad \forall f\subset\partial B. 
\end{equation}
 $\mathring\phi_{\ell}$  ensures $\partial_{j_{f_b}}S$ takes vanishing value at $x_o$ for all boundary face $f_b$. As we discussed in \eqref{eq:appA7}, $\mathring\phi_{\ell_f}$ is equal to the twist angle plus a term proportional to the dihedral angle. { This expression can be viewed as the discrete counterpart of the Ashtekar connection,  $A_a^i = K_a^i + \gamma \Gamma_a^i$ , where  $K_a^i$  corresponds to the dihedral angle and  $\Gamma_a^i$  to the twist angle \cite{PhysRevD.82.084040, Rovelli:2010km}. Thus,} the twist angles depend only on the intrinsic geometry which remains consistent at both the real  and the complex critical points (see \cite{PhysRevD.82.084040} for more details).
As a consequence,  the twist angle in $\mathring\phi_{\ell}$ is identical to that in $\phi_{\ell}$. Combining this observation with \eqref{eq:phiell}, we arrive at the expression
\begin{equation}\label{eq:phiellello}
\phi_\ell=\mathring\phi_\ell+\beta(\mathring\Xi_\ell-\sgn(p_{\ell_f})\sgn(e)\Xi_\ell).
\end{equation}
where $\mathring\Xi_\ell=-\partial_\beta \mathring\phi_\ell$ denotes the dihedral angle compatible with the flat Regge geometry associated with $x_o$. Substituting the result into \eqref{eq:involveXi}, we get
\begin{equation}
\begin{aligned}
S=&S_o+\sum_{f\subset\partial B}i\sgn(p_{\ell_f}) j_f\mathring\phi_{\ell_f}\\
&-i \beta\sum_{f\subset\partial B}(\sgn(e)\Xi_\ell-\sgn(p_{\ell_f})\mathring\Xi_\ell)
\end{aligned}
\end{equation}
 
For a given  real  critical point $x_o=(\mathring j_f,\mathring g_{ve}, \mathring z_{vf})$, the parity transformation 
\begin{equation}
(\mathring j_f,\mathring g_{ve}, \mathring z_{vf})\mapsto (\mathring j_f,\mathring g_{ve}^\dagger{}^{-1}, \frac{\mathring g_{ve}\mathring g_{ve}^\dagger \mathring z_{vf}}{\|\mathring g_{ve}^\dagger \mathring z_{vf}\|^2} )\equiv \tilde x_o
\end{equation}
yields another  real  critical point denoted by $\tilde x_o$. As shown in \cite{Han:2011re}, under the parity transformation, $\mathring\Xi_\ell$ is transformed as
\begin{equation}
\mathring\Xi_\ell\mapsto -\mathring\Xi_\ell.
\end{equation}
In the numerical procedure for searching the complex critical point, the initial point should be chosen from either $x_o$ or $\tilde x_o$ such that the corresponding $\mathring\Xi_\ell$ is closer to $\Xi_\ell$. 


\section{Numerical investigation}\label{sec:num}

%
In the current work, we consider the case with the parameter,
$$GM=2\times 10^5\sqrt{\beta\kappa\hbar},\quad \beta=\frac{1}{10}.$$
In this case, the parameter $\alpha$ takes the value $$\alpha=\frac{\sqrt{3}}{50}\beta\kappa\hbar.$$

With the chosen parameters, we can proceed with the numerical computation step by step. Some keys issues in the numerical computation are introduced as follows. 

\subsection{Fix the {location of } $\partial B$}
{Even though the free parameters as shown above are fixed, we still have the ambiguity on choosing the size of the $B$ region in the entire spacetime. This needs us to fix the location of  $\partial B$.} As outlined in the beginning of Sec. \ref{sec:boundarydata}, this requires us to set the parameters $r_a$ with $a=1,2,3$ and $\theta_i$ with $i=1,2$ for $\partial_-B$. Subsequently, the boundary $\partial_+B$ can be determined accordingly, taking into account the time reversal symmetry.

In our computation, we choose
\begin{equation}
\begin{aligned}
r_1&=\left(\alpha GM/2\right)^{1/3},\\
r_2&=(2 GM+\frac{GM}{100}),\ r_3=(2GM+\frac{GM}{10}).
\end{aligned}
\end{equation}
Here, $r_1$ is chosen as the minimal radial where the collapsing dust ball bounces in the quantum Oppenheimer-Snyder model. $r_2<r_3$ are choose as two radials nearby the horizon. {Consequently, the resulting $B$  region spans from the minimal radius up to just outside the horizon, effectively capturing the region expected to contain the quantum gravity regime. Additionally,  $r_2$  and  $r_3$  are positioned close to the horizon, ensuring that the  $B$  region remains compact and computationally manageable.} 

We select the values of $\theta_1$ and $\theta_2$ to ensure that all the tetrahedra in the flat geometry corresponding to the  real critical point $x_o$ are spacelike. This selection guarantees the applicability of the spacelike EPRL model.  However, it is important to note that the above criterion alone cannot uniquely determine the values of $\theta_1$ and $\theta_2$. 
In the current work, we choose the following values as an example
\begin{equation}\label{eq:criticaltheta}
\begin{aligned}
\theta_1&=\theta_1^{(c)}-10^{-5}= 0.6930807205426226,\\
\theta_2&=\theta_2^{(c)}-10^{-6}= 4.5270907503925455.
\end{aligned}
\end{equation}
with
\begin{equation*}
\begin{aligned}
\theta_1^{(c)}=&\arccosh\left(\frac{9 (r_1-r_2)^2}{2 f(r_1) r_2 (2 r_2-3 r_1)}-1\right),\\
\theta_2^{(c)}=&\arccosh\left(\frac{9}{4f(r_3)}-1\right).
\end{aligned}
\end{equation*}
When we embed the region $B_-$ into Minkowski spacetime using the map $\mathfrak{i}$,  $\theta_1^{(c)}$ and $\theta_2^{(c)}$ are the critical values to ensure the following conditions (see App. \ref{app:derivetheta12c} for detailed derivations):
\begin{itemize}
\item[(1)] All tetrahedra in the transition surface $\mathfrak i(\mathcal T)$ are spacelike;
\item[(2)] The boundaries $\mathfrak i(\partial_-B_1)$ and $\mathfrak i(\partial_-B_2)$ are spacelike;
\item[(3)] The time coordinates $t_2$ and $t_3$ of the spheres $\mathfrak{i}(S_2)$ and $\mathfrak{i}(S_3)$ respectively satisfy $t_2 < t_3$, so that $\mathfrak{i}(S_3)$ occurs in the future of $\mathfrak{i}(S_2)$.
\end{itemize}

Once the boundary $\partial B$ is fixed, the unit $\Delta$ is chosen as the minimal area in the triangulation of $\partial B$. In our case, it is the area of the triangle $\mathfrak p_1\mathfrak p_2\mathfrak p_3$. 
We have
\begin{equation}
\Delta=\|\vec A_{\mathfrak p_1\mathfrak p_2\mathfrak p_3}\|= 264.3604304333458856 \beta\kappa\hbar,
\end{equation}
leading to 
\begin{equation}
t=\frac{1}{264.3604304333458856},
\end{equation}
as the parameter for our asymptotic expansion in \eqref{eq:expandAlambda}.


\subsection{Action at the complex critical point}\label{sec:actionAtC}
Let us use $\sres$ to represent the rescaled action, where the spin foam action is expressed as $\sres/t$. As previously mentioned, $\sres$ takes the form
\begin{equation}\label{eq:sres}
\begin{aligned}
\sres=&\sum_{f_b\subset\partial B}\Bigg\{-\frac{1}{2}\left(j_{f_b}-\frac{\|\vec A_{\ell_{f_b}}\|}{\Delta}\right)^2+j_{f_b}\Big[i\sgn(p_{\ell_{f_b}}) \mathring\phi_{\ell_{f_b}}-\\
&i \chi\, \beta\Big(\sgn(e)\Xi_\ell-\sgn(p_{\ell_f})\mathring\Xi_\ell\Big)+\mathcal S_o^{(f_b)}\Big]\Bigg\}+\\
&\sum_{\text{internal face }f_i}j_{f_i}\mathcal S_o^{(f_i)}.
\end{aligned}
\end{equation} 
Here, we introduce the parameter $\chi\in[0,1]$. When $\chi=1$, $\sres$ represents the action associated with curved-geometry boundary data. Conversely, for $\chi=0$, $\sres$ corresponds to the action associated with the flat-geometry boundary data. 
As illustrated in Fig. \ref{Figure0}, we will solve the critical equations for all $\chi\in [0,1]$, to make the critical point flow from the  real one to the complex one.

By \eqref{eq:SFeq1}, the critical point satisfyies $\partial_{j_i} \sres=0$ and $\partial_{j_b} \sres=0$. In addition, \eqref{eq:sres} indicates that $\sres$ exhibits linearity concerning $j_{f_i}$ and quadratic dependence on $j_{f_b}$. As a consequence, we get the value of $\sres$ at the critical point, denoted by $\sresc$, as
\begin{equation}\label{eq:sresCri}
\sresc=\sum_{f_b\subset\partial B}\frac{\mathfrak b_{f_b}}{2} \left(\frac{2 \|\vec A_{\ell_{f_b}}\|}{\Delta }+\mathfrak b_{f_b}\right)
\end{equation} 
where $\mathfrak b_{f_b}$ is defined as
\begin{equation}\label{eq:bchi}
\begin{aligned}
\mathfrak b_{f_b}=&i\sgn(p_{\ell_{f_b}}) \mathring\phi_{\ell_{f_b}}+\mathcal S_o^{(f_b)}\big|_c\\
&-i \chi\, \beta\Bigg(\sgn(e)\Xi_\ell-\sgn(p_{\ell_f})\mathring\Xi_\ell\Bigg),
\end{aligned}
\end{equation}
with $\mathcal S_o^{(f_b)}\big|_c$ denoting the value of $\mathcal S_o^{(f_b)}$ at the critical point. 

It is worth noting that $\mathcal S_o^{(f_b)}\big|_c$ depends on $\chi$ since the critical points vary with $\chi$.
Indeed, each summand in \eqref{eq:sresCri} is just the value of $-1/2 \left(j_{f_b}-\|\vec A_{\ell_{f_b}}\|/\Delta \right)^2+j_{f_b} \mathfrak{b}_{f_b}$ at its turning point. By substituting \eqref{eq:bchi} into \eqref{eq:sresCri},  $\sres$ at the critical point  reads
\begin{equation}\label{eq:sresp}
\sresc=a\chi^2+b\chi+c
\end{equation}
with
\begin{eqnarray}
a
&=&-\frac{1}{2}\sum_{f_b\subset \partial B}\left(\phi_{\ell_{f_b}}-\mathring\phi_{\ell_{f_b}}\right)^2,\\
b&=&i \sum_{f_b\subset\partial B}\sgn(p_{\ell_{f_b}})\left(\phi_{\ell_{f_b}}-\mathring\phi_{\ell_{f_b}}\right)\times\nonumber\\
&&  \left(i\sgn(p_{\ell_{f_b}}) \mathring\phi_{\ell_{f_b}}+\mathcal S_o^{(f_b)}\big|_{c}+\frac{\|\vec A_{\ell_{f_b}}\|}{\Delta }\right),\\
c&=&\frac{1}{2}\sum_{f_b\subset\partial B}\left(i\sgn(p_{\ell_{f_b}}) \mathring\phi_{\ell_{f_b}}+\mathcal S_o^{(f_b)}\big|_{c}\right)\times\nonumber\\
&&\left(i\sgn(p_{\ell_{f_b}}) \mathring\phi_{\ell_{f_b}}+\mathcal S_o^{(f_b)}\big|_{c}+\frac{2\|\vec A_{\ell_{f_b}}\|}{\Delta }\right).
\end{eqnarray}
Here, \eqref{eq:phiellello} is applied to get the value of $a$ and $b$. It should be remarked that the dependence of $\mathcal S_o^{(f_b)}\big|_{c}$ on $\chi$ leads to the dependence of $b$ and $c$ on $\chi$. Consequently, \eqref{eq:sresp} is not really a quadratic function of $\chi$.

\subsection{Property of  the real  critical point}\label{sec:orientationchange}

As mentioned earlier, we need to find the complex critical point for each set $\{g_\ell^{(\tilde s\bar s)}\}_{\ell\subset\partial B}$ with $\tilde s,,\bar s=\pm 1$. However, the following considerations suggest that we only need to focus on the case where $\tilde s=\bar s$. 

At first, according to the discussion for \eqref{eq:expandAlambda} and \eqref{eq:negativeReS},  the amplitude decreases rapidly as the complex critical point moves away from the  real  critical point $x_o$.  The distance between the complex critical point and $x_o$ is measured by the disparity in boundary conditions corresponding to the complex and real critical points. This observation aligns with the outcome presented in \eqref{eq:sresp}, where the  coefficient of the quadratic term is proportional to the sum of $(\phi_{\ell_{f_b}}-\mathring\phi_{\ell_{f_b}})^2$ for all boundary faces $f_b$. Consequently,  instead of considering all possible $\{g_\ell^{(\tilde s\bar s)}\}_{\ell\subset\partial B}$ with $\tilde s, \bar s=\pm 1$, we can focus specifically on those cases where the associated boundary conditions are closer to that of the  real  critical point $x_o$.

The numerical results demonstrate that $\mathring \phi_\ell$ defined in \eqref{eq:involveXi} satisfies
\begin{equation}
\mathring\phi_{\ell}=-\mathring\phi_{\mathfrak r(\ell)^{-1}}
\end{equation}
and, for each link $\ell_c$ at the corners,
\begin{equation}
\mathring\phi_{\ell_c}=0.
\end{equation}
As a consequence, among the boundary conditions  corresponding to $\{g_\ell^{(\tilde s\bar s)}\}_{\ell\subset\partial B}$ with $\tilde s,,\bar s=\pm 1$, those closer to the boundary conditions given by $x_o$ are the ones satisfying 
\begin{equation}\label{eq:closer}
\phi_{\ell}=- \phi_{\mathfrak r(\ell)^{-1}},\quad \phi_{\ell_c}=0.
\end{equation}
As shown by the equations \eqref{eq:gellbm}, \eqref{eq:grell2}, \eqref{eq:gellbmm} and \eqref{eq:grellm},  the condition \eqref{eq:closer} holds only for $\{g_\ell^{(\tilde s\bar s)}\}_{\ell\subset\partial B}$ with $\tilde s=\bar s$. 

Indeed, given a link $\ell_c$ at the corner, for $\tilde s=-\bar s$, \eqref{eq:phiellc} yields $|\phi_{\ell_c}|=2|\Xi_{\ell_c}|$, relating $|\phi_{\ell_c}|$ to the extrinsic curvature at the corner. 
Given that the corner $S_1$ is located in a highly quantum region, and the choice of $\partial B$ ensures its local intrinsic flatness, the extrinsic curvature nearby $S_1$ is expected to be very large. That implies a large value of $|\phi_{\ell_c}|$ for $\ell_c$ passing through $S_1$. Namely, $\phi_{\ell_c}$ for $\ell_c$ passing through will deviate significantly from $\mathring \phi_{\ell_c}=0$. Consequently, this leads to a highly negative value of the real part of the action at the complex critical point for $\tilde s=-\bar s$, as indicated in \eqref{eq:negativeReS}.

\subsection{Final results}\label{sec:final}
\begin{figure}
\centering
\includegraphics[width=0.45\textwidth]{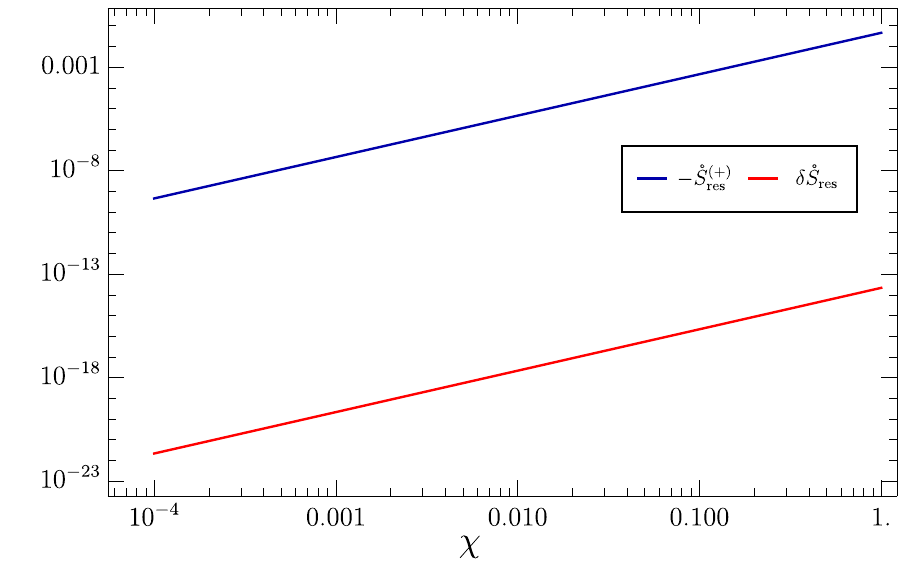}
\caption{The values of $-\sresc^{(+)}$ and $\delta \sresc=\sresc^{(+)}-\sresc^{(-)}$ as a function of the  parameter $\chi\in [0,1]$.}\label{fig:valueSeff}
\end{figure}

As discussed in the last subsection, our attention should be solely directed towards the complex critical point for the boundary data $\{g_\ell^{(++)}\}_{\ell\subset\partial B}$ and $\{g_\ell^{(--)}\}_{\ell\subset\partial B}$. The detailed numerical results are available in \cite{numericalResult}. We denote the values of $\sresc$ associated with these boundary data as $\sresc^{(+)}$ and $\sresc^{(-)}$ respectively. Our numerical investigation reveals that both $\sresc^{(+)}$ and $\sresc^{(-)}$ are real values, as illustrated in Fig. \ref{fig:valueSeff}. According to the numerical results, for $\chi=1$, we have
\begin{equation}\label{eq:result}
\begin{aligned}
\sresc^{(+)}\big|_{\chi=1}=&-0.0458193513442056,\\
\sresc^{(-)}\big|_{\chi=1}=& -0.0458193513442275.
\end{aligned}
\end{equation}
It is interesting that there exist a tiny difference between $\sresc^{(+)}$ and $\sresc^{(-)}$. According to the numerical results in Fig. \ref{fig:valueSeff}, $\ln(-\sresc^{(\pm)})$ exhibits an approximately linear dependence on $\ln(\chi)$. The numerical fitting yields 
\begin{equation}
\begin{aligned}
\ln(-\sresc^{(+)})=&1.9999999999996853 \ln(\chi)\\
&-3.0830487586650438+\epsilon_+(\chi),\\
\ln(-\sresc^{(-)})=&1.9999999999996856 \ln(\chi)\\
&-3.0830487586645638 +\epsilon_-(\chi).
\end{aligned}
\end{equation}
with the residuals satisfying $|\epsilon_{\pm}(\chi)|<2\times 10^{-12}$.

The nature of the geometry corresponding to the complex critical point is still an open issue. It is expected to deviate from the geometry corresponding to the  real critical point $x_o$ with some perturbations.  Let us refer to the geometry associated with $x_o$ as the $x_o$-geometry for convenience. Then, the $x_o$-geometry can provide insight into geometry associated with the complex critical point.

An interesting property of the $x_o$-geometry is that the orientation of the 4-simplices in $B_-$ is dynamical. {Namely, the orientation of 4-simplices in $B_-$ could differ from one 4-simplex to another. }
By computing the deficit angle $\vartheta_{\mathfrak p_a\mathfrak p_{b'}\mathfrak p_{c''}}$ hinged at the internal triangle $\mathfrak p_a\mathfrak p_{b'}\mathfrak p_{c''}\subset B_-$, we obtain
\begin{equation}\label{eq:defectangle}
\begin{aligned}
&|\vartheta_{\mathfrak p_a\mathfrak p_{b'}\mathfrak p_{b''}}|=1.6756169215214705\neq 0,\\
&\forall a,b=1,2,3,4,\ a\neq b.
\end{aligned}
\end{equation}
Let $\Theta_{eve'}$ be the dihedral angle between the 2 tetrahedra $e$ and $e'$ in the 4-simplex $v$. Then, we have
\begin{equation}
\vartheta_{\mathfrak p_a\mathfrak p_{b'}\mathfrak p_{b''}}=\sum_{v}\Theta_{eve'},
\end{equation}
where the sum is taken over all 4-simplices sharing $\mathfrak p_a\mathfrak p_{b'}\mathfrak p_{b''}$.  Equation \eqref{eq:defectangle} implies
\begin{equation}\label{eq:sumdihedral1}
\sum_{v}\Theta_{eve'}\neq 0.
\end{equation}
Furthermore, the fact that $x_o$ is a real critical point implies
\begin{equation}\label{eq:sumdihedral2}
\sum_{v}\sgn(V_v)\Theta_{eve'}=0,
\end{equation}
where $V_v$ represents the volume of the 4-simplex $v$. Consequently, $\sgn(V_v)$ varies among the 4-simplexes sharing the triangles $\mathfrak p_a\mathfrak p_{b'}\mathfrak p_{b''}$. For example, we consider the case where $a=2$ and $b=1$. The triangle $\mathfrak p_2\mathfrak p_{1'}\mathfrak p_{1''}$ is shared by four 4-simplices
\begin{equation}
\begin{aligned}
v_1&=\mathfrak p_2\mathfrak p_{3}\mathfrak p_{4}\mathfrak p_{1'}\mathfrak p_{1''},\quad v_2=\mathfrak p_2\mathfrak p_{3}\mathfrak p_{1'}\mathfrak p_{4'}\mathfrak p_{1''},\\
v_3&=\mathfrak p_2\mathfrak p_{1'}\mathfrak p_{3'}\mathfrak p_{4'}\mathfrak p_{1''},\quad v_4=\mathfrak p_2\mathfrak p_{4}\mathfrak p_{1'}\mathfrak p_{3'}\mathfrak p_{1''}.
\end{aligned}
\end{equation}
We find
\begin{equation}
\begin{aligned}
-\sgn(V_{v_1})=\sgn(V_{v_2})=\sgn(V_{v_3})=\sgn(V_{v_4}).
\end{aligned}
\end{equation}
Indeed, for each triangle $\mathfrak p_a\mathfrak p_{b'}\mathfrak p_{b''}$, its boundary always contains a unique 4-simplex $v_1$ given by $v_1=\mathfrak p_a\mathfrak p_c\mathfrak p_d\mathfrak p_{b'}\mathfrak p_{b''}$, where $c$ and $d$ are neither $a$ nor $b$. The numerical computation confirms that the sign of $V_{v_1}$ is opposite to the signs of all other 4-simplices sharing the triangle. 

The dihedral angles occurring on the transition surface $\mathcal T$ are associated with the triangles in the triangulation of $\mathcal T$. In the current model, the values of these dihedral angles can be classified into two types.
The first type  pertains to the triangles where two points lie on the sphere $S_1$, while the third point lies on the sphere $S_3$. An example is the triangle $\mathfrak p_1 \mathfrak p_2 \mathfrak p_{2''}$. The second type relates to the triangles where only one point is on the sphere $S_1$, and the other two points are on the sphere $S_3$. An example is the triangle $\mathfrak p_1 \mathfrak p_{3''} \mathfrak p_{4''}$. In the current numerical computation, the dihedral angle of the first type takes the value
\begin{equation}
|\Theta_1|=5.6445575060432456,
\end{equation}
and the second type takes
\begin{equation}
|\Theta_2|=0.4095700383700623.
\end{equation}

\section{Summary and outlook}\label{sec:conclusion}

In this work, we investigate the quantum dynamics of the $B$-region in the black-to-white hole transition suggested in \cite{Han:2023wxg}.  To construct the spin foam amplitude, we first introduce a systematic procedure to yield a bulk triangulation from the boundary triangulation in the $B$-region. Then, we define the spin foam amplitude based on the 2-complex dual to the triangulation, where the Thiemann's complexifier coherent state is chosen as the boundary state to resemble the semiclassical boundary geometry. Since the triad fields with different orientations, i.e., $e^a_i$ and $-e^a_i$, yield the same boundary geometry, our work considers the superposition of the coherent states with respect to  both orientations as the boundary state. The superposition switch on the quantum tunneling process involving the change in orientation. 

Moreover, we examine how the amplitude behaves as $t$ approaches $0$, where $t$ is the parameter describing the variance of the coherent state. Within this context, the amplitude can be computed numerically with the method of stationary phase approximation. This method involves solving the critical equation $\delta S=0$, where $S$ represents the effective action in the amplitude. Following the procedure outlined in the paper to solve the critical equation, we obtain our numerical results, as shown in Sec. \ref{sec:num}.

As discussed in Sec. \ref{sec:orientationchange}, the amplitude is dominated by the terms allowing the change of orientation from the past  boundary $\partial_-B$ to the future boundary $\partial_+B$. This result indicates that the black-to-white hole transition should be accompanied by the quantum tunneling process of the change of orientation from  $\partial_-B$ to $\partial_+B$. It  is noted that the change of orientation was also discovered in the loop quantum reduced model \cite{Giesel:2023hys}. 
It will be interesting to explore how the findings in the two models relate to each other. 

In the amplitude, there are two terms that allow for the change of orientation. They correspond to the boundary data $\{g_\ell^{(++)}\}_{\ell\subset\partial B}$ and $\{g_\ell^{(--)}\}_{\ell\subset\partial B}$ as described in Sec. \ref{sec:superposing}. According to the numerical results presented in Sec. \ref{sec:final}, the values of the action at the critical point differ slightly for these two sets of boundary data. The reasons for the difference are still unclear and require future investigation.

{In this work, we apply the conjecture introduced in previous research (see, e.g., \cite{DAmbrosio:2020mut}). In these studies, the region of classical spacetime where QG effects become relevant is divided into three sub-regions, with the  $B$  region—our focus in this work—being one of them. The QG effects in these three sub-regions are addressed using different theories. Specifically, different from the  $B$ region, the other two regions are handled using effective loop-quantum-cosmology-like dynamics. Although the results from loop-quantum-cosmology-like dynamics have gained broad acceptance, a more consistent approach would be to study QG effects across all regions within a unified framework. This is a direction we plan to explore in future work. Indeed, there have already been attempts to apply the spin foam model to cosmological settings \cite{Han:2024ydv}.  
}

The current study focuses solely on determining the value of the effective action at the complex critical point. However, to obtain the final value of the amplitude, additional factors such as the determinant of the Hessian at the complex critical point, as indicated in equation \eqref{eq:expandAlambda}, are required. Furthermore, our computations need to fix the values of parameters such as the black hole mass $M$, the radii $r_1$, $r_2$, and $r_3$, as well as the angles $\theta_1$ and $\theta_2$. {We have successfully obtained the results, demonstrating the feasibility of studying the BH-to-WH transition numerically using the complex critical point method. As the next step, these calculations need to be repeated for various parameter sets. Although the results are expected to be exponentially suppressed, their relative values will provide deeper insights into the BH-to-WH transition within the framework of the spin foam model.} Addressing these issues will be part of our future work. {The primary obstacle in this study is computational power. For the current results, obtaining the two data points in \eqref{eq:result} required approximately 20 days using Mathematica code. To improve computational efficiency, potential strategies such as implementing the code in other programming languages or utilizing more powerful computational resources should be considered.
}

\begin{acknowledgments}
This work is supported by the National Natural Science Foundation of China with Grants No. 12275022. M.H. receives supports from the NSF through grant PHY-2207763, the College of Science Research Fellowship at Florida Atlantic University. MH, DQ and CZ are supported by research grants provided by the Blaumann Foundation. The Government of Canada supports research at Perimeter Institute through Industry Canada and by the Province of Ontario through the Ministry of Economic Development and Innovation. This work benefits from the visitor's supports from Beijing Normal University, FAU Erlangen-N\"urnberg, and the University of Western Ontario. 
\end{acknowledgments}

\onecolumngrid

\appendix

\section{Spin foam amplitude for boundary face with the Thiemann's coherent state}\label{app:sfBdyAction}
Let $f$ be a boundary face surounded by edges $e_0,e_1,\cdots,e_n$, where the order of the edges is compatible with the orientation of $f$. Thus $e_0$ and $e_n$ are the two edges connected to the boundary. The vertices in the boundary of $f$ are labeled as $v_n=e_{n-1}\cap e_n$. The direction of boundary link $\ell$ is defined to go from $e_0$ to $e_n$. The amplitude is a function of the boundary holonomy $h_\ell$, reading
\begin{equation}
A_f(h_{\ell_f})=\tr(Y_\gamma^\dagger g_{e_0 v_1}g_{v_1e_1}Y_\gamma\cdot Y_\gamma^\dagger g_{e_1 v_2}g_{v_2e_2} Y_\gamma\cdot\cdots\cdot Y_\gamma^\dagger g_{e_{n-1} v_n}g_{v_ne_n} Y_\gamma h_\ell^{-1})
\end{equation}
In this work, we are interested in the case where the boundary state is taken as the coherent state \cite{Thiemann:2000bw,Thiemann:2000ca}
\begin{equation}\label{eq:coherentstates1234}
\psi_{g_\ell}(h_\ell)=\sum_{j_\ell}d_{j_\ell}e^{-\frac{t}{2}j_\ell(j_\ell+1)}\sum_{m}D^{j_\ell}_{mm}(g_\ell^{-1} h_\ell),
\end{equation}
with $g_\ell\in\sltc$. Then, we have the spin foam amplitude associated with $f$ as
\begin{equation}
\begin{aligned}
\langle A_f|\psi_{g_{\ell_f}}\rangle=\int\dd\mu_H(h_\ell)A_f(h_\ell)\psi_{g_{\ell_f}}.
\end{aligned}
\end{equation}
Following the derivations for an internal face \cite{Han:2013gna}, we finally get 
\begin{equation}
\langle A_f|\psi_{g_{\ell_f}}\rangle=\sum_{j_f}\int \prod_{f}(d_{j_f})^{|V_f|+1}  \prod_{vf}\dd^2\Omega_{vf}
e^{S_f}.
\end{equation}
where the action $S_f$ is
\begin{equation}\label{eq:actionBoundary}
\begin{aligned}
S_f=&j_f \Bigg[\sum_{\text{ internal edges }e}i\beta \ln( \frac{ \| Z_{s_eef}\|^2}{\| Z_{t_e ef}\|^2} )+\sum_{\text{internal edges }e}2\ln(\frac{\langle Z_{s_e e f},Z_{t_eef}\rangle}{\| Z_{s_eef}\|\| Z_{t_e ef}\| })+\\
&i\beta \ln( \frac{ \| Z_{s_{e_n}e_nf}\|^2}{\| Z_{t_{e_0} e_0f}\|^2} )  +2\ln(\frac{\langle Z_{s_{e_n} e_n f},g_\ell^{-1}Z_{t_{e_0}e_0f}\rangle}{\|Z_{s_{e_n} e_n f}\| \| Z_{t_{e_0}e_0f}\| })\Bigg]-\frac{t}{2}j_f(j_f+1)
\end{aligned}
\end{equation}
with $Z_{vef}=g_{ve}^\dagger z_{vf}$ and $\|Z\|=\sqrt{\langle Z,Z\rangle}$ for $Z\in\mathbb C^2$.

\section{Equation of motion obtained from the spin foam action}\label{app:eombdyaction}

In the spin foam model, the amplitude reads
\begin{equation}
A=\sum_j\int\dd g\dd z e^{S(j,g,z)}.
\end{equation}
The action $S$, in the EPRL model, takes the form (see, e.g., \cite{Han:2013gna,Han:2020fil} and Appendix \ref{app:sfBdyAction})
\begin{equation}\label{eq:actionform}
S(j,g,z)=\sum_{\text{ internal face }f}j_fS_f+\sum_{\text{ boundary face }f}\left(j_fS'_f-\frac{t}{2}j_f(j_f+1)\right)
\end{equation}
where $S_f$ and $S'_f$ are functions depending merely on the $\sltc$ element $g$ and the spinor $z$. 

Applying the Poisson summation formula for the amplitude \cite{Han:2021kll}, one has
\begin{equation}
A=\sum_{k=-\infty}^\infty \int \dd g\dd z\dd x'  e^{S(x',g,z)-2\pi i k x'}\cong \int \dd g\dd z\dd x'  e^{S(x',g,z)},
\end{equation}
where the focus is on the phase in the exponential:
\begin{equation}
S=\sum_{\text{ internal face }f}x'_fS_f+\sum_{\text{ boundary face }f}\left(x'_fS'_f-\frac{t}{2}x'_f(x'_f+1)\right).
\end{equation}

To study the asymptotic behavior of $A$ for $t\to 0$, we rescale $x'$ by introducing $x=t x'$ so that we get
\begin{equation}
S=\frac{1}{t}\left(\sum_{\text{ internal face }f}x_fS_f+\sum_{\text{ boundary face }f}\left(x_fS'_f-\frac{1}{2} x_f^2\right)\right)-\frac{x_f}{2}.
\end{equation}
Then, the asymptotic behavior for $t\to 0$ can be analyzed by the stationary phase approximation, where we define the effective action as
\begin{equation}\label{eq:effectiveactiont20}
S_{\rm eff}=\sum_{\text{ internal face }f}x_fS_f+\sum_{\text{ boundary face }f}\left(x_fS'_f-\frac{1}{2} x_f^2\right)
\end{equation}
The critical point is the solution to the equation of motion $\delta S_{\rm eff}=0$. The properties of the critical point have been well studied in cases where the 2-complex contains no boundary (see, e.g., \cite{Han:2011re,Han:2013gna}). We can apply the results obtained from these studies, with the only exception being the treatment of the boundary faces carrying the coherent state \eqref{eq:coherentstates1234}.  Therefore, our attention should be directed towards the effective action associated with a boundary face $f$
\begin{equation}
\begin{aligned}
S_{\rm eff}^{(f)}
=&j_f \Bigg[\sum_{\text{ internal edges }e}i\beta \ln( \frac{ \| Z_{s_eef}\|^2}{\|Z_{t_e ef}\|^2} )+\sum_{\text{internal edges }e}2 \ln(\frac{\langle Z_{s_e e f},Z_{t_eef}\rangle}{\| Z_{s_eef}\|\|Z_{t_e ef}\| })+\\
&i\beta \ln( \frac{ \| Z_{s_{e_n}e_nf}\|^2}{\| Z_{t_{e_0} e_0f}\|^2} )+2\ln(\frac{\langle Z_{s_{e_n} e_n f},g_\ell^{-1}Z_{t_{e_0}e_0f}\rangle}{\| Z_{s_{e_n} e_n f}\|\|Z_{t_{e_0}e_0f}\| })-\frac{t}{2}j_f^2\Bigg].
\end{aligned}
\end{equation}
Then, \eqref{eq:effectiveactiont20} is a result of the replacement $j_f\mapsto x_f/t$. 

Treating $j_f$ as an integration variable, we need to study the equation $\partial_{j_f}S=0$, indicating that both its real and imaginary part are 0.  Starting with its real part, we get
 \begin{equation}\label{eq:realpart}
\begin{aligned}
\sum_{\text{internal edges }e}  \mathrm{Re} \ln(\frac{\langle Z_{s_e e f},Z_{t_eef}\rangle}{\| Z_{s_eef}\|\|Z_{t_e ef}\|}) +\mathrm{Re}\ln(\frac{\langle Z_{s_{e_n} e_n f},g_\ell^{-1}Z_{t_{e_0}e_0f}\rangle}{\| Z_{s_{e_n} e_n f}\| \|Z_{t_{e_0}e_0f}\| })-\frac{t}{2}j_f=0
\end{aligned}
\end{equation}
We impose the following decomposition for $g_\ell$
\begin{equation}
g_\ell=n_{s_\ell}e^{(-ip+\phi)\tau_3}n_{t_\ell}^{-1},
\end{equation}
where $n_{e_\ell}, n_{t_\ell}\in\sut$, $p,\phi\in \mathbb R$. Then, we could define new variables
\begin{equation}\label{eq:tildeZ}
\tilde Z_{s_{e_n}e_nf}=n_{t_\ell}^{-1}Z_{s_{e_n}e_nf},\quad \tilde Z_{t_{e_0}e_0f}=n_{s_\ell}^{-1}Z_{t_{e_0}e_0f}
\end{equation}
and rewrite \eqref{eq:realpart} as
\begin{equation}\label{eq:realpartp}
\begin{aligned}
\sum_{\text{internal edges }e}\mathrm{Re} \ln(\frac{\langle Z_{s_e e f},Z_{t_eef}\rangle}{\| Z_{s_eef}\| \|Z_{t_e ef}\| }) +\mathrm{Re}\ln(\frac{\langle \tilde Z_{s_{e_n} e_n f},e^{(ip-\phi)\tau_3}\tilde Z_{t_{e_0}e_0f}\rangle}{\|\tilde Z_{s_{e_n} e_n f}\|\| \tilde Z_{t_{e_0}e_0f}\| })-\frac{t}{2}j_f=0
\end{aligned}
\end{equation}
To solve this equation, we impose an additional requirement that the real part of the action itself should vanish. The additional requirement is motivated by the desire to find the critical point where the transition amplitude takes its maximal absolute value. However, a subtle complication arises due to the non-normalization of the employed boundary state. The norm of the boundary state $\psi_g^t$ takes \cite{Zhang:2020mld,Zhang:2021qul}
\begin{equation}
\|\psi_g^t\|=\frac{\sqrt{2\sqrt{\pi}}e^{t/8}\sqrt{|p|}e^{\frac{|p|^2}{2t}}}{t^{3/4}\sqrt{\sinh(|p|)}}.
\end{equation}
Therefore, for effective action corresponding to the transition amplitude  $\langle A_f|\psi_g^t\rangle/\|\psi_g^t\|$, the condition of vanishing its real part results in
\begin{equation}\label{eq:realpartp1}
\begin{aligned}
&\sum_{\text{internal edges }e}\mathrm{Re} \ln(\frac{\langle Z_{s_e e f},Z_{t_eef}\rangle}{\| Z_{s_eef},\|\| Z_{t_e ef}\| }) +\mathrm{Re}\ln(\frac{\langle \tilde Z_{s_{e_n} e_n f},e^{(ip-\phi)\tau_3}\tilde Z_{t_{e_0}e_0f}\rangle}{\| \tilde Z_{s_{e_n} e_n f}\|\| \tilde Z_{t_{e_0}e_0f}\| })\\
&-\frac{|p|^2}{4j_f t}-\frac{t}{4}j_f=0
\end{aligned}
\end{equation}
Comparing this equation with \eqref{eq:realpartp}, we get
\begin{equation}
-\frac{|p|^2}{4j_f t}-\frac{t}{4} j_f=-\frac{t}{2}j_f,
\end{equation}
leading to the area matching condition
\begin{equation}\label{eq:areamatching}
j_f=\frac{|p|}{t}.
\end{equation}
Taking the area matching condition  into account, one can demonstrate 
 \begin{equation}\label{eq:lessthanone}
 \begin{aligned}
 &e^{-tj_f}\left|\left\langle \frac{\tilde Z_{s_{e_n}e_nf}}{\|\tilde Z_{s_{e_n}e_nf} \|},e^{(ip-\phi)\tau_3} \frac{\tilde Z_{s_{e_0}e_0f}}{\|\tilde Z_{s_{e_0}e_0f}\|}\right\rangle^2 \right|\\
 =&\left(\sin\left(\frac{\beta_0}{2}\right) \sin\left(\frac{\beta_n}{2}\right)e^{\frac{p-|p|}{2}}+\cos\left(\frac{\beta_0}{2}\right)\cos\left(\frac{\beta_n}{2}\right)e^{\frac{-|p|-p}{2}}\right)^2-e^{-|p|}\frac{1}{2} \sin ( \beta_0) \sin (\beta_n) [1-\cos (\gamma_0-\gamma_n-\phi )]\\
 \leq &\cos^2(\frac{\beta_n-\beta_0}{2})\leq 1.
 \end{aligned}
 \end{equation}
 Here, the parametrization is applied as follows
\begin{equation}
\begin{aligned}
\frac{\tilde Z_{s_{e_0}e_0f}}{\|\tilde Z_{s_{e_0}e_0f}\|}=&\left\{\sin \left(\frac{\beta_0 }{2}\right) \exp \left(\frac{i \alpha_0 }{2}-\frac{i \gamma_0}{2}\right),\cos \left(\frac{\beta_0}{2}\right) \exp \left(\frac{i \alpha_0 }{2}+\frac{i \gamma_0 }{2}\right)\right\},\\
\frac{\tilde Z_{s_{e_n}e_nf}}{\|\tilde Z_{s_{e_n}e_nf} \|}=&\left\{\sin \left(\frac{\beta_n }{2}\right) \exp \left(\frac{i \alpha_n }{2}-\frac{i \gamma_n}{2}\right),\cos \left(\frac{\beta_n}{2}\right) \exp \left(\frac{i \alpha_n }{2}+\frac{i \gamma_n }{2}\right)\right\}.\\
\end{aligned}
\end{equation}
The result \eqref{eq:lessthanone}, combined with the inequality $
\frac{\langle Z_{s_eef},Z_{t_eef}\rangle^2}{\| Z_{s_eef}\|^2\| Z_{t_eef}\|^2}\leq 1$,
results in the solution of \eqref{eq:realpartp},
\begin{equation}\label{eq:realpartsol}
\begin{aligned}
\frac{Z_{s_eef}}{\|Z_{s_eef}\|}=&e^{i\alpha_{ef}}\frac{Z_{t_eef}}{\|Z_{t_eef}\|},\\
\frac{\tilde Z_{t_{e_0}s_0f}}{\|\tilde Z_{t_{e_0}s_0f}\|}=&\left\{
\begin{array}{ll}
e^{i\frac{\alpha_0-\gamma_0}{2}}(1,0),&p>0,\\
e^{i\frac{\alpha_0+\gamma_0}{2}}(0,1),&p<0,
\end{array}
\right.\\
\frac{\tilde Z_{s_{e_n}s_nf}}{\tilde Z_{s_{e_n}s_nf}}=&\left\{
\begin{array}{ll}
e^{i\frac{\alpha_n-\gamma_n}{2}}(1,0),&p>0,\\
e^{i\frac{\alpha_n+\gamma_n}{2}}(0,1),&p<0.
\end{array}
\right.
\end{aligned}
\end{equation}
When \eqref{eq:realpartsol} is satisfied, we have
\begin{equation}
\begin{aligned}
\mathfrak{P}\left(\left\langle \frac{\tilde Z_{s_{e_n}e_nf}}{\|\tilde Z_{s_{e_n}e_nf} \|},e^{(ip-\phi)\tau_3} \frac{\tilde Z_{s_{e_0}e_0f}}{\|\tilde Z_{s_{e_0}e_0f}\|}\right\rangle^2\right)\Bigg|_{\text{critical point }}=e^{i (\alpha_0-\alpha_n)-\sgn(p)(\gamma_0-\gamma_n-\phi) }
\end{aligned}
\end{equation}
where $\mathfrak{P}(z):=z/|z|$ for $z\in\mathbb C$. In addition, combining \eqref{eq:tildeZ} and \eqref{eq:realpartsol}, we have
\begin{equation}
\frac{Z_{t_{e_0}e_0f}}{\|Z_{t_{e_0}e_0f}\|}=n_{s_\ell} e^{i\frac{\alpha_0-\sgn(p)\gamma_0}{2}}\mathring Z(p),\quad \frac{Z_{s_{e_n}e_nf}}{\|Z_{s_{e_n}e_nf}\|}=n_{t_\ell} e^{i\frac{\alpha_n-\sgn(p)\gamma_n}{2}}\mathring Z(p). 
\end{equation}
where $\mathring Z(p)=(1,0)^T$ for $p>0$ and  $\mathring Z(p)=(0,1)^T$ for $p<0$.  

The above discussion solves the equation $\partial_{j_f}S_{\rm eff}=0$ for a boundary face $f$. Let us summarize the results as follows:

For a boundary face $f$ carrying the coherent state \eqref{eq:coherentstates1234} labelled by $g_\ell=n_{s_\ell}e^{(-ip+\phi)\tau_3}n_{t_\ell}^{-1},$
the area matching condition $j_f=|p|/t$ leads to the following solution of  $\partial_{j_f}S_{\rm eff}=0$,
\begin{equation}\label{eq:djsvanishingFinal}
\begin{aligned}
&\frac{Z_{s_eef}}{\|Z_{s_eef}\|}=e^{i\alpha_{ef}}\frac{Z_{t_eef}}{\|Z_{t_eef}\|},\quad \frac{Z_{t_{e_0}e_0f}}{\|Z_{t_{e_0}e_0f}\|}= e^{i\alpha_{e_0f}}\xi_{s_{\ell}},\quad \frac{Z_{s_{e_n}e_nf}}{\|Z_{s_{e_n}e_nf}\|}= e^{i \alpha_{e_nf}}\xi_{t_{\ell}}, \text{ and, }\\
&\sum_{\text{ internal }e\in\partial f}\beta \ln( \frac{ \| Z_{s_eef}\|}{\| Z_{t_e ef}\|} )+\sum_{\text{internal edges }e}\mathrm{Im}\ln(\frac{\langle Z_{s_e e f},Z_{t_eef}\rangle}{\| Z_{s_eef}\|\| Z_{t_e ef}\| })+\\
&\beta  \ln( \frac{ \| Z_{s_{e_n}e_nf}\|}{\| Z_{t_{e_0} e_0f}\|} )+\mathrm{Im}\ln(\frac{\langle \xi_{s_{\ell}},Z_{t_{e_0}e_0f}\rangle}{\|\xi_{s_{\ell}} \|\|Z_{t_{e_0}e_0f}\|})+\mathrm{Im}\ln(\frac{\langle Z_{t_{e_n}e_nf},\xi_{t_{\ell}}\rangle}{\|Z_{t_{e_n}e_nf}\| \|\xi_{t_{\ell}} \|})+\frac{\sgn(p)}{2}\phi=0,
\end{aligned}
\end{equation}
for $\alpha_{ef}\in\mathbb R$ and $k\in\mathbb Z$, where we define 
\begin{equation}\label{eq:bdyspinor}
\xi_{s_{\ell}}=n_{s_\ell}\mathring Z(p),\  \xi_{t_{\ell}}=n_{t_\ell}\mathring Z(p). 
\end{equation} 

The solutions to the equations $\partial_zS=0$ and $\partial_gS=0$ for boundary faces exhibit same properties as the solutions of the same equations associated with internal faces,
allowing for the direct application of results from previous literature \cite{Han:2011re,Han:2013gna}.

It is worth noting that the first three equations in \eqref{eq:djsvanishingFinal} can be recovered through the usuall model, where the boundary state is selected as $|j_f\xi_{t_{\ell}}\rangle\otimes \langle j_f \xi_{s_{\ell}}|$.  Let $S^{(0)}_f$ denote the value of the action taken at the critical points in that conventional model. Then, the last equation in \eqref{eq:djsvanishingFinal} can be expressed as 
\begin{equation}
S^{(0)}_f+ ij_f \sgn(p)\phi=0.
\end{equation}
This outcome indicates that an alternative choice for the boundary state is the Gaussian state given by
\begin{equation}\label{eq:bdystate2}
\psi_\ell=\sum_{j_f}\exp(-\frac{t}{2}\left(j_f-\frac{|p|}{t}\right)^2+ij_f\sgn(p)\phi)|j_f\xi_{t_{\ell}}\rangle\otimes \langle j_f \xi_{s_{\ell}}|,
\end{equation}
allowing one to derive the same critical equation as in \eqref{eq:djsvanishingFinal}.
The spin foam action associated with the boundary face $f$ carrying the boundary state \eqref{eq:bdystate2} reads
\begin{equation}\label{eq:actionBoundary2}
\begin{aligned}
S_f=&j_f \Bigg[\sum_{\text{ internal }e\in\partial f}i\beta \ln( \frac{ \| Z_{s_eef}\|^2}{\| Z_{t_e ef}\|^2} )+\sum_{\text{internal edges }e}2\,\ln(\frac{\langle Z_{s_e e f},Z_{t_eef}\rangle}{\| Z_{s_eef}\|\| Z_{t_e ef}\| })+\\
&i\beta\ln( \frac{\| \xi_{s_\ell}\|^2}{\|Z_{t_{e_0} e_0f}\|^2} ) +i\beta \ln( \frac{ \| Z_{s_{e_n}e_nf}\|^2}{\| \xi_{t_\ell}\|^2} ) +2\ln(\frac{\langle \xi_{s_{\ell}},Z_{t_{e_0}e_0f}\rangle}{\|\xi_{s_{\ell}} \|\|Z_{t_{e_0}e_0f}\|})\\
& +2 \ln(\frac{\langle Z_{t_{e_n}e_nf},\xi_{t_{\ell}}\rangle }{\|Z_{t_{e_n}e_nf}\| \|\xi_{t_{\ell}} \|})\Bigg]-\frac{t}{2}(j_f-\frac{|p|}{t})^2+i\,\sgn(p)j_f\phi \\
=&j_f \Bigg[\sum_{e\subset\partial f}i\beta \ln( \frac{ \| Z_{s_eef}\|^2 }{\| Z_{t_e ef}\|^2} )+\sum_{e\subset \partial f}2 \ln(\frac{\langle Z_{s_e e f},Z_{t_eef}\rangle}{\| Z_{s_eef}\|\| Z_{t_e ef}\| })\Bigg]-\frac{t}{2}\left(j_f-\frac{|p|}{t}\right)^2+i\,\sgn(p)j_f\phi 
\end{aligned}
\end{equation}
where in the last line we introduce the following convention: when $e=e_0$, the term $Z_{s_eef}$ becomes $\xi_{s_\ell}$, and when $e=e_n$, the term $Z_{t_eef}$ becomes $\xi_{t_\ell}$.

\section{Bridge Continuous Geometry with Coherent States}\label{app:A}
\subsection{Classical phase space}\label{app:A1}
As shown in \cite{Han:2011re}, the value of the spin foam action at the critical point takes the form
\begin{equation}\label{eq:action0}
S=-i\epsilon\sum_{f}\sum_{v \text{ surrounding }f}\sgn(V_4(v)) j_f\Theta_{v}+\text{extra term}
\end{equation}
where $\epsilon=\pm$ is a global sign on the two  complex, $\sgn(V_4(v))$ denote the sign of the 4-volume of the simplex dual to the vertex $v$, and $\Theta_v$ is the 4-D dihedral angle between the two tetrahedra dual to the boundary edges $e$ and $e'$ of $f$ with $e\cap e'=v$. The result is different from the Regge action by the factor $\sgn(V_4(v))$. As we know, the Regge action can be regarded as a discretized version of the Einstein Hilbert action $\int\sqrt{-g}R\dd^4x$. The sign $\sgn(V_4(v))$ factor in  \eqref{eq:action0} makes us believe that $\sqrt{-g}$ should be replaced by the signed volume $\mathfrak e\equiv \det(\se_\mu^I)$ in the classical action corresponding to the spin foam model, where $\se_\mu^I$ is the coframe field. Due to the extra term in \eqref{eq:action0}, we consider the action with the Holst term, 
\begin{equation}\label{eq:SFclassicalaction}
\begin{aligned}
S_G=\frac{1}{2\kappa\beta}\int_{\mathcal M} \se\, \se^\sigma_M\se^\theta_N P^{MN}{}_{ KL}\Omega_{\sigma\theta}^{KL}\dd^4 x
\end{aligned}
\end{equation}
with $\kappa=8\pi G$ and $\beta$ being the Barbero–Immirzi parameter, where $P^{MN}{}_{ KL}=\beta\delta^{[M}_K\delta^{N]}_L+\frac{1}{2}\epsilon^{MN}{}_{ KL}$, and $\Omega^{KL}_{\sigma\theta}$ is the curvature form of the $\mathfrak{sl}(2,\mathbb C)$ connection compatible with $\se_\mu^I$. Note that the integral in \eqref{eq:SFclassicalaction} depends on the orientation of the coordinate system, unlike the case where the action takes $\sqrt{|g|}$ as the volume form. This needs us to fix the orientation of the coordinate system to be the right hand when defining the integral in \eqref{eq:SFclassicalaction}. Then a minus sign should be added if a left hand coordinate system is chosen to do the integration.   

Let $\mathcal M$ be diffeomorphic to $\mathbb R\times \Sigma$ with $\Sigma$ dnoting the spatial manifold. Performing the canonical analysis on the action \eqref{eq:SFclassicalaction}, we get the canonical pairs $(A_a^i,E^b_j)$, where $A_a^i$ is the usual Ashtekar connection but $E^b_j$  becomes
\begin{equation}\label{eq:flux0}
E^b_j=e e^b_j,
\end{equation}
with $e\equiv \det(e_a^i)$.  In this expression and what follows, indices $a,b,\cdots$ denotes spatial tensors, and $i,j,k,\cdots$ represent tensors with value in $\mathfrak{su}(2)$. Under the coordinate transformation from $x$ to $y$, the flux is transformed as 
\begin{equation}
E^b_j\to \det(\frac{\partial(x^1,x^2,x^3)}{\partial(y^1,y^2,y^3)})E^b_j.
\end{equation}
As a consequence, the dual of the flux, i.e., $E^b_j\varepsilon_{bcd}$, is a 2-form. This is in contrast to the case where the flux is $\sqrt{q}e^b_j=|e|e^b_j$, resulting in its dual to be a pseudo 2-form. 

The Poisson bracket between the canonical pairs is 
\begin{equation}\label{eq:poisson0}
\{A_a^i(x), E^b_j(y)\}=\kappa\beta\delta_a^b\delta_j^i\delta^3(x, y).
\end{equation}
Due to the definition  \eqref{eq:flux0} of $E$, the $\delta$-function in the equation is the one such that $\int\delta^3(y_0,y)\dd^3 y=1$ for a right hand coordinate system $y^a$, and $\int\delta^3(y_0,y)\dd^3 y=-1$ for a left hand coordinate system $y^a$.  This ensures the independence of the choice of coordinate systems for the equality 
$\{A_a^i(x),\int E^b_j(y)f^b_j(y)\dd^3 y\}=\kappa\beta f^b_j(x)$, where $f^b_j$ is a smeared function.

\subsection{From the continuous to the discrete phase spaces}\label{app:holonomyflux}

The holonomy $h_\ell$ along an oriented link $\ell:t\mapsto \ell(t)$ with $t\in[0,1]$ is the solution to 
\begin{equation}\label{eq:holonomy}
\partial_t h_{\ell(t)}(A)=h_{\ell(t)}A_a(\ell(t))\dot \ell^a(t),\ h_{\ell(0)}=1,\ h_\ell\equiv h_{\ell(1)}.
\end{equation}
For a link $\ell$, let $S_\ell$ be a 2-surface transversally intersecting $\ell$ at at $p=\ell(1/2)$. 
The covariant fluxes are defined as 
\begin{equation}\label{eq:flux}
\begin{aligned}
p^i_{s,\ell}(A,E):=&\frac{-1}{\beta\kappa\hbar }\tr\left[-2 \tau^i\int_{S_\ell}  h(\rho^s_\ell(\sigma) ) E^c(\sigma) n_c(\sigma) h(\rho^s_\ell(\sigma)^{-1})\dd^2 \sigma\right]\\
p^i_{t,\ell}(A,E):=&\frac{1}{\beta\kappa\hbar}\tr\left[-2\tau^i\int_{S_\ell} h(\rho^t_\ell(\sigma) ) E^c(\sigma)n_c(\sigma) h(\rho^t_\ell(\sigma)^{-1})\dd^2\sigma \right]
\end{aligned}
\end{equation} 
where $\rho_\ell^s(x)$ is a path starting from the source of $\ell$, traveling along $\ell$ until 
reaching $p$, and then running in $S_\ell$ to reach $x\in S_\ell$, and $\rho_\ell^t(x)$ is a similar path but starting from the target of $\ell$. 
In order to provide a clearer understanding of the definition in \eqref{eq:flux}, let us clarify how to do the integration.  To begin with, a right hand coordinate system $\{x^0,\sigma^1,\sigma^2\}$ is chosen in a region containing the surface $S_\ell$. It is required that $(\sigma^1,\sigma^2)$ could be chosen as a coordinate on $S_\ell$. With this coordinate, $n_c$ is defined as the conormal with direction compatible with the orientation of $S_\ell$, satisfying $n_c(\partial/\partial x^0)^c=\pm 1$. 
Without loss of generality, the orientation of $S_\ell$ is chosen to be compatible with $\ell$ in what follows. In the general case where the orientation of $S_\ell$ is allowed to be incompatible with $\ell$, one only needs the replacements $p^i_{s,\ell}\mapsto \sgn(\dot\ell^a n_a)p^i_{s,\ell}$ and $p^i_{t,\ell}\mapsto \sgn(\dot\ell^a n_a)p^i_{t,\ell}$ to ensure the validity of all subsequent conclusions.

According to \eqref{eq:flux}, one has
\begin{equation}\label{eq:fluxrelatedbyholonomy}
p^i_{s,\ell}\tau_i=-h_\ell p^i_{t,\ell}\tau_i h_\ell^{-1}. 
\end{equation}
The non-vanishing Poisson brackets between the covariant fluxes and the holonomy are 
\begin{equation}\label{eq:Poissonalgebra}
\begin{aligned}[]
\{h_\ell,p^i_{s,\ell}\}&=-\tau^ih_\ell,\\
\{h_\ell,p^i_{t,\ell}\}&=h_\ell\tau^i,\\
\{p^i_{s,\ell},p^j_{s,\ell}\}&=-\epsilon_{ijl}p^l_{s,\ell},\\
\{p^i_{t,\ell},p^j_{t,\ell}\}&=-\epsilon_{ijl}p^l_{t,\ell},
\end{aligned}
\end{equation}
where $\tau_k:=-i\sigma_k/2$ with $\sigma_k$ being the Pauli matrix. 
In the quantum theory, a wave function $\psi(A)$ is given by a cylindrical function
\begin{equation}
\psi(A)=\psi_\gamma(\{h_\ell(A)\}_{\ell\in E(\gamma)})
\end{equation}
where $\gamma$ is a graph embedded in $\Sigma$ comprising links $\ell\in E(\gamma)$ and $\psi_\gamma$ is a complex function on $\sut^{|E(\gamma)|}$ with  $|E(\gamma)|$ denoting the number of elements in $E(\gamma)$. The action of the flux operators on the wave functions are 
\begin{equation}\label{eq:operatoraction}
\begin{aligned}
\hat p^i_{s,\ell}\psi(A)&=i\frac{\dd}{\dd t}\Big|_{t=0}\psi_\gamma(h_{\ell'}(A),\cdots,e^{t\tau_i}h_\ell(A),\cdots),\\
\hat p^i_{t,\ell}\psi(A)&=i\frac{\dd}{\dd t}\Big|_{t=0}\psi(h_{\ell'}(A),\cdots,h_\ell(A)e^{-t\tau_i},\cdots),
\end{aligned}
\end{equation}
which implements the Poisson algebra \eqref{eq:Poissonalgebra} in such a way that $[\cdot,\cdot]=i\{\cdot,\cdot\}$.

\subsection{Coherent state}\label{app:coherentstate}
In this section, we focus on a graph of a single link $\ell$ for our calculation. Given $g_\ell\in\sltc$ which is parametrized by 
\begin{equation}
g_\ell=e^{-i\vec p_\ell\cdot\vec\tau}u_\ell,
\end{equation}
the coherent state labeled by $g_\ell$, as shown in \eqref{eq:coherentstates1234}, is given by
\begin{equation}\label{eq:coherentstate}
\psi_{g_\ell}^t(h_\ell)=\sum_jd_je^{-\frac{t}{2}j(j+1)}\sum_mD^j_{mm}(h_\ell g_\ell^{-1} ).
\end{equation}
For a unit vector $\vec v=(\cos\alpha\sin\eta,\sin\alpha\sin\eta,\cos\eta),$
we introduce $n(\vec v)\in\sut$ 
such that 
\begin{equation}\label{eq:np}
n(\vec v)\tau_3 n(\vec v)^{-1}=\vec v\cdot\vec\tau.
\end{equation}
Using these notions, we can decompose $g_\ell$ as
\begin{equation}\label{eq:decompose}
g_\ell=n(\vec p_\ell/p_\ell) e^{(-ip_\ell+\phi_\ell )\tau_3} n(\vec{p'}_\ell/p_\ell)^{-1}, 
\end{equation}
where $p_\ell\in\mathbb R$ is given by $|p_\ell|=\|\vec p_\ell\|$, and $\vec{p'}_\ell$ as well as $\phi_\ell $ is defined by 
\begin{equation}
u_\ell =n(\vec p_\ell/p_\ell ) e^{\phi_\ell \tau_3} n(\vec{p'}_\ell/p_\ell)^{-1}.
\end{equation}
It can be verified that 
\begin{equation}
\begin{aligned}
u_\ell ^{-1} (\vec p_\ell \cdot\vec\tau ) u_\ell =\vec{p'}_\ell\cdot\vec\tau.
\end{aligned}
\end{equation}
%

The geometry interpretation for the parameters $\vec p_\ell $, $\vec{p'}_\ell$ and $\phi_\ell $ can be deduced from the expectation values of operators. We have \cite{Zhang:2020qxw}
\begin{equation}
\begin{aligned}
\frac{\langle\psi_{g_\ell}^t|\hat p_{s,\ell}^k|\psi_{g_\ell}^t\rangle}{\|\psi_{g_\ell}^t\|^2}&\cong \frac{1}{t}p_\ell ^k\\
\frac{\langle\psi_{g_\ell}^t| \hat p_{t,\ell}^k|\psi_{g_\ell}^t\rangle}{\|\psi_{g_\ell}^t\|^2}&\cong-\frac{1}{t}p'_\ell{}^k,\\
\frac{\langle \psi_{g_\ell}^t| h_\ell|\psi_{g_\ell}^t\rangle}{\|\psi_{g_\ell}^t\|^2}&\cong n(\vec p_\ell /p_\ell)e^{\phi_\ell \tau_3}n(\vec{p'}_\ell/p_\ell)^{-1}.
\end{aligned}
\end{equation}
In terms of the expection values, $g_\ell$ is expressed as
\begin{equation}\label{eq:sl2cgell}
g_\ell=\exp(-it\langle \hat p_{s,\ell}^k\rangle\tau_k)\langle h_\ell\rangle. 
\end{equation}

\subsection{Compute $g_\ell$ from the intrinsic and extrinsic geometry}\label{app:Adc}

The Ashtekar connection can be written in terms of the spin connection $\Gamma_a^i$ and the extrinsic curvature for $K_a^i$ as
\begin{equation}
A_a=\Gamma_a+\beta K_a. 
\end{equation}
Given a link $\ell:[0,1]\ni t\mapsto \ell_t\in\Sigma$, we consider $G_{\ell_t}$ as the parallel transport of the spin connection, i.e., solution to
\begin{equation}\label{eq:spinholo}
\partial_tG_{\ell_t}=G_{\ell_t}\Gamma_a(\ell_t)\dot \ell^a_t,\ G_{\ell_0}=1.
\end{equation}
Decomposing $h_{\ell_t}$ into
\begin{equation}\label{eq:hdecom}
h_{\ell_t}=V_{\ell_t}G_{\ell_t},
\end{equation}
we could obtain the equation for $V_{\ell_t}$, 
\begin{equation}\label{eq:equV}
\partial_t V_{\ell_t}=\beta V_{\ell_t}G_{\ell_t} K_a(\ell_t)\dot\ell^a_tG_{\ell_t}^{-1},\ V_{\ell_0}=1. 
\end{equation}

Let the link $\ell$ be a geodesic with respect to the intrinsic geometry given by the triad $e^a_i=\pm \sqrt{|\det E|}^{-1} E^a_i$. A straightforward calculation shows that
\begin{equation}
\partial_t(G_{\ell_t} (e_a^i(\ell_t)\dot\gamma^a_t\tau_i)G_{\ell_t}^{-1})=0
\end{equation}
leading to the following property for $G_\ell\equiv G_{\ell_1}$
\begin{equation}\label{eq:propertyGe}
G_\ell (e_a^i(\ell_1)\dot\ell^a_1\tau_i)G_\ell^{-1}=e_a^i(\ell_0)\dot\ell^a_0\tau_i.
\end{equation}
Let $n_s$ and $n_s$ be given by \eqref{eq:np} for $\vec v$ being $-\dot\ell^a_0e_a^i(\ell_0)/\|\dot\ell_0\|$ and $-\dot\ell^a_1e_a^i(\ell_1)/\|\dot\ell_1\|$ respectively. They satisfy
\begin{equation}\label{eq:nsntgamma}
n_s\tau_3 n_s^{-1}=-\frac{\dot\ell^a_0e_a^i(\ell_0)\tau_i}{\|\dot\ell_0\|},\quad n_t\tau_3 n_t^{-1}=-\frac{\dot\ell^a_1e_a^i(\ell_1)\tau_i}{\|\dot\ell_1\|}.
\end{equation}
Taking advantage of $n_s$ and $n_t$ and applying \eqref{eq:propertyGe}, we get that $G_\ell$ takes the following form,
\begin{equation}\label{eq:expressionG}
G_\ell=n_s e^{\theta_\ell\tau_3} n_t^{-1},
\end{equation}
for some real number $\theta_\ell$.

In order to calculate $V_{\ell_t}$, we assume that the extrinsic curvature $K_{ab}$ takes the form, 
\begin{equation}\label{eq:assumeK}
K^{ab}(t)=\frac{\xi(t)}{\|\dot\ell_t\|^2}\dot\ell^a_t\dot\ell^b_t
\end{equation}
along the link $\ell$, for a field $\xi$ on $\ell$. Indeed, \eqref{eq:assumeK} is assumed because the extrinsic curvature in the Regge calculus takes the same form with a distributional field $\xi$. It is easy to verify that 
\begin{equation}\label{eq:xitrk}
\xi(t)=K^{ab}(t)n_a(t)n_b(t)\equiv K^{nn}(t),
\end{equation}
where $n_a(t)=q_{ab}(t)\dot\ell^b_t/\|\dot\ell_t\|$ with $q_{ab}=\delta_{ij}e^i_ae^j_b$ being the spatial metric.
The assumption \eqref{eq:assumeK} leads to
\begin{equation}
\dot\ell_t^aK_a^i(t)=\dot\ell_t^aK_a{}^b(t)e_b^i(t)=\xi(t)\dot\ell^b_te_b^i(t).
\end{equation}
As a consequence, we get
\begin{equation}\label{eq:GKGconstant}
G_{\ell_t} \left(\dot\ell^a_tK_a(t)\right) G_{\ell_t}^{-1}=\xi(t)\dot\ell^a_0e_a^i(\ell_0)\tau_i.
\end{equation}
This equation, together with \eqref{eq:equV}, results in $V_\ell\equiv V_{\ell_1}$ as follows
\begin{equation}\label{eq:expressionV}
V_\ell=\exp(\beta\left(\int_\ell\xi(s)\dd s\right)\frac{\dot\ell^a_0e_a^i(\ell_0)}{\|\dot\ell_0\|}\tau_i ),
\end{equation}
where we have change the variable of integration to the curve length $s$.  Combining the equations \eqref{eq:hdecom}, \eqref{eq:expressionG} and \eqref{eq:expressionV}, we get
\begin{equation}\label{eq:classicalholonomy}
\begin{aligned}
h_\ell=\exp(\beta\left(\int_\ell\xi(s)\dd s\right)\frac{\dot\ell^a_0e_a^i(\ell_0)}{\|\dot\ell_0\|}\tau_i )n_s e^{\theta_\ell\tau_3} n_t^{-1}=n_s\exp(\left[\theta_\ell-\beta\left(\int_\ell\xi(s)\dd s\right)\right]\tau_3)  n_t^{-1}.
\end{aligned}
\end{equation}
This expression for $h_\ell$ ensures 
\begin{equation}\label{eq:holoelltangent}
h_\ell\, e_a^i(\ell_1)\dot\ell^a_1\tau_i\, h_\ell^{-1}=e_a^i(\ell_0)\dot\ell^a_0\tau_i,
\end{equation}
which shares a similar form as \eqref{eq:fluxrelatedbyholonomy}, but has a different sign on the right hand side. 

Now let us come to the fluxes $p^i_{t,\ell}$ and $p^i_{s,\ell}$ associated with a surface $S_\ell$ intersecting $\ell$ at $x_o=\ell_{t_o}$. We introduce $h_{\ell[0,t_o]}$ as the holonomy along the segment $[\ell_0,x_o]\subset\ell$, and $h_{\ell[t_o,1]}$ as the one along $[x_o,\ell_1]\subset \ell$. By definition of the fluxes in \eqref{eq:flux}, we have
\begin{equation}
\begin{aligned}
h_{\ell[0,t_o]}^{-1} (p^i_{s,\ell}\tau_i) h_{\ell[0,t_o]}=-h_{\ell[t_o,1]} ( p^i_{t,\ell}\tau_i) h_{\ell[t_o,1]}^{-1}=\frac{-1}{\kappa\beta }\int_{S_\ell}  h(\rho_{x_o}(\sigma) ) E^c(\sigma) n_c(\sigma) h(\rho_{x_o}(\sigma)^{-1})\dd^2 \sigma
\end{aligned}
\end{equation} 
where $\rho_{x_o}(\sigma)$ is the path starting at $x_o$ and running in $S_\ell$ till $\sigma\in S_\ell$.  

To align with our discussion in \eqref{app:coherentstate}, we introduce $p_\ell$ as we did in \eqref{eq:decompose}. Specifically, $p_\ell\in\mathbb R$  satisfies
\begin{equation}
(p_\ell)^2=\|\vec p_{s,\ell}\|^2=\|\vec p_{t,\ell}\|^2
\end{equation}
Let us choose $S_\ell$ appropriately  to make  the following relation holds
\begin{equation}\label{eq:fluxet}
\frac{-1}{\kappa\beta }\int_{S_\ell}  h(\rho_{x_o}(\sigma) ) E^c(\sigma) n_c(\sigma) h(\rho_{x_o}(\sigma)^{-1})\dd^2 \sigma
=-\sgn(e)  |p_\ell | \frac{\dot\ell^a_{t_0}e_a^i(x_o)\tau_i}{\|\dot\ell_{t_o}\|}.
\end{equation}
Indeed, \eqref{eq:fluxet} is inspired by the facts that 
for a general $S_\ell$, one has
\begin{equation}
\frac{-1}{\kappa\beta }\int_{S_\ell}  h(\rho_{x_o}(\sigma) ) E^c(\sigma) n_c(\sigma) h(\rho_{x_o}(\sigma)^{-1})\dd^2 \sigma\cong \sgn(e) |p_\ell | \frac{n_c(x_o)e^c_i(x_o)\tau^i}{\|n(x_o)\|}
\end{equation}
and that one can choose a $S_\ell$ ensuring
\begin{equation}
q^{ab}(x_o)n_b(x_o) \propto \frac{\dot\ell^a_{t_o}}{\|\dot\ell_{t_o}\|}.
\end{equation}
Then, we get
\begin{equation}
p^i_{s,\ell}\tau_i=-h_{\ell[0,t_o]} \left( \sgn(e) |p_\ell|\frac{ \dot\ell^a_{t_o}e_a^i(x_o)\tau_i}{\|\dot\ell_{x_o}\|}\right)h_{\ell[0,t_o]}^{-1}=- \sgn(e) |p_\ell| \frac{\dot\ell^i_0 e^i_a(\ell_0)\tau_i}{\|\dot\ell_0\|},
\end{equation}
and 
\begin{equation}
p^i_{t,e}\tau_i=h_{\ell[t_o,1]}^{-1} \left( \sgn(e) |p_\ell |\frac{ \dot\gamma^a_{t_o}e_a^i(x_o)\tau_i}{\|\dot\ell_{t_o}\|}\right)h_{\ell[t_o,1]}=\sgn(e)  |p_\ell| \frac{\dot\ell^a_1e_a^i(\ell_1)\tau_i}{\|\dot\ell_1\|}.
\end{equation}
Define $n(\vec p_{s,\ell}/p_\ell)$ and $n(-\vec p_{t,\ell}/p_\ell)\in \sut$ by
\begin{equation}
\begin{aligned}
n(\vec p_{s,\ell}/p_\ell) \tau_3 n(\vec p_{s,\ell}/p_\ell)^{-1}&= \frac{\ p_{s,\ell}^i\tau_i}{p_\ell},\\
n(-\vec p_{t,\ell}/p_\ell) \tau_3 n(-\vec p_{t,\ell}/p_\ell)^{-1}&=-\frac{p_{t,\ell}^i\tau_i}{p_\ell}.
\end{aligned}
\end{equation}
We get
\begin{equation}
\begin{aligned}
n(\vec p_{s,\ell}/p_\ell) \tau_3 n(\vec p_{s,\ell}/p_\ell)^{-1}&=-\sgn(e)\sgn(p_\ell) \frac{\dot\ell_0^ae_a^i(\ell_0)\tau_i}{\|\dot\ell_0\|},\\
n(-\vec p_{t,\ell}/p_\ell)  \tau_3 n(-\vec p_{t,\ell}/p_\ell) ^{-1}&=-\sgn(e)\sgn(p_\ell)\frac{\dot\ell_1^a e_a^i(\ell_1)\tau_i}{\|\dot\ell_1\|},
\end{aligned}
\end{equation}
leading to 
\begin{equation}
\begin{aligned}
h_\ell =n(\vec p_{s,\ell}/p_\ell)  \exp(\left[\tilde \theta_\ell-\beta\,\sgn(e)\sgn(p_\ell) \left(\int_\ell\xi(s)\dd s\right)\right]\tau_3) n(-\vec p_{t,\ell}/p_\ell) ^{-1},
\end{aligned}
\end{equation}
where \eqref{eq:classicalholonomy} is applied and $\tilde\theta_\ell$ is given by
\begin{equation}
n_s e^{\theta_\ell \tau_3}n_t^{-1}=n(\vec p_{s,\ell}/p_\ell) \exp(\tilde \theta_\ell \tau_3) n(-\vec p_{t,\ell}/p_\ell) ^{-1}.
\end{equation}
In this expression, $\tilde \theta_\ell$ has the geometric meaning of the twist angle (see, e.g., \cite{Rovelli:2010km} for the definition), and $\int_\ell\xi(s)\dd s$ is related to the extrinsic curvature.

Due to \eqref{eq:xitrk}, the integration for $\xi(s)$ can be alternatively rewritten as
\begin{equation}
\int_\ell \xi(s)\dd s=\int_\ell K^{nn}(s)\dd s.
\end{equation}
Given a cylinder $\mathscr{C}$ surrounding the link $\ell$, we could foliate $\mathscr{C}$ by slices perpendicularly intersecting $\ell$. Let $A_s$ denote the slice located at $\ell_s$, and $n_s^a$, the unit normal of $A_s$. With the orientation of $A_s$ chosen to be compatible with $\ell$, we have
\begin{equation}\label{eq:xiK}
\int_\ell \xi(s)\dd s=\int_\ell \left[\lim_{A_s\to 0}\left(\int_{A_s}\sqrt{q}n_s^a\epsilon_{abc}\dd x^b\dd x^c\right)^{-1}\int_{A_s}\sqrt{q} K^{nn} n_s^a\epsilon_{abc}\dd x^b\dd x^c\right]\dd s.
\end{equation}
where $q$ is the determinant of the metric $q_{ab}$. 
If the cylinder is choose such that the area of $A_s$ are the same for all $s$, \eqref{eq:xiK} becomes 
\begin{equation}
\int_\ell \xi(s)\dd s=\lim_{\|A\|\to 0}\|A\|^{-1}\int_{\mathscr{C}} \sqrt{q} K^{nn}\dd ^3 x,
\end{equation}
where $\|A\|=\int_{A_s}\sqrt{q}n_s^a\epsilon_{abc}\dd x^b\dd x^c$ denotes the area of arbitrary $A_s$.

\subsection{Geometric meaning of $\int_\ell \xi(s)\dd s$}
Embed our spatial manifold $\Sigma$ into a spacetime $(M,g_{\mu\nu})$. Let $n_\mu$ be the unit conormal field of $\Sigma$, i.e., $n_\mu n^\mu=-1$. 
We have
\begin{equation}\label{eq:Kmunu}
\begin{aligned}
K_{\mu\nu}=q^\sigma_\mu\nabla_\sigma n_\nu
\end{aligned}
\end{equation}
where $q^\nu_\mu=\delta^\nu_\mu+n_\mu n^\nu$ is the projection that projects tensors to their spatial part.   

Given a link $\ell$ lying in the spatial manifold $\Sigma$, we consider $\mathfrak g^\mu{}_\nu\in$GL(3) as the parallel transport of the Levi-Civita connection $\Gamma^\mu_{\sigma\nu}$ along $\ell$. Namely, $\mathfrak g^\mu{}_\nu(\ell_t)$ is the solution to
\begin{equation}\label{eq:holonomyGamma}
\partial_t \mathfrak g(\ell_t)^\mu{}_\nu=\mathfrak  g(\ell_t)^\mu{}_\alpha\Gamma^\alpha_{\sigma\nu}(\ell_t)\dot\ell_t^\sigma.
\end{equation}
This equation together with \eqref{eq:Kmunu} leads to
\begin{equation}\label{eq:extrinsicCurvature}
K_{t\nu}(\ell_t)\equiv \dot\ell_t^\mu K_{\mu\nu}(\ell_t)=\partial_t\left(n_\mu(\ell_t)(\mathfrak  g(\ell_t)^{-1})^\mu{}_\alpha \right)\mathfrak  g(\ell_t)^\alpha{}_\nu.
\end{equation}
Let $e^\mu_I$ be the vielbein field in $M$ which takes $e^\mu_0= -n^\mu$ and $e^\mu{}_i=\delta^\mu_aE^a_i/\sqrt{|\det E|}$ for $i=1,2,3$ on the hypersurface $\Sigma$. The spin connection associated to $e^\mu_I$ is given by
\begin{equation}
\nabla_\mu e^\nu_I=:\omega_\mu ^J{}_Ie^\nu_J.
\end{equation}
The holonomy $\mathfrak p^I{}_J$ of the spin connection $\omega_\mu^I{}_J$ is given by
\begin{equation}\label{eq:holonomyP}
\partial_t \mathfrak p(\ell_t)^I{}_J=\mathfrak p(\ell_t)^I{}_K\omega_\mu^K{}_J(\ell_t)\dot\ell^\mu.
\end{equation}
Along the link $\ell$, we have
\begin{equation}
e^\nu_J(\ell_t)=(\mathfrak g(\ell_t)^{-1})^\nu{}_\mu\mathfrak p(\ell_t)^I{}_J e^\mu_I(\ell_0). 
\end{equation}
This equation motivates us to contract $e^\nu_I$ and  $\mathfrak p^{-1}$ with the both sides of \eqref{eq:extrinsicCurvature}, resulting in
\begin{equation}\label{eq:KI4D}
K_{t\nu}(\ell_t) e^\nu_I(\ell_t) (\mathfrak p(\ell_t)^{-1})^I{}_J=\partial_t\Big(n_\nu(\ell_t)(\mathfrak g(\ell_t)^{-1})^\nu{}_\alpha e^\alpha_J(\ell_0)\Big).
\end{equation} 
It is useful to recognize the geometric meaning of the right hand side. By definition, $n_\nu(\ell_t)(\mathfrak g(\ell_t)^{-1})^\nu{}_\alpha=\tilde n_\alpha(t)$ is the vector obtained by parallel transporting $n_\mu(\ell_t)$ to the starting point $\ell_0$ of $\ell$. Then, $\tilde n_\alpha(t) e^\alpha_J(\ell_0)$ is the inner product of $\tilde n_\alpha(t)$ with the veilbein at $\ell_0$, giving the information of the direction of $\tilde n(t)$ with respect  to $e^\alpha_J(\ell_0)$. Due to $e^\alpha_0(\ell_0)=n^\alpha(\ell_0)$, one has $$\left|\tilde n_\alpha(t)e^\alpha_0(\ell_0)\right|=\cosh\Big(\text{the dihedral angle between }\tilde n_\alpha(t) \text{ and } n^\alpha(\ell_0)\Big).$$

In the formula \eqref{eq:KI4D}, the holonomy $\mathfrak p^{-1}$ is the one of the 4-D spin connection. To interpret the geometry meaning of $\int_\ell\xi(s)\dd s$, the 3-D connection and extrinsic curvature should be incorporated. 

By the definition of $e^\nu_I$, we get
\begin{equation}
e_I^\nu n_\nu=:n_I=(1,0,0,0). 
\end{equation}
Let $q^I_J$ be the projection operator
\begin{equation}
q^I_J=\delta^I_J+n^I n_J.
\end{equation}
We introduce
\begin{equation}
\begin{aligned}
\tilde\omega_\mu^I{}_J&:=\omega_\mu^K{}_Lq^L_Jq^I_K,\\ 
\tilde K_\mu^I{}_J&=\omega_\mu^K{}_L\left(q^L_Jn^In_K+n^Ln_J q^I_K\right).
\end{aligned}
\end{equation}
so that 
\begin{equation}
\omega_\mu^I{}_J=\tilde \omega_\mu^I{}_J-\tilde K_\mu^I{}_J,\\
\end{equation}
Substituting this equation into \eqref{eq:holonomyP}, we have
\begin{equation}
\begin{aligned}
\mathfrak p^I{}_J&=\mathfrak p_1^I{}_K\mathfrak p_2^K{}_J,\\
\partial_t \mathfrak p_2(\ell_t)^I{}_J&=\mathfrak p_2(\ell_t)^I{}_K\tilde\omega_\mu^K{}_J(\ell_t)\dot\ell^\mu_t,\\
\partial_t \mathfrak p_1(\ell_t)^I{}_J&=-\mathfrak p_1(\ell_t)^I{}_K (\mathfrak p_2(\ell_t)\tilde K_\mu(\ell_t)\dot\ell^\mu_t\mathfrak  p_2(\ell_t)^{-1})^K{}_J. 
\end{aligned}
\end{equation}

The following proposition which will be introduced without proof, turns out to be helpful for our further computation. 
\begin{pro}
Given a $\sltc$ holonomy $g$ with respect to some $\mathfrak{sl}(2,\mathbb C)$ connection $A$, i.e., 
\begin{equation}
\partial_t g(\ell_t)=g(\ell_t) A_\mu(\ell_t)\dot\ell^\mu_t,
\end{equation}
the $SO(1,3)$ correspondence of $g$, namely the representation $g^I{}_J$ of $g$ on $\mathbb R^{1,3}$,  will be the holonomy of the corresponding $\mathfrak{so}(1,3)$ connection $A^I{}_J$, i.e. 
\begin{equation}
\partial_t g(\ell_t)^I{}_J=g(\ell_t)^I{}_KA_\mu^K{}_J(\ell_t)\dot\ell^\mu_t.
\end{equation}
\end{pro}

Let $\sigma_I=(\mathbbm{1}_2,\vec\sigma)$ and $\tilde \sigma_I=(-\mathbbm{1}_2,\vec\sigma)$ be the Pauli 4-vectors.  The $\mathfrak{sl}(2,\mathbb C)$ connection $\tilde\omega_\mu$ corresponding to $\tilde\omega_\mu^I{}_J$ is given by 
\begin{equation}
\tilde \omega_\mu=\frac{1}{4}\tilde \omega_\mu^{IJ}\sigma_I\tilde \sigma_J. 
\end{equation}
Then, we could construct the $\sltc$ holonomy  corresponding  to $\mathfrak p_2^I{}_J$, which is denoted by $\mathfrak p_2$. Indeed, $\mathfrak p_2$ is just holonomy $G_{\ell_t}$ given in \eqref{eq:spinholo}, due to
\begin{equation}\label{eq:identifyp2G}
\dot\ell^\mu_t\tilde \omega_\mu=-\frac{1}{2}\dot\ell^\mu_t\omega_\mu^{mn} \epsilon_{mn}{}^k\tau_k=-\frac{1}{2}\dot\ell^a_t\Gamma_a^{mn} \epsilon_{mn}{}^k\tau_k=\dot\ell^a_t\Gamma_a^k\tau_k.
\end{equation}
The $\mathfrak{sl}(2,\mathbb C)$ correspondence of $\tilde K_\mu^I{}_J$ is
\begin{equation}
\begin{aligned}
\tilde K_\mu=&\frac{1}{4}\tilde K_\mu^{IJ}\sigma_I\tilde\sigma_J=\frac{1}{2}\omega^{l0}_\mu\sigma_l.
\end{aligned}
\end{equation}
Thus, the $\sltc$ holonomy corresponding to $\mathfrak p_1$ satisfies
\begin{equation}\label{eq:p1}
\partial_t \mathfrak p_1(\ell_t)=-\mathfrak p_1(\ell_t) (\mathfrak p_2(\ell_t)\tilde K_\mu(\ell_t)\dot\ell^\mu \mathfrak p_2(\ell_t)^{-1})=-\mathfrak p_1(\ell_t)\left[\mathfrak p_2 (\ell_t)\left(\frac{1}{2}\omega^{l0}_t(\ell_t)\sigma_l\right) \mathfrak p_2(\ell_t)^{-1}\right].
\end{equation}

Turning to the left-hand side of \eqref{eq:KI4D}, we get
\begin{equation}
K_{t\nu}e^\nu_I=-\dot\ell_t^\mu n_\nu\nabla_\mu e^\nu_I=-\omega_t^0{}_I
\end{equation}
leading to
\begin{equation}\label{eq:kesigmafinal}
K_{t\nu}e^\nu_I(\mathfrak p^{-1})^I{}_J\tilde\sigma^J=\mathfrak  p_1^{-1\dagger} \mathfrak p_2\omega_t^{i0} \sigma_i \mathfrak p_2^{-1}  \mathfrak p_1^{-1}= \partial_t(\mathfrak p_1^{-1\dagger}\mathfrak p_1^{-1}).
\end{equation}
where \eqref{eq:p1} is applied to get the last equality. Combining the result \eqref{eq:kesigmafinal} with \eqref{eq:KI4D} results in
\begin{equation}
\partial_t(\mathfrak p_1(\ell_t)^{-1\dagger}\mathfrak p_1(\ell_t)^{-1})=\partial_t(n_\nu(\ell_t)(\mathfrak g(\ell_t)^{-1})^\nu{}_\alpha e^\alpha_J(\ell_0)\tilde\sigma^J)
\end{equation} 
implying 
\begin{equation}\label{eq:ppdagger}
\mathfrak p_1(\ell_t)^{-1\dagger}\mathfrak p_1(\ell_t)^{-1}=n_\nu(\ell_t)(\mathfrak g(\ell_t)^{-1})^\nu{}_\alpha e^\alpha_J(\ell_0)\tilde\sigma^J
\end{equation} 
where we have considered the initial condition that both sides takes value $\mathbbm{1}_2$ at $\ell_0$. This equation implies
\begin{equation}\label{eq:ppinverse}
\mathfrak p_1(\ell_t)\mathfrak p_1(\ell_t)^\dagger=-\tilde n^\alpha(t)e_\alpha^J(\ell_0)\sigma_J
\end{equation}
where we introduced the convention $\tilde n_\alpha(t)=n_\nu(\ell_t)(\mathfrak g(\ell_t)^{-1})^\nu{}_\alpha$. Since $\mathfrak p_1(\ell_t)\mathfrak p_1(\ell_t)^\dagger$ is purely boost,  \eqref{eq:ppinverse} results in $\tilde n^\alpha(t)e_\alpha^0(\ell_0)=\tilde n^\alpha(t) n_\alpha(\ell_0) <0$ and
\begin{equation}\label{eq:generalp}
\mathfrak p_1(\ell_t)=\exp(-\frac{1}{2}\arccosh(-n_\alpha(\ell_0) \tilde n^\alpha(t)) \frac{\tilde n^\alpha(t)e_\alpha^i(\ell_0)\sigma_i}{\|\tilde n^\alpha(t)e_\alpha^i(\ell_0)\|}) u(t)
\end{equation}
for some element $u(t)\in \sut$. 

To relate $\mathfrak p_1$ to $V_{\ell_t}$ given in \eqref{eq:equV}, we need the relation between $\tilde K_\mu\dot\ell_t^\mu$ and the extrinsic curvature $K_a^i\dot\ell^a_t$
\begin{equation}
\dot\ell_t^\mu\tilde K_\mu=\frac{1}{2}\dot\ell_t^\mu\omega_\mu^{l0}\sigma_l=i\dot\ell_t^aK_a^l\tau_l.
\end{equation}
Since we have identified $\mathfrak p_2$ with the holonomy $G_{\ell_t}$, the equation for $\mathfrak p_1$ now becomes 
\begin{equation}
\partial_t\mathfrak p_1(\ell_t)=-i \mathfrak p_1(\ell_t) G_{\ell_t} K_a(\ell_t)\dot\ell_t^aG_{\ell_t}^{-1}.
\end{equation}
Focusing on the special case considered in \eqref{eq:assumeK} and applying \eqref{eq:GKGconstant}, we could get 
\begin{equation}\label{eq:expressionV0}
\mathfrak p_1(\ell_t)=\exp(-\left(\int_{\ell[0,t]}\xi(s)\dd s\right)\frac{\dot\ell^a_0e_a^i(\ell_0)}{\|\dot\ell_0\|}\frac{\sigma_i}{2} ),
\end{equation}
where $\ell[0,t]\subset \ell$ denotes the segment of $\ell$ with the parameter taking values in $[0,t]$.  Comparing this result with \eqref{eq:generalp}, we get $u(t)=1$ and
\begin{equation}\label{eq:relationp}
\begin{aligned}
&\int_{\ell[0,t]}\xi(s)\dd s= \arccosh(-n_\alpha(\ell_0) \tilde n^\alpha(t)) ,\quad \frac{\dot\ell^a_0e_a^i(\ell_0)}{\|\dot\ell_0\|}= \frac{\tilde n^\alpha(t)e_\alpha^i(\ell_0)}{\|\tilde n^\alpha(t)e_\alpha^i(\ell_0)\|},\\
\text{ or    }&\\
&\int_{\ell[0,t]}\xi(s)\dd s=-\arccosh(-n_\alpha(\ell_0) \tilde n^\alpha(t)),\quad \frac{\dot\ell^a_0e_a^i(\ell_0)}{\|\dot\ell_0\|}=-\frac{\tilde n^\alpha(t)e_\alpha^i(\ell_0)}{\|\tilde n^\alpha(t)e_\alpha^i(\ell_0)\|}.
\end{aligned}
\end{equation}
Here we considered the fact that the vectors involved are unit vectors. By definition of the dihedral angle \cite{Barrett:2009mw}, \eqref{eq:relationp} implies that $\int_\ell\xi(s)\dd s$ is the signed dihedral  angle  between $n_\alpha(\ell_0)$ and $\tilde n^\alpha(1)$, i.e., 
$$\int_\ell\xi(s)\dd s=\mp\left(-\arccosh(-n_\alpha(\ell_0) \tilde n^\alpha(1))\right)$$
where the sign depends on whether the hypersurface $\Sigma$ is `convex', i.e., $\dot\ell^a_0e_a^i(\ell_0)\propto \tilde n^\alpha(t)e_\alpha^i(\ell_0)$, or not convex, i.e., $\dot\ell^a_0e_a^i(\ell_0)\propto -\tilde n^\alpha(t)e_\alpha^i(\ell_0)$. In the Regge calculus, the angle $-\arccosh(-n_\alpha(\ell_0) \tilde n^\alpha(1))$ becomes the dihedral angle between the 4-D normals of the two adjacent tetrahedra dual to the points $\ell_0$ and $\ell_1$. Using $\Theta_\ell$ to denote this angle, we have
\begin{equation}\label{eq:intKandDihedral}
\int_\ell\xi(s)\dd s=\left\{
\begin{array}{cl}
-\Theta_\ell,& \Sigma \text{ is convex,}\\
\Theta_\ell,& \Sigma \text{ is not convex.}
\end{array}
\right.
\end{equation}

\subsection{Compatibility between the spin foam action and the  coherent state }\label{eq:appA7}
Consider a 2-complex $\mathcal K$ and the spin foam model on it. Given a non-degenerate critical point, one could construct a flat simplical geometry dual to the 2-complex \cite{Han:2011re}. 
In this dual simplical geometry, each 4-simplex is convex.  Given a boundary triangle $f$ surrounded by tetrahedra  $\tau_1,\cdots,\tau_n$ in the dual simplicial geometry, the spin foam action associated with $f$ takes the following value at the critical point \cite{Han:2011re}, 
\begin{equation}\label{eq:sfactionflat}
S_f=-i\epsilon\, \sgn(V_4)\beta j_f\sum_{k}\Theta_{k,k+1}-ij_f \theta_f,
\end{equation}
where $\epsilon=\pm$ and $\sgn(V_4)=\pm$ are some sign factor, and $\Theta_{k,k+1}$ are the dihedral angle between $\tau_k$ and $\tau_{k+1}$.
Since $\tau_k$ and $\tau_{k+1}$ share a 4-simplex denoted by $T_{k,k+1}$, the dihedral angle between them can be well-defined as the Lorentzian angle between the outwards normals of $\tau_k$ and $\tau_{k+1}$. In addition, since the dual geometry is flat, we could also calculate the dihedral angle, denoted by $\Theta_f$, between the tetrahedral $\tau_1$ and $\tau_n$, by treating $\tau_1\cup\tau_n$ as the boundary of the simplicial complex $T_{1,2}\cup T_{2,3}\cup\cdots\cup T_{n-1,n}$. Let us investigate the relation between the term $\sum_{k}\Theta_{k,k+1}$ in \eqref{eq:sfactionflat} and the dihedral angle $\Theta_f$ between $\tau_1$ and $\tau_n$. 

\begin{figure}
\includegraphics[width=0.5\textwidth]{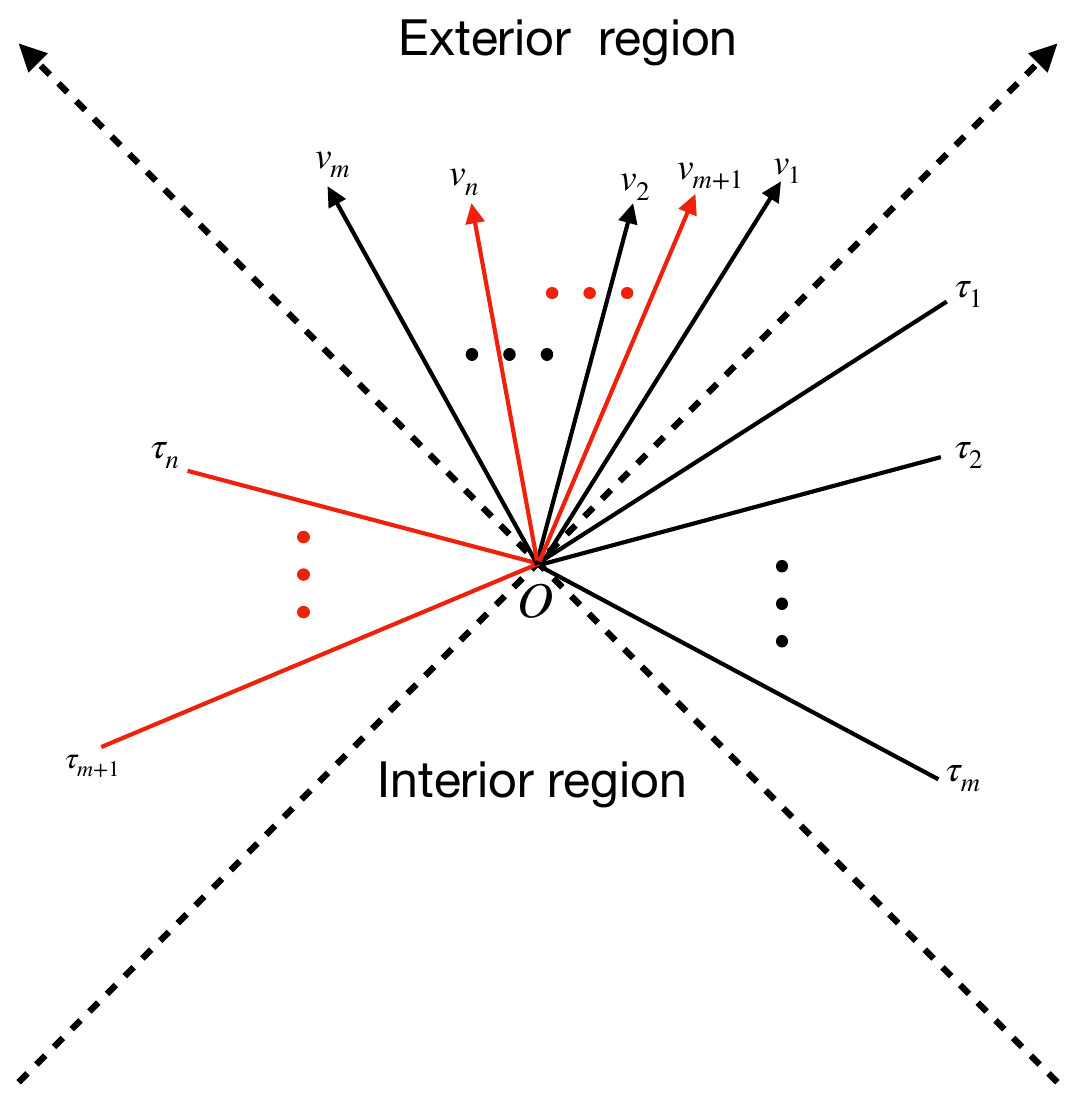}
\caption{The intersection of the normal plane with the simplicial complex. The lines without arrow are the intersection of the tetrahedra with the normal plane and the arrowed lines are the corresponding future pointing normals.}\label{fig:dihedrals}
\end{figure}

Since the dual geometry is flat, the normals of the tetrahedra are coplanar. The plane containing all of the normals are referred to as the normal plane. The normal plane intersects the triangle $f$ at a point, denoted by $O$. When this normal plane intersects the tetrahedron $\tau_k$, it generates a line originating from $O$. In Fig. \ref{fig:dihedrals}, we illustrate a general situation of the intersection between the normal plane and the simplicial complex. Here, we assume that the intersection fills the region beneath $\tau_1\cup\tau_k$.

The direction of the line corresponding to the tetrahedron $\tau_i$ is specified by the coordinates $(t_i, x_i)$, given by
\begin{equation}
(t_i,x_i)=\left\{
\begin{array}{lc}
(\sinh(\theta_i),\cosh(\theta_i)),& i\leq m\\
(\sinh(\theta_i),-\cosh(\theta_i)),& i> m\\
\end{array}
\right.
\end{equation}
Thus the future pointing unit normal vector of $\tau_i$ is 
\begin{equation}
v_i=\left\{
\begin{array}{lc}
(\cosh(\theta_i),\sinh(\theta_i)),& i\leq m,\\
(\cosh(\theta_i),-\sinh(\theta_i)),& i> m.\\
\end{array}
\right.
\end{equation}
For the tetrahedra $\tau_k$ and $\tau_{k+1}$ contained in $T_{k,k+1}$, their dihedral angle $\Theta_{k,k+1}$ is given by 
\begin{equation}
\Theta_{k,k+1}=\left\{
\begin{array}{lc}
\arccosh(-v_k\cdot\eta\cdot v_{k+1})=\left|\theta_n-\theta_{k+1}\right|,& k\neq m,\\
-\arccosh(-v_{m}\cdot\eta\cdot v_{m+1})=-\left|\theta_m+\theta_{m+1}\right|,& k= m.
\end{array}
\right.
\end{equation}
Since $\theta_1>\theta_2>\cdots>\theta_m$ and $\theta_n>\theta_{n-1}>\cdots>\theta_{m+1}$, we have
\begin{equation}
\Theta_{k,k+1}=\left\{
\begin{array}{lc}
\theta_{k}-\theta_{k+1},& k< m,\\
-\left|\theta_m+\theta_{m+1}\right|,& n= m,\\
\theta_{k+1}-\theta_k,& k> m,\\
\end{array}
\right.
\end{equation}
leading to
\begin{equation}\label{eq:sumoverdihedral}
\sum_{\tau_k}\Theta_{k,k+1}=(\theta_1+\theta_n)-\left|\theta_m+\theta_{m+1}\right|-(\theta_m+\theta_{m+1}).
\end{equation}
Since the 4-simplex $T_{m,m+1}$ shared by $\tau_m$ and $\tau_{m+1}$ is convex, we have 
$\theta_m+\theta_{m+1}<0$. Therefore, \eqref{eq:sumoverdihedral} becomes
\begin{equation}
\sum_k\Theta_{k,k+1}=\theta_1+\theta_n=\left\{
\begin{array}{cc}
-\arccos(-v_1\cdot\eta\cdot v_n)=\Theta_f,&\theta_1+\theta_n\leq 0,\\
\arccos(-v_1\cdot\eta\cdot v_n)=-\Theta_f,&\theta_1+\theta_n>0.
\end{array}
\right.
\end{equation}
Note that $\theta_1+\theta_n\leq 0$ implies that $\tau_1\cup\tau_n$ is convex, and $\theta_1+\theta_n> 0$, implies that $\tau_1\cup\tau_n$  is not convex. We thus get
\begin{equation}
\sum_k\Theta_{k,k+1}=\left\{
\begin{array}{cl}
\Theta_f,& \tau_1\cup\tau_n \text{ is convex},\\
-\Theta_f,&\tau_1\cup\tau_n \text{ is not convex}.
\end{array}
\right.
\end{equation}

In addition, $\tau_1\cup\tau_n$, as the boundary of the simplicial complex, carries the extrinsic curvature of the form given in \eqref{eq:assumeK}, where $\xi(t)$ becomes a distribution and the curve $\ell$ is considered to be the straight line perpendicular to the triangle $f$. In accordance with  \eqref{eq:intKandDihedral}, taking into account the geometric interpretation of $\Theta_\ell$ within that expression, we get
\begin{equation}\label{eq:relationxiTheta}
\int_\ell\xi(s)\dd s=-\sum_k\Theta_{k,k+1}.
\end{equation}

\section{Proof on the extended spin foam data solving the critical equation}\label{app:proofExtended}
According to \eqref{eq:pxiBp}, $\partial_+B$ carries the same boundary data as in $\partial_-B$. The  spin foam data in $B_-$, which satisfies the boundary condition on $\partial_-B$, is replicated to obtain the data in $B_+$. Consequently, the extended data in the entire region $B$ remains compatible with the boundary condition on $\partial B$. Moreover, the replication ensures that the extended data solve the equations associated with all $v, e, f$ except for those located on the transition surface $\mathcal{T}$. These $v, e, f$ that are not located on the transition surface $\mathcal{T}$ will be referred to as the interior objects. According to this discussion, we are allowed to focus only on the equations that involve objects crossing the transition surface $\mathcal{T}$.

At first, $\partial_{j_f}S=0$ in \eqref{eq:SFeq1} implies $\mathrm{Re}(S_f)=0$, where $\mathrm{Re}(S_f)$ denotes the real part of $S_f$. 
As shown in literature \cite{Han:2013gna}, $\mathrm{Re}(S_f)=0$ is equivalent to
\begin{equation}\label{eq:criticaleq1}
\frac{Z_{s_ee f}}{\|Z_{s_eef}\|}=e^{i\alpha^f_{e}}\frac{Z_{t_e ef}}{\|Z_{t_eef}\|},
\end{equation}
for some real number $\alpha^f_{e}$, implying that  $Z_{vef}$ is parallel to $Z_{v'ef}$.
Exploiting this result, we can simplify the last two equations in \eqref{eq:SFeq1} to (see, e.g., \cite{Han:2013gna} for the derivation)
\begin{equation}\label{eq:criticaleqs}
\begin{aligned}
&\frac{g_{ve}Z_{vef}}{\langle Z_{vef},Z_{vef}\rangle}=\frac{g_{ve'}Z_{ve'f}}{\langle Z_{ve'f},Z_{ve'f}\rangle},\\
&\sum_{f \text{ at } e}j_f\epsilon_{ef}(v)\frac{\langle Z_{vef},\vec\sigma Z_{vef}\rangle}{\langle Z_{vef},Z_{vef}\rangle}=0,
\end{aligned}
\end{equation}
Here, the first equation involves two edges $e$ and $e'$ satisfying $e\cap e'=v$ and $e,e'\in\partial f$. In the last equation, $\epsilon_{ef}(v)$ is define by 
\begin{equation}\label{eq:orientationep}
\epsilon_{ef}(v)=\left\{
\begin{aligned}
1&\quad  v=t_e, \\
-1&\quad v=s_e.
\end{aligned}
\right.
\end{equation}
Since the equations in \eqref{eq:criticaleqs} do not involve objects crossing $\mathcal{T}$, their validity can be directly derived from the fact that the extended data solve the equations associated with the interior objects. Thus, we only need to prove the validity of $\partial_{j_f}S=0$ for all faces $f$ crossing $\mathcal T$.

By definition \eqref{eq:action}, we have
\begin{equation}\label{eq:partialj}
\begin{aligned}
\partial_{j_f}S=&\sum_{e\in\partial f}\Bigg[\ln(\frac{\langle Z_{s_e e f},Z_{t_eef}\rangle^2}{\langle Z_{s_eef},Z_{s_e ef} \rangle\langle Z_{t_e ef},Z_{t_e ef}\rangle })+i\gamma \ln( \frac{ \langle Z_{s_eef},Z_{s_e ef}\rangle}{\langle Z_{t_e ef},Z_{t_e ef}\rangle} )\Bigg].
\end{aligned}
\end{equation}
The edges $e\in\partial f$ comprise of those entirely lie in $B_\pm$ and these crossing $\mathcal T$. 
For edges that entirely lie in $B_-$, we apply \eqref{eq:criticaleq1} to get 
\begin{equation}\label{eq:equaforeinBm}
\begin{aligned}
\ln(\frac{\langle Z_{s_e e f},Z_{t_eef}\rangle^2}{\langle Z_{s_eef},Z_{s_e ef} \rangle\langle Z_{t_e ef},Z_{t_e ef}\rangle })=&\ln(e^{i\alpha_{e}^f })\\
\ln( \frac{ \langle Z_{s_eef},Z_{s_e ef}\rangle}{\langle Z_{t_e ef},Z_{t_e ef}\rangle} )=&\ln(\frac{\|Z_{s_e e f}\|^2}{\|Z_{t_e e f}\|^2}).
\end{aligned}
\end{equation}
For the edge $e'=\mathfrak r(e)$ that entirely lie in $B_+$, the compatibility of the orientation of the face $f$ leads to $s_{e'}=\mathfrak r(t_e)$ and $t_{e'}=\mathfrak r(t_e)$. This fact, together with \eqref{eq:extendingdata} and \eqref{eq:equaforeinBm}, results in 
\begin{equation}\label{eq:equaforeinBp}
\begin{aligned}
\ln(\frac{\langle Z_{s_{e'} e' f},Z_{t_{e'}e'f}\rangle^2}{\langle Z_{s_{e'}e'f},Z_{s_{e'} {e'}f} \rangle\langle Z_{t_{e'} {e'}f},Z_{t_{e'} {e'}f}\rangle })=&\ln(e^{-i\alpha_{e}^f })\\
\ln( \frac{ \langle Z_{s_{e'}e'f},Z_{s_{e'} e'f}\rangle}{\langle Z_{t_{e'} e'f},Z_{t_{e'} e'f}\rangle} )=&\ln(\frac{\|Z_{t_e e f}\|}{\|Z_{s_e e f}\|}). 
\end{aligned}
\end{equation}
As a consequence of \eqref{eq:equaforeinBm} and \eqref{eq:equaforeinBp}, for each edge $e_o$ lying entirely in $B_-$ we have
\begin{equation}\label{eq:critical1}
\begin{aligned}
&\sum_{e\in \{e_o,\mathfrak r(e_o)\}}\Bigg[\ln(\frac{\langle Z_{s_e e f},Z_{t_eef}\rangle^2}{\langle Z_{s_eef},Z_{s_e ef} \rangle\langle Z_{t_e ef},Z_{t_e ef}\rangle })+i\gamma \ln( \frac{ \langle Z_{s_eef},Z_{s_e ef}\rangle}{\langle Z_{t_e ef},Z_{t_e ef}\rangle} )\Bigg]=0. 
\end{aligned}
\end{equation}
Finally let us consider the edges $e$ crossing $\mathcal T$. The equation \eqref{eq:extendingdata} and the fact $e=\mathfrak r(e)^{-1}$ for such edges lead to
\begin{equation}\label{eq:critical2}
\begin{aligned}
\ln(\frac{\langle Z_{s_e e f},Z_{t_eef}\rangle^2}{\langle Z_{s_eef},Z_{s_e ef} \rangle\langle Z_{t_e ef},Z_{t_e ef}\rangle })+i\gamma \ln( \frac{ \langle Z_{s_eef},Z_{s_e ef}\rangle}{\langle Z_{t_e ef},Z_{t_e ef}\rangle} )=0. 
\end{aligned}
\end{equation}
Substituting \eqref{eq:critical1} and \eqref{eq:critical2} into \eqref{eq:partialj}, we finally get $\partial_{j_f}S=0$ for $f$ crossing $\mathcal T$.

\section{Derivative the critical values $\theta_1^{(c)}$ and $\theta_2^{(c)}$ given in \eqref{eq:criticaltheta}}\label{app:derivetheta12c}
Let us do our derivation in the flat geometry constructed using the data introduced at the end of Sec. \ref{sec:valuepxi}.

At first, we need all tetrahedra in $\mathfrak i(\mathcal T)$ are spacelike which can be ensure by satisfying the condition
\begin{equation}\label{eq:condition1}
\frac{t_3-t_1}{r_3-3r_1}=\frac{1}{3}-\epsilon,
\end{equation}
with $\epsilon>0$. Indeed, for $\epsilon\ll 1$, we can compute the dihedral angles between adjacent tetrahedra in $\mathfrak i(\mathcal T)$. They takes one of the two values
$\Delta_1$ and $\Delta_2$ given by 
\begin{equation}\label{eq:dihedralep2}
\begin{aligned}
\mathrm{cosh}(\Delta_1)= \frac{3 r_1^2+3 r_1 r_3-4 r_3^2}{2 \sqrt{3} \sqrt{\left| 3 r_1^4-4 r_3^2 r_1^2+r_3^4\right| }}+O(\epsilon)
\end{aligned}
\end{equation}
and 
\begin{equation}\label{eq:dihedralep1}
\mathrm{cosh}(\Delta_2)= \frac{\sqrt{\frac{2}{3}} r_3}{3 \sqrt{\left| 2 r_1^2-\frac{2 r_3^2}{3}\right| }}\frac{1}{\sqrt{\epsilon}}+O(\sqrt{\epsilon})
\end{equation}
According to this two equations, $\Delta_2$ approaches $\infty$ as $\epsilon=0$, implying that when $\epsilon=0$ one of the adjacent tetrahedra with dihedral angle $\Delta_2$ becomes null. 
Moreover, the boundary $\partial_-B_1$ and $\partial_-B_2$ are expected to be spacelike, leading to 
\begin{equation}\label{eq:condition2}
\frac{t_2-t_1}{r_2-3r_1}<\frac{1}{3},\quad \text{and}\quad \frac{t_3-t_2}{r_3-r_2}<\frac{1}{3}.
\end{equation}

Taking advantage of \eqref{eq:condition1} and \eqref{eq:condition2} and considering $t_1=0$ in the construction, we finally obtain 
\begin{equation}\label{eq:t3}
t_3=\left(\frac{1}{3}-\epsilon\right)(r_3-3r_1),\quad t_2=\frac{1}{3}\left(r_2-3r_1\right)-\delta\left(r_3-3r_1\right)
\end{equation}
with $\epsilon$ and $\delta$ satisfying
\begin{equation}\label{eq:inequality1}
0<\delta<\epsilon.
\end{equation}
Since $S_3$ has to be in the future of $S_2$, we need $t_3>t_2$, which leads to
\begin{equation}\label{eq:inequality2}
\epsilon-\delta <\frac{r_3-r_2}{3(r_3-3 r_1)}.
\end{equation}
Here, we use the fact that  $r_3>3r_1$ for a large black hole. 

According to \eqref{eq:a12aaa}, the values of $a_1$ and $a_2$ can be written in terms of $\delta$ and $\epsilon$ as 
\begin{equation}\label{eq:slopeepsilondelta}
\begin{aligned}
a_1=&\frac{1}{3}-\frac{2r_1/3+\delta(r_3-3r_1)}{r_2-r_1},\\
a_2=&\frac{1}{3}-\frac{(\epsilon-\delta)(r_3-3 r_1)}{r_3-r_2}.
\end{aligned}
\end{equation}
The values of $\theta_1$ and $\theta_2$, according to \eqref{eq:theta1} and \eqref{eq:theta2}, can also be related to $\delta$ and $\epsilon$. With this relation and the inequalities \eqref{eq:inequality1} and \eqref{eq:inequality2}, we finally get 
\begin{equation}\label{eq:conditions}
\begin{aligned}
\frac{2}{f(r_1)}-1<&\cosh (\theta_1)<\frac{2}{f(r_1)}\left(1-\left(\frac{r_2-3r_1}{3(r_2-r_1)}\right)^{2}\right)^{-1}-1\\
\frac{2}{f(r_3)}\left(1-\left(\frac{r_2-r_1}{3(r_3-r_1)}\right)^2\right)^{-1}-1<& \cosh (\theta_3)<\frac{9}{4f(r_3)}-1
\end{aligned}
\end{equation}
where we used $f(r_3),f(r_1)>0$.


%

\end{document}